\documentclass[longauth]{aa}  
\usepackage{graphicx}
\usepackage{xcolor}
\usepackage[varg]{txfonts}
\usepackage{natbib}
\usepackage{amssymb}
\usepackage{amsmath}
\usepackage[rightcaption]{sidecap}
\sidecaptionvpos{figure}{b}

\pdfoutput=1
\usepackage[colorlinks=true,linkcolor=bluea,allcolors=blue]{hyperref}

\begin{document}

   \title{Deciphering the imprint of AGN feedback in Seyfert galaxies:}
   \subtitle{Nuclear-scale molecular gas deficits}

 \author{S.~Garc\'{\i}a-Burillo\inst{1}
	   \and	
	    E.~K.~S.~Hicks\inst{2,3}
	   \and
	   A.~Alonso-Herrero\inst{4} 
	   \and
	   M.~Pereira-Santaella\inst{5}
	   \and
	   A.~Usero\inst{1}
	   \and
	   M. Querejeta\inst{1}
	    \and
	   O.~Gonz\'alez-Mart\'{\i}n\inst{6}	
	   \and
	   D.~Delaney\inst{2, 3}
	   \and	  
	   C.~Ramos Almeida\inst{7, 8}
	    \and
	    F.~Combes\inst{9} 
	    \and
	    D.~Angl\'es-Alc\'azar\inst{10, 11}
	    \and
	    A.~Audibert\inst{7,8}
	    \and
	    E.~Bellocchi\inst{12,13}
	    \and
	    R.~I.~Davies\inst{14}
	    \and
	    T.~A.~Davis\inst{15}
	    \and
	    J.~S.~Elford\inst{15}
	    \and
	    I.~Garc\'{\i}a-Bernete\inst{16}	    
	    \and
	    S.~H\"onig\inst{17}
	    \and
%
	   A.~Labiano\inst{18}
	   \and
	   M.~T.~Leist\inst{19}
	   \and
	   N.~A.~Levenson\inst{20}
	   \and
	   E.~L\'opez-Rodr\'{\i}guez\inst{21}	   
	   \and
	    J.~Mercedes-Feliz \inst{10}	   
	   \and
	   C.~Packham\inst{19, 22}
	  \and
	   C. Ricci\inst{23, 24}
	   \and
	   D.~J.~Rosario\inst{25}
	   \and
	   T.~Shimizu\inst{14}
	   \and
	   M.~Stalevski\inst{26, 27}
	   \and
	   L.~Zhang\inst{19}
	   }
   \institute{
     Observatorio Astron\'omico Nacional (OAN-IGN)-Observatorio de Madrid, Alfonso XII, 3, 28014-Madrid, Spain \email{s.gburillo@oan.es} 		  
        \and
     Department of Physics \& Astronomy, University of Alaska Anchorage, Anchorage, AK 99508-4664, USA
   \and
     Department of Physics, University of Alaska, Fairbanks, Alaska 99775-5920, USA   
     \and
       Centro de Astrobiolog\'{\i}a (CAB), CSIC-INTA, Camino Bajo del Castillo s/n, 28692 Villanueva de la Ca\~nada, Madrid, Spain 
      \and  
       Instituto de F\'{\i}sica Fundamental (IFF), CSIC, Serrano 123, 28006, Madrid, Spain      
       \and    
   Instituto de Radioastronom\'{\i}a y Astrof\'{\i}sica (IRyA-UNAM), 3-72(Xangari), 8701, Morelia, Mexico
      \and 
   Instituto de Astrof\'{\i}sica de Canarias, Calle V\'{\i}a L\'actea, s/n, E-38205 La Laguna, Tenerife, Spain
       \and
   Departamento de Astrof\'{\i}sica, Universidad de La Laguna, E-38205, La Laguna, Tenerife, Spain
      \and  
     LERMA, Observatoire de Paris, Coll\`ege de France, PSL University, CNRS, Sorbonne University,  Paris
     \and
     Department of Physics, University of Connecticut, 196 Auditorium Road, U-3046, Storrs, CT 06269-3046, USA
    \and
     Center for Computational Astrophysics, Flatiron Institute, 162 Fifth Avenue, New York, NY 10010, USA     
    \and     
    Departmento de F\'{\i}sica de la Tierra y Astrof\'{\i}sica, Fac. de CC F\'{\i}sicas, Universidad Complutense de Madrid, E-28040 Madrid, Spain    
    \and
    Instituto de  F\'{\i}sica de Part\'{i}culas y del Cosmos IPARCOS,   Fac. de CC F\'{\i}sicas, Universidad Complutense de Madrid, E-28040 Madrid, Spain
    \and    
    Max-Planck-Institut f\"ur Extraterrestrische Physik, Garching, Germany   
   \and   
   Cardiff Hub for Astrophysics Research \& Technology, School of Physics \& Astronomy, Cardiff University, Queens Buildings, The Parade, Cardiff, CF24 3AA,
UK   
   \and 
   Department of Physics, University of Oxford, Keble Road, Oxford OX1 3RH, UK
    \and
   School of Physics \& Astronomy, University of Southampton, Hampshire SO17 1BJ, Southampton, UK    
   \and
   Telespazio UK for the European Space Agency (ESA), ESAC, Camino Bajo del Castillo s/n, 28692 Villanueva de la Ca\~nada, Spain   
    \and
    The University of Texas at San Antonio, One UTSA Circle, San Antonio, TX 78249, USA   
   \and    
    Space Telescope Science Institute, 3700 San Martin Drive Baltimore, Maryland 21218, USA
    \and  
   Kavli Institute for Particle Astrophysics \& Cosmology (KIPAC), Stanford University, Stanford, CA 94305, USA
    \and
   National Astronomical Observatory of Japan, National Institutes of Natural Sciences (NINS), 2-21-1 Osawa, Mitaka, Tokyo 181-8588, Japan    
   \and 
      Instituto de Estudios Astrof\'isicos, Facultad de Ingenier\'ia y Ciencias, Universidad Diego Portales, Av. Ej\'ercito Libertador 441, Santiago, Chile            
    \and
      Kavli Institute for Astronomy and Astrophysics, Peking University, Beijing 100871, People's Republic of China
    \and
   School of Mathematics, Statistics and Physics, Newcastle University, Newcastle upon Tyne, NE1 7RU, UK 
    \and  
    Astronomical Observatory, Volgina 7, 11060 Belgrade, Serbia
     \and
    Sterrenkundig Observatorium, Universiteit Ghent, Krijgslaan 281-S9, Ghent, B-9000, Belgium}
    

    \date{Received: April, 2024; Accepted:--, --}

 
  \abstract{
  
  We use a sample of 64 nearby ($D_{\rm L}$=7-45~Mpc) disk galaxies including 45 active galactic nuclei (AGN) and 19 non-AGN, that have high spatial resolution (median value $\simeq36$~pc) multiline CO interferometer observations obtained at millimeter wavelengths with the Atacama Large Millimeter Array (ALMA) and/or Plateau de Bure Interferometer (PdBI) to study the distribution of cold molecular gas in their circumunuclear disks (CND; $r\leq200$~pc). We analyze whether the concentration and  normalized radial distribution of cold molecular gas  change as a function of the X-ray luminosity in the 2--10 keV range ($L_{\rm X}$) to decipher the imprint left by AGN feedback. We also look for similar trends in the concentration and normalized radial distribution of the hot molecular gas as well as the hot-to-cold-molecular gas mass ratio  in a subset of 35 galaxies using near infrared (NIR) integral field spectroscopy data obtained for the H$_2$ 1-0 S(1) line. We find a significant turnover in the distribution of the cold molecular gas concentration as a function of the X-ray luminosity with  a breakpoint which divides the sample into two branches: (1) the `AGN build-up branch' ($L_{\rm X}\leq10^{41.5\pm0.3}$erg~s$^{-1}$) and (2) the `AGN feedback branch' ($L_{\rm X}\geq10^{41.5\pm0.3}$erg~s$^{-1}$) . Lower luminosity AGN and non-AGN of the AGN build-up branch show high cold molecular gas concentrations and centrally  peaked radial profiles on nuclear ($r\leq50$~pc) scales.   Higher luminosity AGN of the  AGN feedback branch, show a sharp decrease in the concentration of molecular gas and flat or inverted radial profiles. The cold molecular gas concentration index ($CCI$), defined as the ratio of surface densities at  $r\leq50$~pc ($\Sigma^{\rm gas}_{\rm 50}$) and $r\leq200$~pc $(\Sigma^{\rm gas}_{\rm 200}$), namely $CCI \equiv$~log$_{\rm 10}(\Sigma^{\rm gas}_{\rm 50}/\Sigma^{\rm gas}_{\rm 200}$), spans a factor $\simeq$4-5 between the galaxies lying at the high end of the AGN build-up branch and the galaxies showing the most extreme nuclear-scale molecular gas deficits in the AGN feedback branch. The concentration and radial distributions of the hot molecular gas in our sample follow qualitatively similar but less extreme trends  as a function of the X-ray luminosity. As a result, we find higher values of the hot-to-cold molecular gas mass ratios on nuclear  scales in the highest luminosity AGN sources of the AGN feedback branch. These observations confirm, on a three times larger sample, previous  evidence found in the context of the Galaxy Activity Torus and Outflow Survey (GATOS) by \citet{GB21} that the imprint of AGN feedback on the CND-scale distribution of molecular gas is more extreme in higher luminosity Seyfert galaxies of the local universe.
}

  \maketitle
%

\section{Introduction}\label{intro}

The optical spectral properties of active galactic nuclei (AGN) are used to classify them into two categories: type 1 AGN show both broad and narrow line 
regions (BLR and NLR), while type 2 AGN only show NLR. In the AGN unifying theories the central engines of type 2 
objects are thought to be hidden behind large amounts of obscuring material located in a 'torus' or disk of a few pc-size  \citep[e.g.,][]{Ant85, Ant93, Kro88, Urr95}. Being a key player in the last steps of the AGN fueling process, observations of tori and their surrounding 
regions can therefore shed light on our understanding of the feeding and feedback 
cycle of the gas in AGN \citep[e.g.,][]{Ram17, Sto19, Combes21, Har24}. 

Molecular tori are the expected launching sites of multi-scale and multi-phase outflows, a manifestation of the AGN 
feedback process. There is increasing multi-wavelength observational evidence for the existence of outflows that can extend from tens 
of parsecs to several kiloparsecs in different AGN samples \citep[e.g.,][]{Mor05, Mor15, Alat11, Aal12, Aal20, Com13, GB14, GB19, Que16b, Fio17, Alo18, Alo19, Alo23, Aud19, Flu19, Flu21, Min19, Lut20, Dom20, Vei20, Gar21, Ven21, Ram22, Uli24}.  Galaxy formation models and numerical simulations  predict that AGN feedback is a necessary ingredient required to explain the observable
properties of massive galaxies \citep[e.g.,][]{Schaye15,Ang17,Wei17, Pill18, Dav19}. However, quantitatively establishing the potential impact of the feedback of nuclear 
activity in their host galaxies remains a matter of debate even in high luminosity AGN \citep[e.g.,][]{Har23, Har24}. The impact predicted by 
numerical simulations is likely cumulative over the multiple short-lived episodes that are expected during the lifetime of the AGN \citep{Pio22}. 
The  recent study  published by \citet{Eli21}, based on kpc-scale observations of the Extragalactic Data base for Galaxy Evolution-Calar Alto Legacy Integral Field Area (EDGE-CALIFA) sample \citep{San12, Bol17}, found  that the molecular gas fractions ($\Sigma_{\rm gas}/\Sigma_{*}$) of central AGN regions are a factor of $\simeq$2 lower than those in non-AGN star-forming regions. However, AGN variability, linked to the chaotic nature of the black hole feeding process, tends naturally to eliminate any significant correlation between the global (kpc-scale) molecular gas content  and the AGN luminosity  \citep{Izu16b, Ros18, Ang21,GB21, Ram22}.  The search for a clear observational ‘smoking gun’  signature of AGN feedback on the distribution and kinematics of molecular gas should be conducted on spatial scales closer to the central engine (tens to hundreds of parsecs),  considering the short flickering timescales ($<10^{5}$~yrs) associated with individual nuclear activity episodes \citep{Mart04,Hic14,Sch15, Kin15}.

 The observations of the circumnuclear disk (CND) of the Seyfert~2 galaxy NGC~1068 obtained with the Atacama Large Millimeter Array (ALMA)  unveiled the dust continuum and molecular line emission 
 in the CO(6--5) line from a 10~pc-diameter disk or torus  of $M_{\rm gas}^{\rm torus}\sim1\times10^5$~M$_\sun$ \citep{GB16}. This result has been confirmed by further ALMA images revealing the many `faces' of the torus obtained in a set of different 
 molecular lines showing the different layers of the disk extending from 7 to up to 30\,pc scales  \citep{Gal16, Ima18, Ima20,  
 Imp19, GB19}.  These new high-resolution images showed that the launching of the molecular outflow is shaping not only the morphology of the 
 torus but also its kinematics.  The wide-angle AGN wind launched from the accretion disk of NGC~1068 is  impacting a sizeable 
 fraction of the gas inside the torus, driving a dusty molecular outflow, identified by vertical X-shape gas protrusions. The outflowing torus
 picture, also advocated to explain the ALMA observations  of NGC~1377 \citep{Aal17, Aal20} and NGC~5643 \citep{Alo18}, fits the 
 scenarios  predicted by radiation-driven wind models and numerical simulations \citep[e.g.,][]{Eli06, Wad12, Wad15, Wad16, Cha16, Cha17,   Hoe17, Will19, Ven20, Will20}. In this scenario, inflowing and outflowing components can coexist in different regions of the 
 torus, a  manifestation of the complex  nature of the obscuring material around AGN \citep{Ram17, Hoe19, Gar24}. On larger 
 spatial scales, the ALMA images of NGC~1068 published by \citet{GB14} and \citet{GB19} resolved  its CND as a ringed disk  of 
  $D\simeq400$~pc-size with  a marked deficit of molecular gas in its central $\simeq130$~pc-region.  This highly-contrasted ring 
  morphology, reminiscent of a cavity, reflects the accumulation of molecular gas on the working surface of the AGN-driven ionized gas 
  wind of NGC~1068. In agreement with this picture, the kinematics of  molecular gas show strong deviations from circular motions modelled as a three-dimensional outflow that reflects the effects of AGN feedback.

 The case of NGC~1068 is by no means unique. The existence of a cavity or lacuna in the inner 500~pc of the cold molecular gas distribution of the Seyfert~2 galaxy  NGC~2110  was  described by \citet{Ros19}. The cavity, oriented in the direction of the radio jet and filled with warm molecular gas and an ionized gas wind, was also recently analyzed  by \citet{Per23}, pointing to a causal relationship similar to that favored to account for the nuclear-scale molecular gas deficit  in NGC~1068.  A deficit of molecular gas  attributed to gas removal by the AGN wind was also identified in the CND of the Seyfert~2 galaxies NGC~5728  \citep{Shi19}, ESO428-G014 \citep{Fer20}, and NGC~7172 \citep{Alo23}.

 Significantly large (diameters $\geq$20-50 pc) and massive ($\sim 10^5-10^7M_\odot$) tori have been detected in cold molecular gas and dust in other local Seyferts,  low-luminosity AGN, and compact obscured nuclei \citep{Aal17, Aal19, Alo18, Alo19, Alo20, Alo23, Izu18, Izu23, Aud19, Aud21, Com19, GB21, Tri22}. In particular, \citet{GB21}  used ALMA observations of the molecular gas obtained in  the CO(3--2) and HCO$^+$(4--3) lines  with  spatial resolutions $\simeq0.1\arcsec$ ($7-13$~pc) in the CND of 19 nearby AGN, which belong to the core sample of the  Galaxy Activity Torus and Outflow Survey (GATOS) and the southern hemisphere low luminosity AGN of the Nuclei of Galaxies (NUGA) sample \citep{Com19}.   The GATOS core sample  was  selected from the {\it Swift}/BAT catalog of AGN \citep{Bau13} with
distances   $<28\,$Mpc and  luminosities  $L_{\rm AGN}$(14-150 keV) $\ge 10^{42}\,{\rm erg~s}^{-1}$ to purposely complement the lower luminosity range covered by the NUGA survey \citep{Com19}. The galaxies analyzed by \citet{GB21} share the presence of AGN ionized outflows and/or radio jets launched from the vicinity of dusty molecular disks which have median diameters and molecular gas masses  of $\sim42$~pc and $\sim6\times10^5$~M$_{\sun}$, respectively. These compact disks  tend to show an orientation perpendicular to the AGN outflow axes, as would be expected from an equatorial torus geometry.

 \citet{GB21} analyzed the concentration of cold molecular gas (index-I in the notation used in their paper). They defined the concentration index as the ratio of gas surface densities derived at $r<50$~pc (a region characteristic of the tori and their surroundings) and  $r<200$~pc (a region where molecular gas reservoirs are expected to build up and potentially reveal the smoking gun signature of AGN feedback), namely  index-I$\equiv$$\log (\Sigma_{\rm H_2}^{\rm 50pc}/\Sigma_{\rm H_2}^{\rm 200pc})$. Their findings indicated that there is a statistically significant trend pointing to a decrease in molecular gas concentration indices in the higher luminosity and higher Eddington ratio sources of their sample beyond a turnover X-ray luminosity $L_{\rm 2-10keV}\simeq10^{42}$erg~s$^{-1}$, suggesting the existence of two  `branches' in the parameter space defined by the concentration index and the AGN luminosity. AGN with luminosities below the turnover point show higher molecular gas concentration indices. This result would agree with the theoretical prediction that outflows in the highest luminosity AGN could partially clear the material around the black hole on the scales of the tori and their surrounding regions. The existence of a qualitatively similar trend was first found by \citet{Ric17b}, recently confirmed by \citet{Ric22} and \citet{Ric23b}, who found that the covering factor measured at X-ray wavelengths decreases with the Eddington ratio based on a survey of AGN in the local Universe,  an indication that radiative feedback is an efficient mechanism capable of regulating the distribution of the gas and dust within the torus and its surrounding region preferentially  in the most extreme AGN \citep[see also][]{Ezh17, Gon17, Zhu18, Tob19, Tob21, Gar22}. The latter is consistent with evidence from analytical models suggesting that IR-driven dusty winds can develop most favorably within a certain range of  gas column densities and Eddington ratios \citep[e.g.,][]{Ven20, Alo21, Gar22}.

 More recently, \citet{Elf24} studied how the concentration of molecular gas as defined by \citet{GB21} changes as a function of $L_{\rm X}$ for 16 objects in the  mm-Wave Interferometric Survey of Dark Object Masses project (WISDOM) sample. In their analysis they included a sizeable fraction ($\simeq$63$\%$) of early-type galaxies (ETG: 6 E and 4 SO Hubble types) that have
X-ray luminosities well below the turnover identified in the distribution of concentration indices derived for the spiral galaxy sample studied by \citet{GB21}. Six of the ten ETG analyzed by \citet{Elf24} showed  concentration indices slightly below the range found by \citet{GB21} for very low-luminosity objects expected to be ruled by secular evolution processes rather than by AGN feedback.  As we  argue in Sect.~\ref{summary}, this discrepancy is not surprising, as we do not expect to find a direct evolutionary link that could connect bulge-dominated ETG to disk-dominated spiral galaxies in the scenario described by \citet{GB21} and this paper. In contrast, the few spiral  galaxies in \citet{Elf24} with X-ray luminosities higher than 10$^{42}$~erg~s$^{-1}$ closely follow the relation found by \citet{GB21}.

\section{Aims of this paper}

To verify the results of  \citet{GB21} on a more robust statistical basis, we have selected for this paper a larger sample of nearby ($D_{\rm L}$=7-45~Mpc) 45 AGN and 19 non-AGN galaxies that have high-resolution multiline CO observations obtained at millimeter wavelengths by the ALMA and/or Plateau de Bure Interferometer (PdBI) arrays to study the distribution of cold molecular gas at CND scales ($r\leq200$~pc) in our sample. We explore whether the concentration and normalized radial distribution of cold molecular gas  change as function of the X-ray luminosity to quantify the imprint left by AGN feedback on our sources. We purposely include in our analysis the data from a subset of 19 non-AGN galaxies used as a comparison sample. Unlike the analysis of \citet{GB21}, restricted to the use of data obtained from a single CO line (the 3--2 transition), in this work we use the available data from 3 different CO transitions (the 1--0, 2--1, and 3--2 rotational lines) to investigate the dependence of molecular gas concentration on the CO line. We also use in our analysis near Infrared (NIR) integral field spectroscopy data obtained for the H$_2$ 1-0 S(1) line available for a subset of 35 targets of the CO-based sample. 
These observations allow us to complete using a larger sample the pioneering work of \citet{Hic13}, who  studied the differences in  H$_2$ surface brightness profiles between quiescent and Seyfert galaxies and found evidence of  higher central concentration in H$_2$ surface brightness profiles in AGN. The new data are used to derive  the concentration and normalized radial distribution of the hot molecular gas phase as well as the hot-to-cold-molecular gas mass ratio as a function of the X-ray luminosity. 
Taken together, these observations provide the basis for an evolutionary scenario describing the feeding and feedback cycle of the gas in AGN that can be compared with the predictions of state-of-the-art numerical simulations.

The paper is organized as follows. We present in Sect.~\ref{sample} the galaxy sample. Sect.~\ref{observations} describes the ALMA and NIR  observations used in this work. Sect.~\ref{cold-H2} studies the distribution of cold molecular gas derived from CO performed  by analyzing how the concentration and the radial distribution of this component changes as function of the X-ray luminosity in our sample, as detailed  in Sects.~\ref{CO-ratios}, \ref{CO-dep}, \ref{CO-profiles}, and \ref{non-param}.  Sect.~\ref{hot-H2} presents a similar analysis for the hot molecular gas component probed by the NIR observations of  our sample, as detailed in Sects.~\ref{NIR-ratios} and \ref{NIR-profiles}. 
Section~\ref{hot-cold} describes how the hot-to-cold molecular gas mass ratio changes as a function of the X-ray luminosity in our sample. A tentative scenario describing an evolutionary link connecting the galaxies analyzed in this work  is presented in Sect~\ref{discussion}. The main conclusions of the paper are summarized in Sect.~\ref{summary}.

\section{The sample}\label{sample}

Table~\ref{Tab1} lists the main observational parameters of the sample used in this work.  
We selected a sample of AGN and non-AGN spiral galaxies\footnote{With the exception of one elliptical galaxy classified as non-AGN, NGC~5845, our sample only includes disk galaxies; the median value of the Hubble stage T parameter for our sample is 2.2, which corresponds to a Sab spiral galaxy type; see Table~\ref{Tab1}.} that have high-resolution CO(3-2), CO(2-1), and/or CO(1-0) observations obtained with ALMA and/or the PdBI arrays from different surveys. We consider a galaxy to be an AGN if it is well detected in hard X-rays and have a corresponding Seyfert/LINER classification according to NED. However, six of the galaxies detected in hard X-rays have $L_{\rm X}\leq10^{40}$erg.~s$^{-1}$ but no optical identification of an AGN (see Table~\ref{Tab1} for details). For these sources a non-negligible contamination by X-ray binaries cannot be excluded. The combined sample includes the GATOS survey \citep{GB21,Alo21} the NUGA survey \citep{GB03, Gar12, Com19}, a selection of targets of the Physics at High Angular resolution in Nearby GalaxieS survey \citep[PHANGS;][]{Ler21}, the WISDOM sample \citep{Oni17, Elf24} \footnote{We do not expect to find a direct evolutionary link  that could connect ETG to spiral (disk) galaxies,  as argued in Sect.~\ref{discussion}, therefore, we have only included in our combined sample the WISDOM sources that are not classified as ETG.}, the Luminous Local AGN with Matched Analogs  survey \citep[LLAMA;][]{Ros18}, as well as case studies of individual galaxies like M51 \citep{Sch13}, NGC1068 \citep{GB19}, and NGC7172 \citep{Alo23}. We also used the CO(1--0) images obtained in a sample of 12 AGN observed with ALMA by \citet{Ric23a}. The AGN targets are selected to lie at a range of distances
$D_{\rm L}=7-45$~Mpc.  We set the upper limit  to the distance to image with enough spatial resolution the distribution of molecular gas in the circumnuclear disks of our targets. In particular, the median value of the spatial resolutions of the CO observations used in this work is $36$~pc and the corresponding value of the maximum recoverable scales of the images is $\simeq400$~pc. We also used a comparison sample of non-AGN drawn from the LLAMA and PHANGS surveys, with a range of distances similar to the AGN sources: $D_{\rm L}=7-41$~Mpc.   The CO sample comprises 57 different images of 45 AGN  as well as maps  of 19 non-AGN galaxies. In our AGN sample 7 of the galaxies have images obtained in more than one CO transition. The latter includes  NGC~6300, NGC~5643, NGC~7314, NGC~4388, and NGC~3227, which have data available from the 3--2 and 1--0 CO transitions, as well as NGC7172 and NGC~5728, which have data from the 3--2, 2--1, and 1--0 CO transitions. The availability of these  data  allows us to study the dependence of the trends analysed in this paper on the particular CO transition used to image the distribution of molecular gas in these sources.

Near-IR K-band (or H+K-band) integral field spectroscopy data are available for 46 of the galaxies of our sample, seven of which are non-AGN.  Of these galaxies 35  have well detected H$_{\rm 2}$ 1–0 S(1) 2.12~$\mu$m (SNR$>$ 3) out to a radius of 200~pc.  Two AGN (NGC~1365 and NGC~628) and four non-AGN (IC~5332, NGC~1079, NGC~3621, and NGC~5845) were found to have no detectable H$_{\rm 2}$ emission within this region and thus no concentration index was calculated for these targets.  While NGC~4579, NGC~4501, and NGC~7465 have well detected H$_{\rm 2}$ emission out to a radius of 50~pc, available data only cover 65$\%$, 62$\%$, and 38$\%$, respectively, of the deprojected 200~pc radius aperture and thus only upper limits on the concentration index are obtained.  In the majority of the 35 galaxies with well detected H$_{\rm 2}$ emission (24 galaxies) the data provide full coverage of the 200~pc radius deprojected aperture and in all other cases at least 70$\%$ of the aperture is measured and an aperture correction was applied assuming an azimuthally symmetric flux distribution.

Hard X-ray luminosities in the $2-10\,$keV band  (hereafter $L_{\rm X}$) are adopted as a proxy for the AGN luminosity.  For the majority of our sources we take $L_{\rm X}$ from \citet{Ric17a} rescaled to the assumed distances, except for five sources where we used archival observations to compute X-ray luminosities (see details in Table~\ref{Tab1} and Appendix~\ref{Lx}).

\begin{figure}[tb!]
  \centering
    \includegraphics[width=8.5cm]{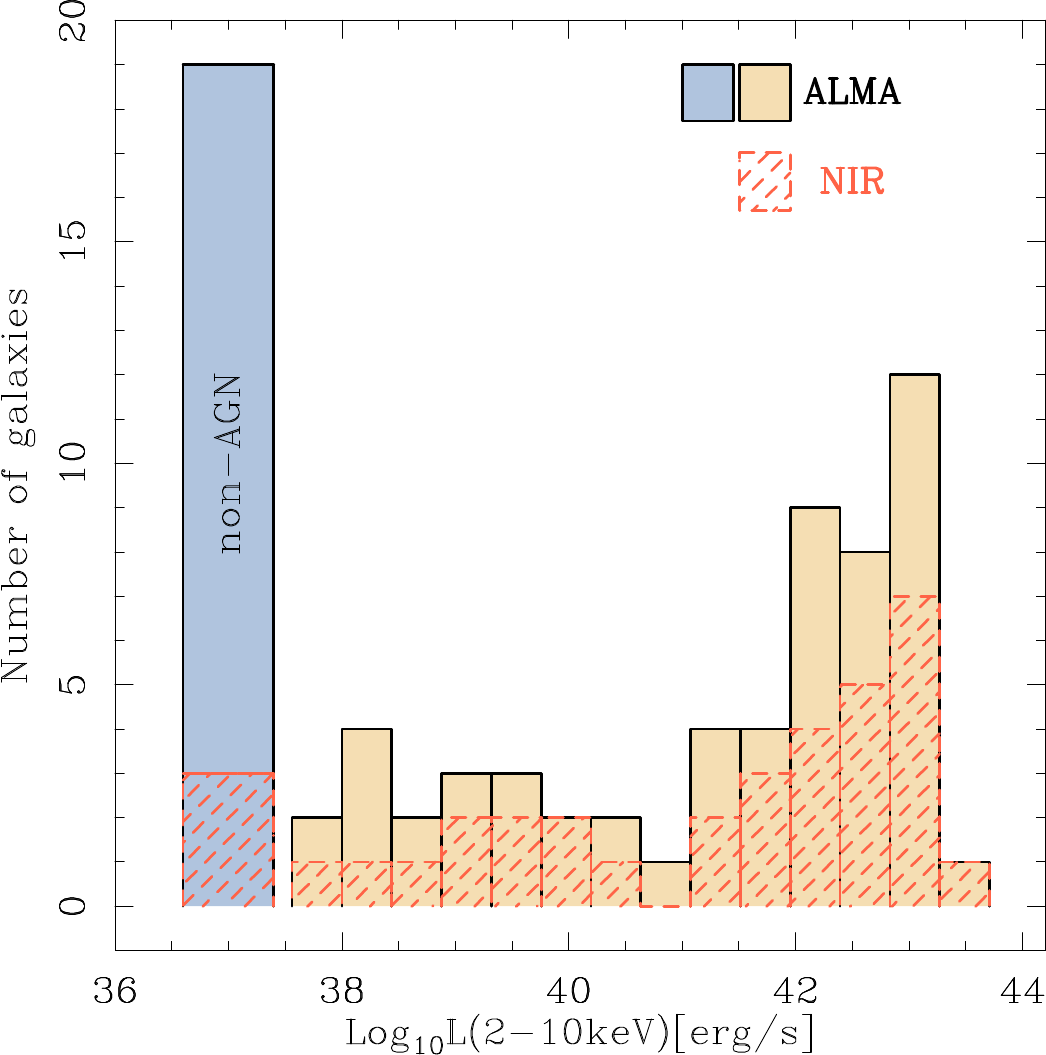}
       \caption{Histograms showing the distribution of independent data sets  obtained from the targets of the combined ALMA+PdBI and NIR data samples as a function of the  AGN luminosities measured at X-ray wavelengths ($L_{\rm X}$) in the 2–10 keV band (in log units).
For the sake of comparison galaxies classified as non-AGN populate the histogram arbitrarily located  at log($L_{\rm X})$$=37$~erg~s$^{\rm -1}$.}  
   \label{histo-samples}
    \end{figure}

 Figure~\ref {histo-samples} shows the distribution of independent data sets  of AGN and non-AGN galaxies of our combined CO and NIR sample as a function of  $L_{\rm X}$.  AGN targets are evenly distributed as a function of  $L_{\rm X}$ below the turnover luminosity ($L_{\rm X}$~$\simeq$~$10^{42.0}$erg~s$^{-1}$), a regime for which \citet{GB21} found evidence of systematically high values of the concentration of molecular gas derived from CO(3-2). In this luminosity  range ($10^{37.5}$erg~s$^{-1}$~$\leq$~$L_{\rm X}$~$\leq$~$10^{42.0}$erg~s$^{-1}$) the CO sample  includes 27  galaxy images and NIR data from 15 targets. Beyond the turnover luminosity  our CO sample includes a similar number of galaxy images: 30 CO and 17 NIR maps.  Galaxies beyond the turnover are concentrated  in a comparatively smaller luminosity range: $10^{42.0}$erg~s$^{-1}$~$\leq$~$L_{\rm X}$~$\leq$~$10^{43.5}$erg~s$^{-1}$. This uneven distribution of the sources as a function $L_{\rm X}$ ensures that the abrupt transition to lower molecular gas  concentration indices, which appears beyond the turnover luminosity \citep{GB21}, can be sufficiently  sampled.


  \begin{figure*}[tb!]
  \centering
    \includegraphics[width=8.5cm]{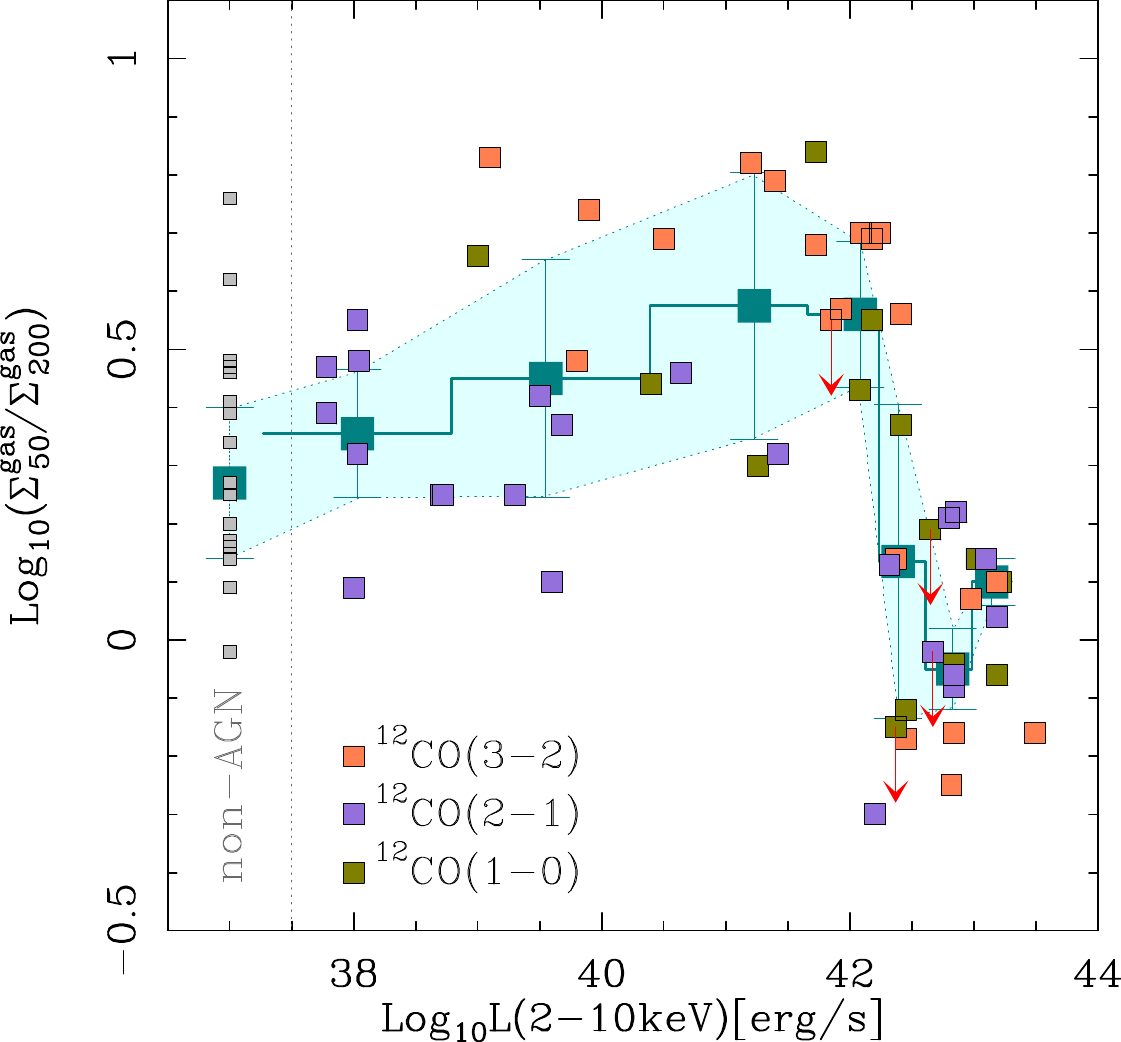}
    \includegraphics[width=8.5cm]{ 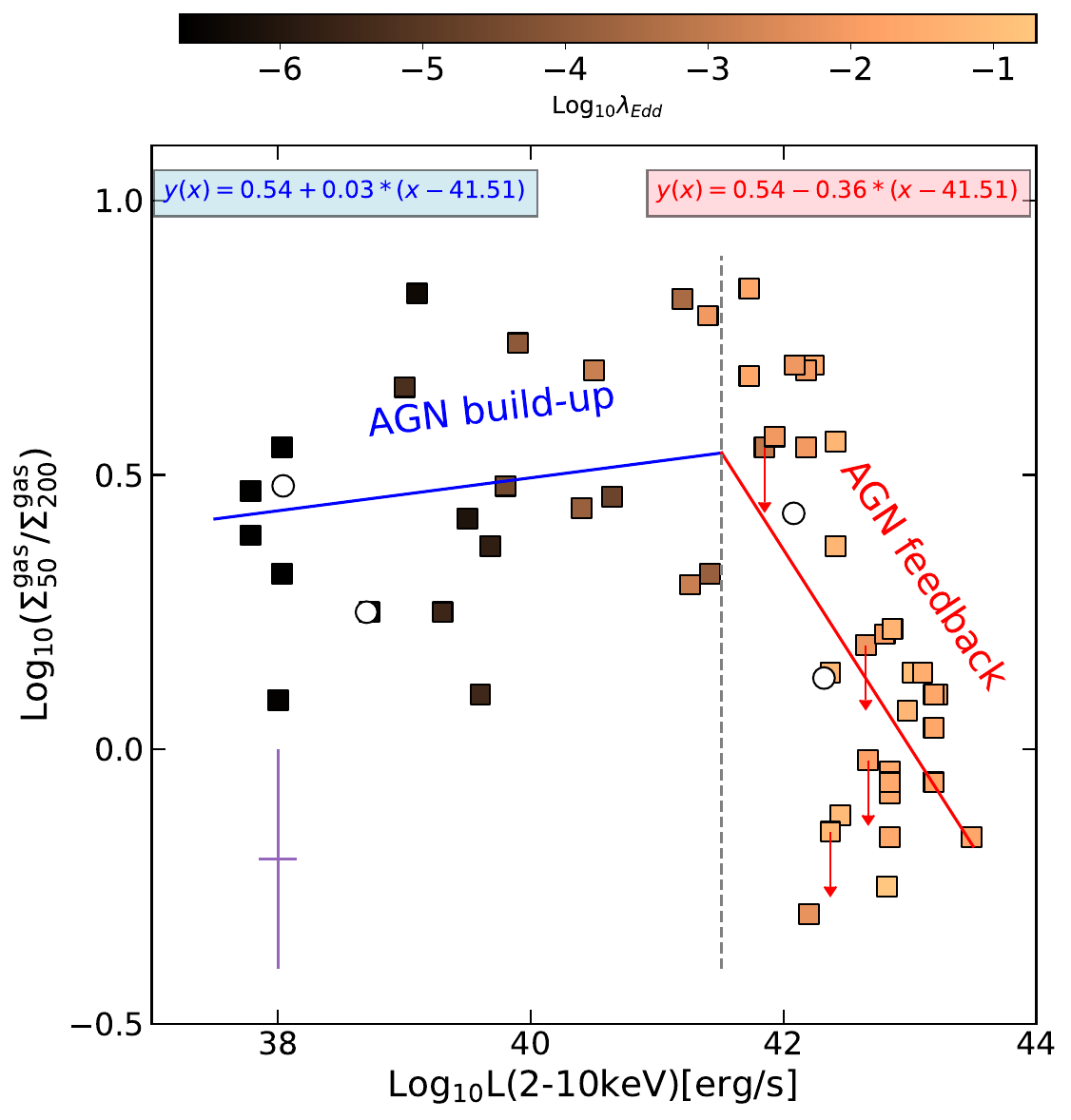}     
       \caption{{\em Left panel}:~concentration of `cold' molecular gas measured for the central regions of galaxies of the sample derived  from
the ratio (in log units) of the average H$_{\rm 2}$ surface densities estimated from CO line emission at two spatial scales: $r\leq50$pc ($\Sigma^{\rm gas}_{\rm 50}$) and 
$r\leq200$pc ($\Sigma^{\rm gas}_{\rm 200}$).  These spatial scales are representative of the nuclear and circumnuclear disk regions, respectively, as defined in Sect.~\ref{CO-ratios}. The concentration of molecular gas is shown as a function of the AGN luminosities measured at X-ray wavelengths ($L_{\rm X}$) in the 
2–10 keV band (in log units).  Symbols (small squares) are color-coded  to identify the CO line used to estimate the surface densities for each AGN target. Concentration indices for non-AGN targets are displayed as (gray) square symbols located arbitrarily at log($L_{\rm X})$$=37$~erg~s$^{\rm -1}$. The big (green) squares stand for the median value of the concentration index estimated  for the 7 $L_{\rm X}$  bins that cover the range of AGN luminosities spanned by the sample, as  defined in  Sect.~\ref{CO-ratios},  as well as for the non-AGN galaxies in the comparison sample. Errorbars for each L$_{\rm X}$ bin represent the median absolute deviations around the median values. The green-shaded area illustrates the uncertainties around the median values. {\em Right panel}:~same as {\em left panel} but showing the two-branch linear solution found by the MARS algorithm to fit the distribution of concentration indices as a function of $L_{\rm X}$ for the AGN build-up phase (blue straight line) and the AGN feedback phase (red straight line). Symbols are color coded as a function of the Eddington ratio ($\lambda_{\rm Edd}$) except for the four galaxies for which this information is not available (empty circles). Uncertainties on the molecular gas mass ratios and on the AGN luminosities are $\pm0.2$~dex and $\pm0.15$~dex, respectively. Red arrows identify upper limits.}  
   \label{conc-CO}
    \end{figure*}

\section{Observations}\label{observations}

\subsection{CO observations}\label{CO-data}

Details on the reduction, imaging and post-processing of the CO data sets of the different surveys used in this work are included in the references listed in Sect~\ref{sample}. The CO-based `cold' molecular gas masses were derived from the following expression, which includes the mass of Helium ($M_{\rm gas}^{\rm CO}=1.36\times M({\rm H_2})$):

\begin{equation}
\frac{M_{\rm gas}^{\rm CO}}{M_{\sun}}  = 1.07 \times 10^4 \times \left (\frac{\nu_{\rm 10}^2}{\nu_{\rm ij}^2} \right) \times \left (\frac{S_{\rm CO({\rm i-j})}}{{\rm Jy~km~s^{-1}}} \right) \times R_{\rm ij}^{\rm -1} \times \left (\frac{\alpha_{\rm CO}}{2\times10^{20}} \right) \times  \left (\frac{D}{\rm Mpc}\right)^2.
\label{mass-CO}
\end{equation}

In Eq.~\ref{mass-CO}, $\nu_{\rm ij}$ and $S_{\rm CO_{\rm ij}}$ are  the frequency and the flux of the CO(i-j) transition (in Jy~km~s$^{-1}$-units),  $R_{\rm ij}$ is the assumed CO(i-j)-to-CO(1-0) line ratio (in $T_{\rm mb}$-units), $\alpha_{\rm 
CO}$ is the assumed CO(1-0)-to-H$_{\rm 2}$  conversion factor in units of mol~cm$^{-2}$~(K~km~s$^{-1}$)$^{-1}$, and $D$ is the distance in Mpc. Mass surface densities for the two representative regions used in this work ($r\leq50$~pc and $r\leq200$~pc) are derived by deprojecting the CO images using the $PA$ and $i$ values listed in Table~\ref{Tab1}.

The estimates of the molecular gas masses derived  from the different CO lines used in this work are subject to uncertainties related to the assumed $R_{\rm ij}$ ratio and  to the adopted CO--to--H$_2$ conversion factor \footnote{We assume common values of $R_{\rm ij}$ and $\alpha_{\rm CO}$ for the two spatial scales used in this work: $R_{\rm 31}$=2.9 and $R_{\rm 21}$=2.5 measured by \citet{Vit14} in NGC~1068, and a Milky-Way value of  $\alpha_{\rm CO}=2\times10^{20}$mol~cm$^{-2}$~(K~km~s$^{-1}$)$^{-1}$ \citep{Bol13}.}.  
In line with the values typically assumed in the literature \citep[e.g.,][]{Gen10,GB21} we estimate that the latter may contribute up to $\sim$0.12-0.13~dex each if we allow them to explore the typical range of values seen in the observations of different populations of galaxies on nuclear spatial scales and, also, if we assume that these uncertainties are  uncorrelated. Under these extreme hypotheses the total uncertainty would  amount to $\pm$0.20~dex. However, we emphasize that the analysis of the concentration indices and normalized radial distributions of molecular gas in the galaxies of the sample defined in Sect.~\ref{cold-H2} would be insensitive to any galaxy-to-galaxy changes in the global conversion factors. Furthermore, any potential change of these conversion factors, which are here assumed to be the same for the two spatial scales analysed in this work  ($r\leq50$~pc and $r\leq200$~pc), would likely reinforce the trends found in Sect.~\ref{cold-H2} \citep[see discussion in Sect.~\ref{CO-ratios} and also related discussion in Sect.~8 of][]{GB21}.


\begin{table}[]
\caption{Median values of the concentration indices for the cold and hot molecular gas and hot-to-cold ratios.}
\centering
\resizebox{.5\textwidth}{!}{ 
\begin{tabular}{ccc} 
\hline
\hline
\noalign{\smallskip} 
  log$_{\rm 10}$$L_{\rm X}$ bin   [CO]     &  number of galaxies & median $CCI$  \\
 \noalign{\smallskip} 
 \hline  
   erg~s$^{-1}$      &  --- & log units  \\  
\noalign{\smallskip}  
\hline
\hline
 \noalign{\smallskip} 
      bin-1:~[42.85--43.50]   &  9 &  0.10$\pm$0.04      \\
      bin-2:~[42.65--42.85]   &  8 &  -0.05$\pm$0.07      \\     
      bin-3:~[42.20--42.65]  &  8 &  0.13$\pm$0.27      \\     
      bin-4:~[41.73--42.20]   &  8 &  0.56$\pm$0.13     \\     
      bin-5:~[40.40--41.73]   &  8 &  0.58$\pm$0.23      \\
      bin-6:~[39.00--40.40]   &  8 &  0.45$\pm$0.20      \\
      bin-7:~[37.80--39.00]  &  8 &  0.36$\pm$0.11      \\     
      non-AGN bin   &  19 &  0.27$\pm$0.13      \\                
  \noalign{\smallskip}
  \hline
  \hline
 \noalign{\smallskip}
    log$_{\rm 10}$$L_{\rm X}$ bin   [NIR]     &   Number of galaxies  & median $HCI$  \\
  \noalign{\smallskip}   
    \hline  
   erg~s$^{-1}$      &  --- & log units  \\  
\noalign{\smallskip}  
\hline
\hline
 \noalign{\smallskip} 
      bin-1:~[42.84--43.50] &  8 &   0.16$\pm$0.18      \\
      bin-2:~[42.37--42.82]  &  6 &  0.24$\pm$0.05      \\     
      bin-3:~[41.73--42.24]  &  6 &  0.62$\pm$0.25      \\     
      bin-4:~[39.68--41.42]  &  6 &  0.56$\pm$0.17      \\     
      bin-5:~[37.78--39.60]  &  6 &  0.46$\pm$0.07      \\
        non-AGN bin   & 3 &  ----      \\                
   \noalign{\smallskip} 
\hline 
\hline
 \noalign{\smallskip}
    log$_{\rm 10}$$L_{\rm X}$ bin   [hot/cold]     &   Number of galaxies &  hot/cold ratio concentration  \\
  \noalign{\smallskip}   
    \hline  
   erg~s$^{-1}$      &  --- & log units  \\  
\noalign{\smallskip}  
\hline
\hline
 \noalign{\smallskip} 
      bin-1:~[42.98--43.50]  &  7 &  0.35$\pm$0.14      \\
      bin-2:~[42.80--42.86]  &  7 &  0.21$\pm$0.10      \\     
      bin-3:~[42.37--42.67]  &  7 &  0.34$\pm$0.05      \\     
      bin-4:~[41.73--42.24]  &  7 &  0.06$\pm$0.16      \\     
      bin-5:~[39.60--41.42]  &  7 &  -0.08$\pm$0.08      \\
      bin-6:~[37.78--39.30]  &  7 &  -0.09$\pm$0.19      \\        
 \noalign{\smallskip} 
\hline 
\hline
\end{tabular}}
\tablefoot{Column (1) identifies the range of $L_{\rm X}$ luminosities (in log units) used to define the 
different bins that group the CO and NIR data sets of the galaxies of the sample. Column (2) lists the number of 
independent data sets in each $L_{\rm X}$ bin. Column (3) lists the median values and the median 
absolute deviations of the concentration indices ($CCI$ and $HCI$) and the concentration of the hot-to-cold ratio (shown in 
Fig.~\ref{hot-to-cold}) for each  $L_{\rm X}$ bin.}
\label{Tab3} 
\end{table}

\subsection{NIR observations}\label{NIR-data}

All NIR Integral Field Unit (IFU) observations were obtained with ESO SINFONI \citep{Eis03, Bon04} with the exception of NGC~7465, which was observed with OSIRIS at the W. M. Keck Observatory \citep{Lar06}.  Adaptive optics (AO) was employed in the acquisition of the vast majority of the data sets (31 in natural guide and 2 in laser guide star mode), and of those data sets without AO all were AGN and only one lacked a H$_2$ 2.12~$\mu$m detection (see Table~\ref{Tab5} for details).  Reduction of the SINFONI data was performed using the SINFONI Data Reduction Software or a custom package SPRED developed at MPE \citep{Abu06}.  The OSIRIS data reduction pipeline was used for the single OSIRIS data set.  In all cases reduction of the data followed standard reduction steps needed for near-IR spectra as well as additional routines necessary to reconstruct the data cube.  For improved subtraction of OH sky emission lines the routines {\tt mxcor} and {\tt skysub} \citep{Dav07} were used, which correlate reconstructed object and sky cubes spectrally and shift them such that their OH line wavelengths match before subtracting the scaled sky frame.  Nearby standard stars (A- or B-type) were observed close in time to the science frames and used for telluric correction and flux calibration.  The typical flux calibration uncertainty is accurate to at least 10$\%$ in all data sets.

The emission-line flux distribution was extracted by fitting the H$_{\rm 2}$ 2.12~$\mu$m line profile using our custom code {\tt LINEFIT} \citep{Dav11}.  This code fits the emission line in the spectrum of each spatial element, or spaxel, with an unresolved line profile (a sky line) convolved with a Gaussian, as well as a linear function to the line-free continuum.  The point spread function (PSF) and the galaxy center, assumed to be coincident with the AGN location, are determined through fitting of the CO 2.29 $\mu$m bandhead.  The CO bandhead was fit at each spaxel with a broadened late-type stellar template found to be well matched with our galaxy spectra.  The intrinsic K-band CO equivalent width is assumed to be constant to within 20$\%$ at 12~{\AA}, and a deviation from this value therefore represents dilution from the non-stellar continuum \citep[e.g.,][]{Dav07, Hic13, Bur15}.  This non-stellar continuum is assumed to represent the unresolved AGN-associated dust emission which we then utilize to establish the galaxy center and characterize the PSF.

Under the assumptions of local thermal equilibrium and an excitation temperature of 2000 K, the mass of `hot' molecular gas component is obtained from \citep[e.g.,][]{Sco82, Rif14}:

%


\begin{equation}
\frac{M({\rm H_2})}{M_{\sun}} = 5.0766 \times 10^{13} \left (\frac{F_{\rm H_2}}{\rm ergs^{-1} cm^{-2}}\right) \left (\frac{D}{\rm Mpc}\right)^2.
\label{hot-eq}
\end{equation}

As for the cold molecular gas component when we derive the hot molecular gas surface densities for $r\leq50$~pc and $r\leq200$~pc we adopt the same conversion factors for the two spatial scales  and also correct for the mass of Helium using the expression $M_{\rm gas}^{\rm NIR}=1.36\times M({\rm H_2})$. As argued in Sect.~\ref{NIR-ratios}, we note that assuming common conversion factors for both spatial scales would underestimate the trends found in Sect~\ref{hot-H2}.

\section{Distribution of cold molecular gas}\label{cold-H2}
\subsection{Concentration indices}\label{CO-ratios}

Following the methodology adopted by \citet{GB21} we explore below if a simple parameterization of the radial distribution of molecular gas can reveal the imprint of AGN feedback in the galaxies of the sample. To
accomplish this, we first derived the average molecular gas surface densities measured from the different CO lines using two spatial scales representative of the nuclear  ($r\leq50$~pc) and CND environments ($r\leq200$~pc). In our derivation we used the position angles ($PA$) and inclination angles listed in Table~\ref{Tab1}. With this definition,  nuclear scales probe the region typically occupied by the molecular tori and their immediate surroundings \citep{GB21, Alo21}. We normalize the nuclear scale gas surface densities ($\Sigma^{\rm gas}_{\rm 50}$) by the corresponding densities measured on CND scales ($\Sigma^{\rm gas}_{\rm 200}$). In our estimate of $\Sigma^{\rm gas}_{\rm 50}$ and $\Sigma^{\rm gas}_{\rm 200}$ we adopted the same values of the conversion factors ($R_{\rm 31}$, $R_{\rm 21}$, and $\alpha_{\rm CO}$) for the two spatial scales.
This normalization counterbalances any potential bias  associated with the different overall molecular gas content of the galaxies of the sample and it therefore serves as a measurement of the concentration of  molecular gas, which is the physical parameter relevant to the present analysis.  Furthermore, the normalized radial distribution is by definition insensitive to any potential galaxy-to-galaxy change in the global conversion factors.

  \begin{figure*}[tb!]
  \centering
    \includegraphics[width=0.78\textwidth]{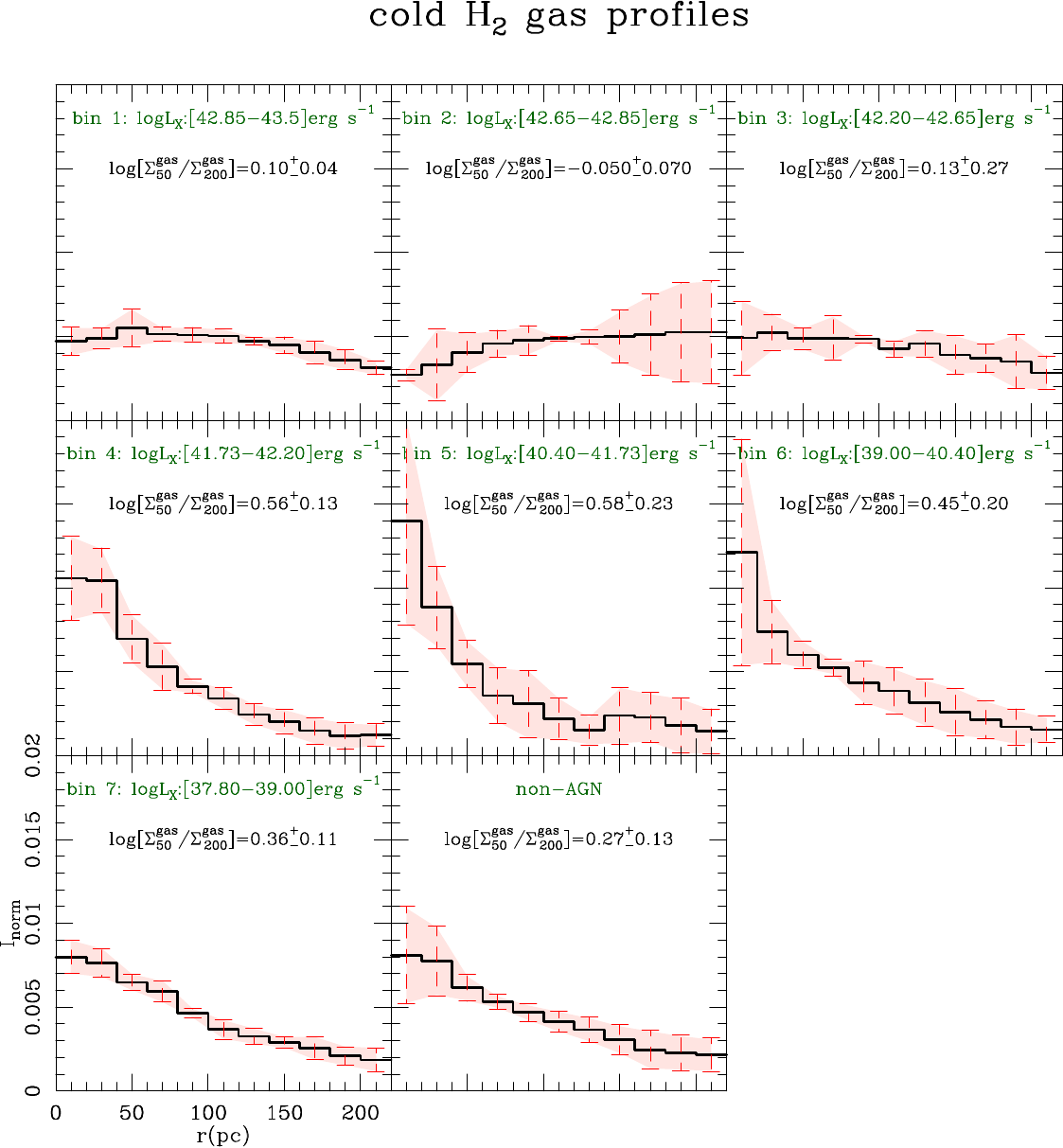}
           \caption{Panels show the average normalized radial distributions ($I_{\rm norm}(r)$) of `cold' molecular gas derived from CO out to $r=210$~pc for the 7 $L_{\rm X}$ bins  defined in  Sect.~\ref{CO-data}. We also show the $I_{\rm norm}$(r) profile representative of the non-AGN galaxy subset. Bins 1--4 and 5--7 correspond to the AGN feedback and build-up branches, respectively. Histograms and errorbars for each bin represent the median and the median absolute deviation values derived from the distribution of normalized individual galaxy profiles sampled with a radial binning $\Delta r=20$~pc. The red-shaded areas illustrate the uncertainties around the median values. The median value of the concentration index of molecular gas and its associated uncertainty are shown in each panel. The common scales adopted for all bins in the x and y axes are shown in the lower left panel.}  
   \label{histo-median-all}
    \end{figure*}

The left panel of Figure~\ref{conc-CO} illustrates how the concentration index of cold molecular gas, defined as 
the ratio (in logarithmic units) of the average H$_{\rm 2}$ surface densities derived from CO line emission at the two spatial scales representative of the nuclear  and CND scales changes as a function of 
$L_{\rm X}$ in the galaxies of the sample. The concentration index (hereafter referred to as $CCI$) is therefore defined as:

\begin{equation}
  CCI \equiv~{\rm log}_{\rm 10}(\Sigma^{\rm gas}_{\rm 50}/\Sigma^{\rm gas}_{\rm 200}). 
\end{equation}  

  Table~\ref{Tab2} lists the values of $CCI$ for the sample.
A visual inspection of Fig.~\ref{conc-CO} indicates that the distribution of median values of the concentration index for the 7 luminosity bins listed in Table~\ref{Tab3} \footnote{The binning used along the $L_{\rm X}$ axis is designed to have a similar number of 8-9 independent CO data sets per luminosity bin.} shows evidence of a turnover at  $L_{\rm X}
\simeq10^{41.5}$erg~s$^{-1}$. AGN with $L_{\rm X}\leq10^{41.5}$erg~s$^{-1}$ and non-AGN galaxies are located along a moderately ascending sequence, hereafter referred to as the `AGN build-up branch', 
which is characterized by median values of $CCI$ that monotonically increase with $L_{\rm X}$ from $\simeq+0.27$$\pm$0.13 to $\simeq+0.58$$\pm$0.23. However, we note that given the associated error bars, the reported increase in $CCI$ among the galaxies of the AGN build-up branch is not statistically significant.  AGN of higher luminosities lie at a descending 
sequence, hereafter referred to as the `AGN feedback branch', which is characterized by median values of $CCI$ that sharply decrease from  $\simeq+0.56$$\pm$0.13 down to $\simeq-0.05$$\pm$0.07. The 
concentration of molecular gas in the sample expands a statistically significant 0.63~dex range between the galaxies lying at the high end of the `AGN build-up branch' and the galaxies showing the most extreme nuclear-scale molecular gas deficits in the `AGN feedback branch'. As illustrated in Fig.~\ref{conc-CO-bis}, if we adopt the region defined by the outer corona of the CND ($50$~pc$\leq r \leq 200$~pc) to 
normalize  the molecular surface density of the nuclear region, the values of this alternative definition of the concentration of cold molecular gas expand a larger 0.72~dex range 
in our sample, equivalent to an overall  factor $\geq$5 difference.  We note that although galaxies along the AGN feedback branch tend to have higher Eddington ratios compared to the galaxies in the AGN build-up branch (as shown in the right panel of Fig.~\ref{conc-CO}), the corresponding trend of $CCI$ as a function of the Eddington ratio, which is available for fewer sources in our sample,  is less statistically significant than as a function of $L_{\rm X}$ (see Fig.~\ref{Edd_ratio}).

To further confirm statistically the existence of two branches grouping the galaxies of the sample in the parameter space of Fig.~\ref{conc-CO}, we used the Multivariate Adaptive Regression 
Splines ({\tt MARS}) fit routine available from the {\tt Rstudio} package \citep{Rcorteam}. The MARS algorithm gives the position of the breaking points (cut points) for a linear regression using multiple slopes. To this aim we 
derived the {\tt MARS} fit on 100 Monte Carlo realisations of the $CCI$  versus $L_{\rm X}$ relation taking into account the data uncertainties in both axes due to the assumed conversion factors ($\pm0.2$~dex), and on the AGN luminosities ($\pm0.15$~dex). The result of the fit is 
displayed in the right panel of Fig.~\ref{conc-CO}. From this simulation we find a turnover for the distribution at log$_{\rm 10}L_{\rm X}=41.51$$\pm0.31$~erg~s$^{-1}$ with a fit value of $CCI=0.53$$\pm$0.09, 
which similarly splits the sample into the two branches identified above based on the trend shown by the median values of $CCI$ as a function of $L_{\rm X}$. The power-law index found for the AGN buildup branch, $N=+0.03$$\pm$0.05, is consistent with a flat trend for $L_{\rm X}
\leq10^{41.5}$erg~s$^{-1}$, confirming that the differences in $CCI$ values are not statistically significant in this luminosity regime. In contrast, the power-law index found for the AGN feedback branch, $N=-0.36$$\pm$0.08, shows  a statistically significant evidence ($\simeq4.5\sigma$) of a descending trend of the concentration of molecular gas 
for galaxies lying beyond the turnover X-ray luminosity.

The fit of {\tt MARS} for the distribution of $CCI$ values shown in the lower panel of Fig.~\ref{conc-CO-bis}, which uses  the outer corona region of the CND  to 
normalize  the molecular surface densities, finds similar values for the turnover  at 
log$_{\rm 10}L_{\rm X}=41.54$$\pm0.25$~erg~s$^{-1}$  with a fit value of $CCI=0.64$$\pm$0.10, as well as power-law indices $N=+0.04$$\pm$0.06 and $N=-0.43$$\pm$0.09, for the AGN build-up 
and feedback branches respectively. The latter implies that the power-law index along the AGN feedback branch differs from zero by a statistically significant factor $\simeq~4.8\sigma$.

  \begin{figure}[tb!]
  \centering
    \includegraphics[width=6cm]{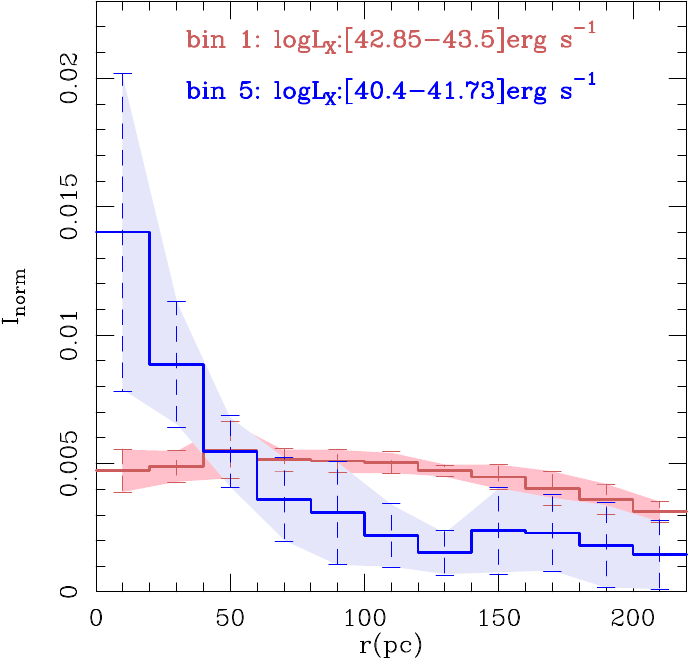}
    \includegraphics[width=6cm]{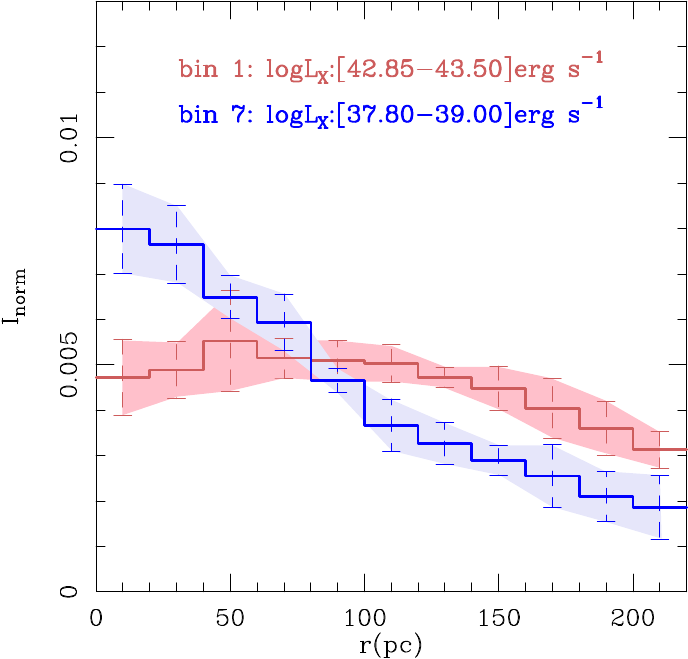}
       \caption{{\em Upper panel}:~Comparison between the average radial distributions of molecular gas for bins 1 and 5, which correspond to the subsets of galaxies showing the lowest and the highest concentration indices of molecular gas in the sample respectively. {\em Lower panel}:~same as {\em upper panel} but comparing  the average radial distributions of molecular gas for bins 1 and 7, which represent  the highest and the lowest ends of the $L_{\rm X}$ distribution in the sample respectively.}  
   \label{histo-median-bin1-5-7}
    \end{figure}

\subsection{Dependence of trends on CO transitions and conversion factors}\label{CO-dep}

The picture drawn from the different CO lines  is overall consistent within the uncertainties regarding the existence of the two branches described above.  In particular, along the AGN feedback branch, where the number of independent data sets ($N$) from the three transitions is comparable ($N$=13, 11, and 11 for the  CO(3--2), CO(2--1), and CO(1--0) lines, respectively) the derived $CCI$ values concur to show a similar monotonic decline with $L_{\rm X}$ beyond the turnover (see discussion in Appendix~\ref{CO-lines}). We can therefore conclude that for the typical range of physical conditions probed by the different CO transitions ($n_{\rm H_2}$$\simeq$a few 10$^{2-5}$~cm$^{-3}$, $T_{\rm K}$$\simeq$10-100~K) the decreasing trend of $CCI$ along the AGN feedback branch  reported in Fig.~\ref{conc-CO} shows a weak dependence on the particular line selected.

While a sizeable fraction of the data along the AGN build-up branch has been obtained only in the CO(2--1) line ($N$=6, 13, and 3 for the  CO(3--2), CO(2--1), and CO(1--0) lines, respectively), the range of $CCI$ values derived from the different transitions is comparable within the errors. However, confirmation of whether the trend of $CCI$ with $L_{\rm X}$ along the AGN build-up branch is either moderately increasing or flat would require obtaining more data in the CO(3--2) and CO(2--1) lines.

 While the values of $R_{\rm 31}$, $R_{\rm 21}$ for the galaxies in our sample are not constrained, we nevertheless expect that molecular gas will be  comparatively  less excited on nuclear scales at low AGN luminosities. Significant changes in the value  $\alpha_{\rm CO}$ have only been described within galaxies on kpc-scales  between their central kpc-region and their outer 
($r>0.2R_{\rm 25}$) disks \citep[e.g.,][]{San13,Yas23}.  In any case, this is well beyond the spatial scales relevant to this work.  The recent work of \citet{Ten23} found that the value of $\alpha_{\rm CO}$ in galaxy centers decreases with  CO opacity and shows a strong anticorrelation with the gas velocity dispersion. Moreover, if anything,  $\alpha_{\rm CO}$ values would probably have to be lowered on the nuclear ($r\leq50$~pc) as opposed to the CND-scales ($r\leq200$~pc) of the highest luminosity AGN galaxies in our sample where gas is expected to be more strongly irradiated and live in a dynamically perturbed environment characterized by an enhanced turbulence. Overall, our adopted choice of common values of the conversion factors for the nuclear and CND scales would be therefore conservative, underestimating the differences between the concentration of molecular gas measured for  AGN build-up and AGN feedback branch galaxies shown in Fig.~\ref{conc-CO}.

  \begin{figure*}[tb!]
  \centering
    \includegraphics[width=13.2cm]{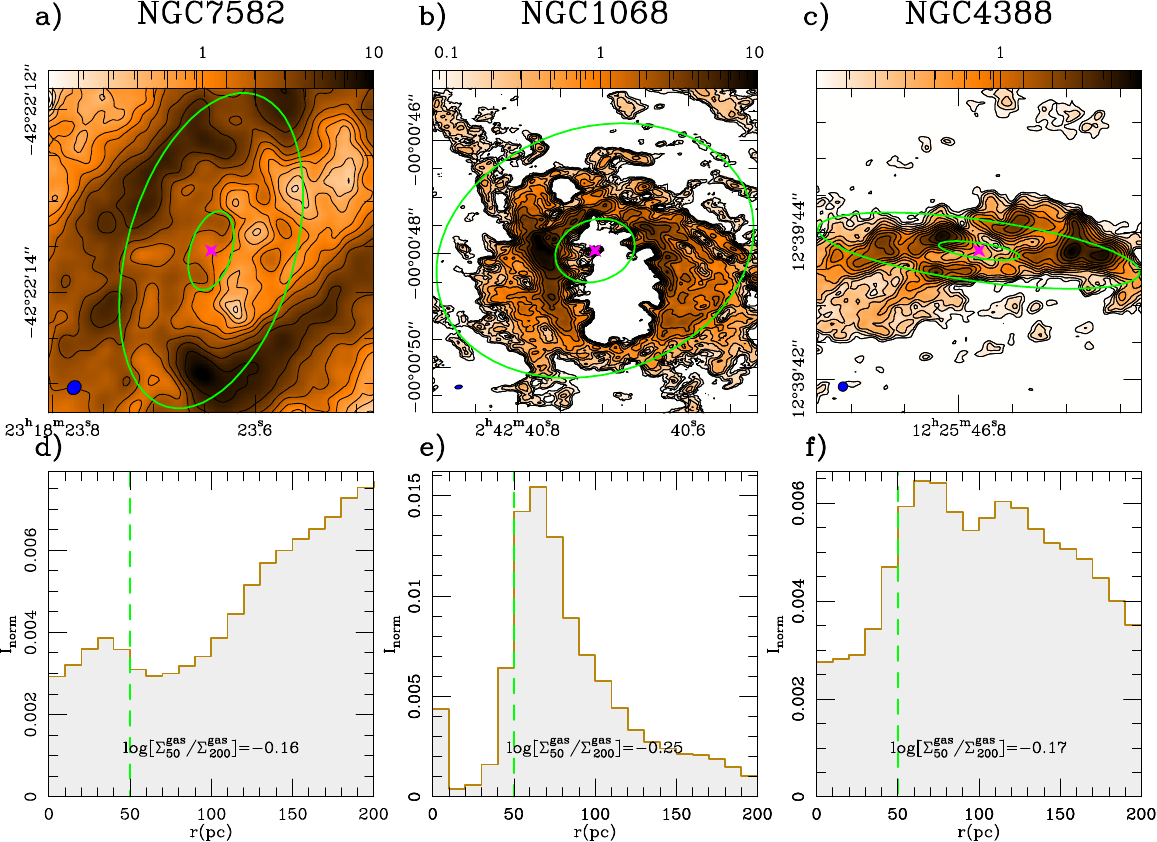}
       \caption{{\em Upper panels}:~CO velocity-integrated intensity maps derived for 3 representative targets selected from the galaxies of $L_{\rm X}$ bin~1 (NGC~7582; {\em a)}), bin 2 (NGC~1068; {\em b)}),  and bin~3 (NGC~4388; {\em c)}), as defined in Sect.~\ref{CO-ratios}. Contour levels have a logarithmic spacing from 2.5$\sigma$ co to 90$\%$ of the peak CO intensity inside the displayed field of view, which corresponds to 400~pc$\times$400~pc. The (green) ellipses identify the circular regions in the deprojected planes of the galaxies extending out to $r\leq50$~pc and $r\leq200$~pc. The (blue) filled ellipses  at the bottom left corners represent the beam sizes of the observations. {\em Lower panels}:~Normalized radial distributions of molecular gas derived out to $r=210$~pc for the galaxies shown in the {\rm upper panels}. The inner $r\leq50$~pc-region is delimited by the dashed line. The estimated molecular gas concentration indices are displayed in panels {\em d)}-to-{\em f)}.}
   \label{maps-bins1-2-3}
    \end{figure*}

  \begin{figure*}[tb!]
  \centering
    \includegraphics[width=13.2cm]{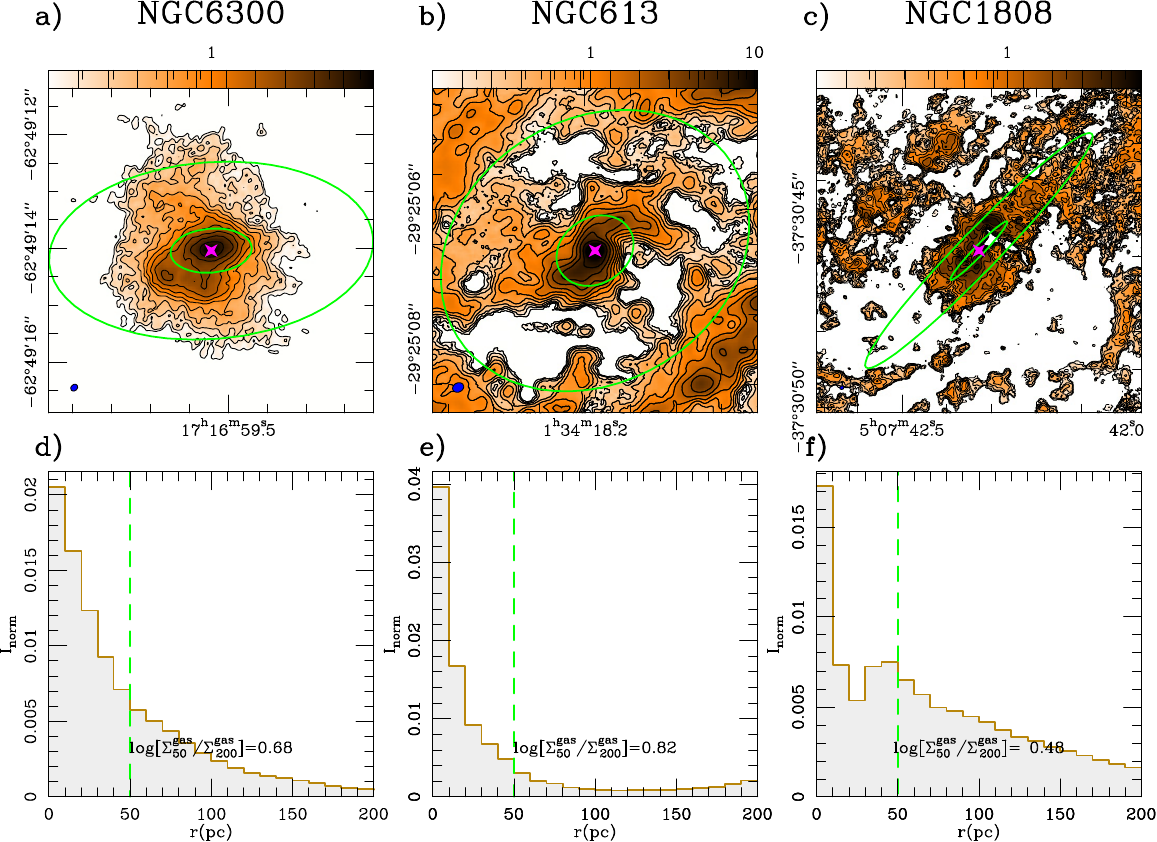}
       \caption{Same as the {\em upper} and  {\em lower} panels of Fig.~\ref{maps-bins1-2-3} but for  3 representative targets selected from the galaxies of $L_{\rm X}$ bin~4 (NGC~6300; {\em a)}), bin 5 (NGC~613; {\em b)}),  and bin~6 (NGC~1808; {\em c)}). Symbols as in Fig.~\ref{maps-bins1-2-3}.}  
   \label{maps-bins4-5-6}
    \end{figure*}

  \begin{figure}[tb!]
  \centering
    \includegraphics[width=9cm]{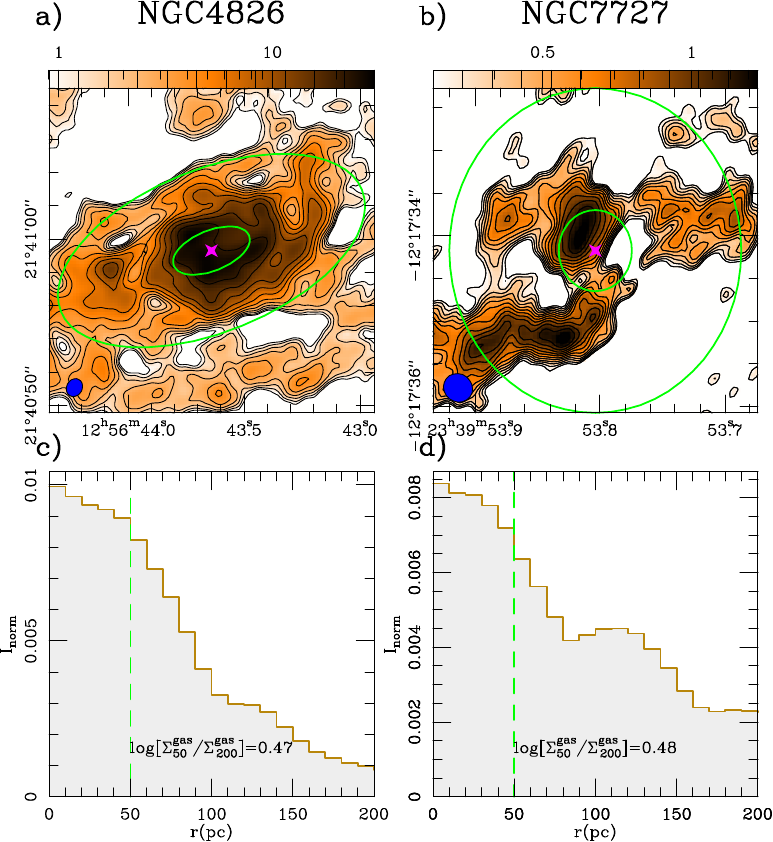}
       \caption{Same as the {\em upper} and  {\em lower} panels of Fig.~\ref{maps-bins1-2-3} but for 2 representative targets  selected from the galaxies of $L_{\rm X}$ bin~7 (NGC~4826; {\em a)}) and from the subset of non-AGN  (NGC~7727; {\em b)}). Symbols as in Fig.~\ref{maps-bins1-2-3}.}  
   \label{maps-bins7-8}
    \end{figure}

\subsection{Radial distributions}\label{CO-profiles}

We have shown in Sect.~\ref{CO-ratios} that the distribution of $CCI$ values for the galaxies of the sample shows a turnover at log$_{\rm 10}L_{\rm X}\simeq41.5$$\pm0.3$~erg~s$^{-1}$. We explore below in more detail how the  shape of the radial profiles of cold molecular gas changes as a function of $L_{\rm X}$ in the sample. To this aim, we derived the radial profiles of cold molecular gas out to $r=210$~pc for each target using  a radial binning $\Delta r=20$~pc, which corresponds to about half of the median value of the spatial resolutions of the CO observations used in this work. Individual profiles are further normalized by the CO flux integrated out to $r=210$~pc for each target ($I_{\rm norm}(r)$). Finally, we derived the  median values of $I_{\rm norm}(r)$ and their associated median absolute deviations for the 7 $L_{\rm X}$ bins  defined in  Sect.~\ref{CO-ratios}. 

Figure.~\ref{histo-median-all} shows the  $I_{\rm norm}(r)$ profiles derived from CO in the sample.  As expected, the profiles show remarkable differences that fairly reflect the trends shown by the $CCI$ values as function of $L_{\rm X}$. Profiles appear as flat or inverted (i.e., with lower $I_{\rm norm}(r)$ values at small radii)  for the highest X-ray luminosities  of the galaxies of the AGN feedback branch grouped in $L_{\rm X}$ bins 1, 2, and 3. In contrast, galaxies in the peak phase of the AGN build-up branch, grouped in  $L_{\rm X}$ bins 5, and 6, show highly concentrated molecular gas distributions. Furthermore, the lowest luminosity AGN and the non-AGN targets, grouped in  bin 7 and the non-AGN bin respectively, show  comparatively less concentrated radial profiles of molecular gas.

Figure ~\ref{histo-median-bin1-5-7} shows that these differences are statistically significant when we compare the profiles of the highest luminosity AGN feedback branch targets (bin 1) with those of the highest luminosity AGN build-up phase targets (bin 5). A similar conclusion can be drawn when we compare the radial profiles of the galaxies belonging to the two extreme $L_{\rm X}$ classes, namely bins 1 and 7. 
In particular, Kolmogorov-Smirnov (KS)  tests performed to reject the null hypothesis that the two pairs of radial profiles corresponding to bins 1 and 5, as well as to bins 1 and 7, are drawn from the same distribution yield consistently low $p$-values in either case  ($p$-values$\leq$0.001).

Figures~\ref{maps-bins1-2-3}, ~\ref{maps-bins4-5-6}, and ~\ref{maps-bins7-8} show the 2-dimensional and  normalized radial distributions of the cold molecular gas in the inner $r\leq200$~pc central regions of 8 galaxies representative of  the different AGN and non-AGN bin categories.  The CO images help visualize directly the noticeable progression in the morphology of the distribution of the gas as a function of $L_{\rm X}$. The latter is characterized by a marked nuclear-scale ($r\leq50$~pc) deficit  in the AGN feedback branch targets belonging to bins 1, 2, and 3  (NGC~7582, NGC~1068, and NGC~4388, respectively), which is reminiscent of a ring-like distribution. This is in stark contrast with the highly-concentrated nuclear-scale distribution of the gas in the targets of bins 4, 5 and 6 (NGC~6300, NGC~613, and NGC1808, respectively). The presence of a contrasted two-arm CO trailing nuclear spiral structure in  NGC~613 and NGC~1808 has been interpreted as smoking gun evidence of ongoing AGN feeding in these targets \citep{Aud19, Aud21}. This particular morphology would appear at an evolutionary stage where the gas is piling up on nuclear scales under the influence of negative gravity torques inside the Inner Lindblad Resonance (ILR) regions of the stellar bars \citep {Wad94, But96, Pee06, Com14}. The distribution of cold molecular gas in the lower luminosity and non-AGN targets, shown in Fig.~\ref{maps-bins7-8} (NGC~4826 and NGC~7727, respectively) is less centrally concentrated on average.

  \begin{figure}[tb!]
   \centering
   	\includegraphics[width=6.76cm]{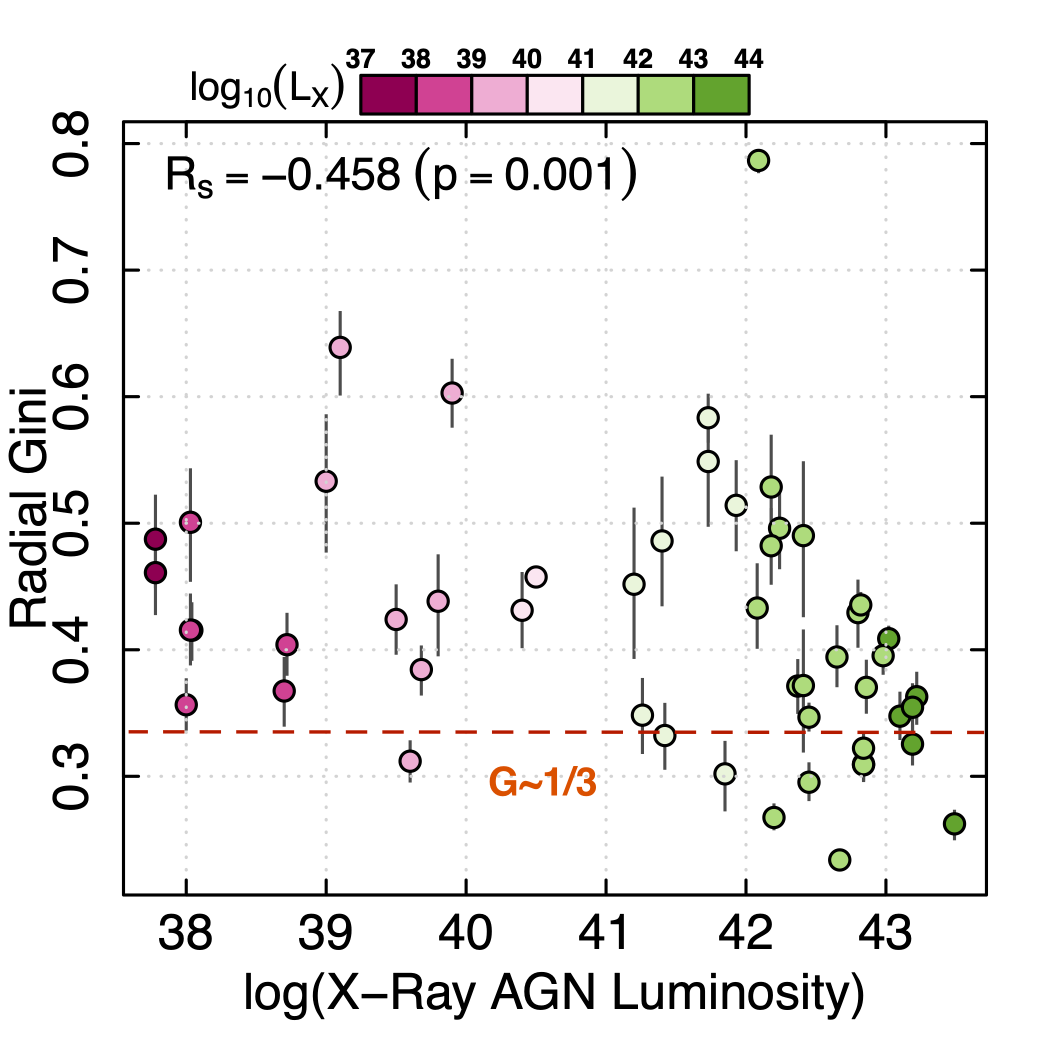}
       	\includegraphics[width=6.76cm]{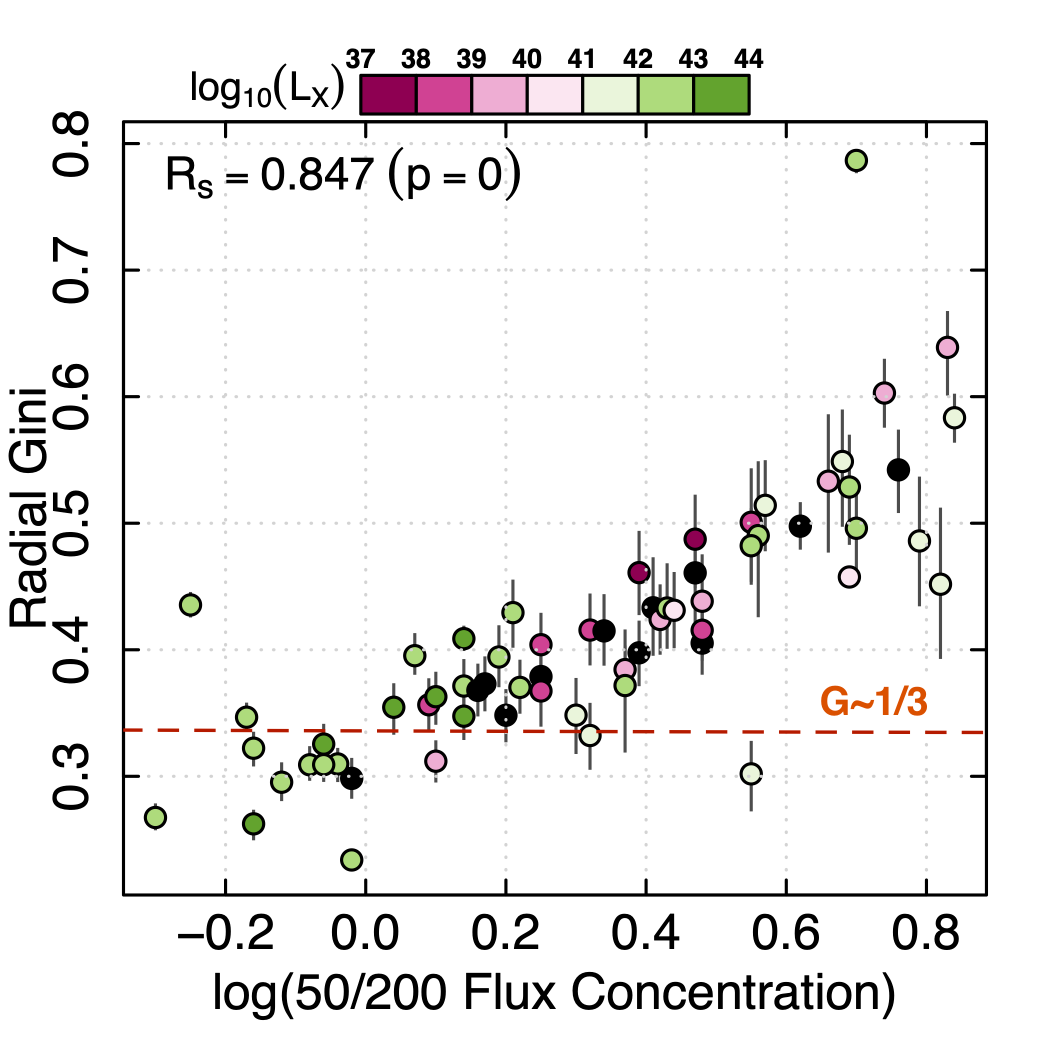}
	\includegraphics[width=6.76cm]{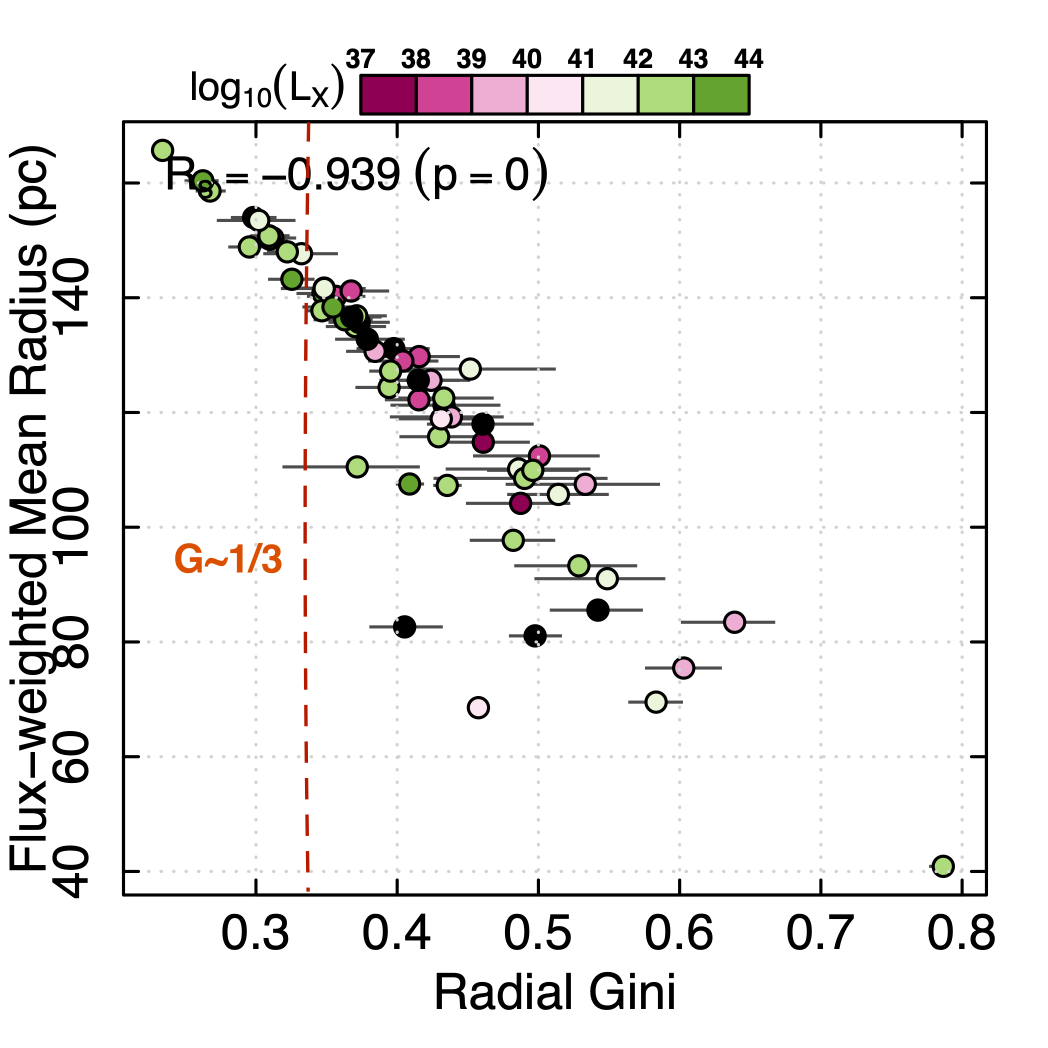}	
       \caption{Panels show the Gini coefficients derived for the radial profiles obtained from the CO distributions in the galaxies  of the sample as function of  $L_{\rm X}$ ({\em upper panel}) and the  
       concentration  index of cold molecular gas ({\em middle panel}). Flux-weighted mean radii of the normalized distribution of cold molecular gas  as function of the Gini coefficients of the radial profiles derived for the galaxies of our sample ({\em lower panel}). Errorbars account for the estimated uncertainties on the Gini coefficients. We indicate the estimated correlation coefficients ($R_{\rm s}$) and their associated $p$-values for the 3 distributions. Gini coefficient values are color-coded as a function of $L_{\rm X}$ (black color for non-AGN galaxies). }  
   \label{Gini}
    \end{figure}

\subsection{Non-parametric analysis of the radial profiles}\label{non-param}

We confirm these results with an alternative, non-parametric method that builds on the Gini coefficient, $G$. This statistic, originally an econometric tool to measure income or wealth (in)equality, has been adopted in astronomy to e.g. quantify the (in)equality of the flux/brightness distribution in pixel sets \citep{Abr03, Lot04, Dav22}. Here, we use $G$ in a different manner to assess the shape of our CO radial profiles. In short, CO flux and radius in our case respectively play the roles of people and income in the original application of $G$. More precisely, we measure the inequality in the distribution of radius$^2$, so that a flat radial profile corresponds to $G\approx1/3$, as expected for a uniform distribution. This is basically a normalization prescription that does not affect the results.

We calculate $G$ for each galaxy in our sample over the $r$=0--200~pc range with the \verb|DescTools R| package. Associated 99\% confidence intervals are derived from bootstrap percentiles. Fig.~\ref{Gini} (upper panel) shows $G$ as a function of $L_X$ for the X-ray detected AGN galaxies. To first order, the overall distribution resembles that of $CCI$ in Fig.~\ref{conc-CO}: a virtually flat trend up to $L_X\approx10^{41.5}$~erg~s$^{-1}$, at which point a noticeable drop sets in. This brings X-ray brightest galaxies close to the $G\approx1/3$ value that, as indicated above, corresponds to a flat CO radial profile. The excellent agreement between the $G$ and $CCI$ trends stems from the tight correlation between the two statistics (Spearman coefficient, $R_s$ of $\sim0.85$), which we show in Fig.~\ref{Gini} (middle panel). Fig.~\ref{Gini} (lower panel) displays an even tighter anticorrelation between $G$ and the flux-weighted mean radius ($R_s\sim-0.94$), which help us visualize the meaning of $G$ in this context. I.e., high $G$ values are linked with the build-up of nuclear concentrations of gas, whereas $G$ decreases as the mean radius of the CO light shifts outwards. In Fig.~\ref{Gini} (lower panel), $G\approx1/3$ corresponds to mean radii of $\lesssim 140$~pc, which is close to the expectation for a radial profile flat over the $r$=0--200~pc range (2/3$\times200\approx133$~pc).

  \begin{figure*}[tb!]
  \centering
    \includegraphics[width=8cm]{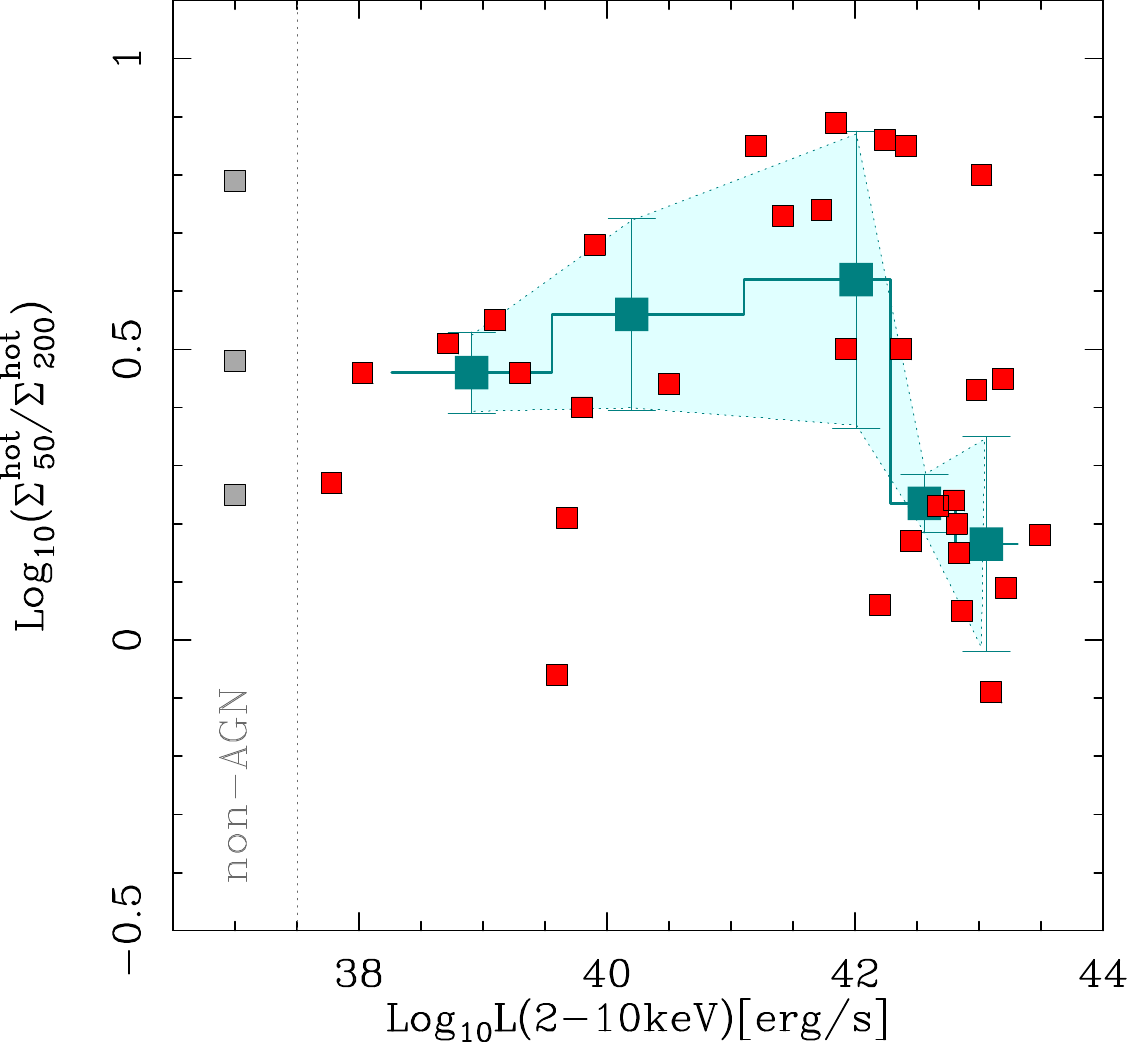}
     \includegraphics[width=8cm]{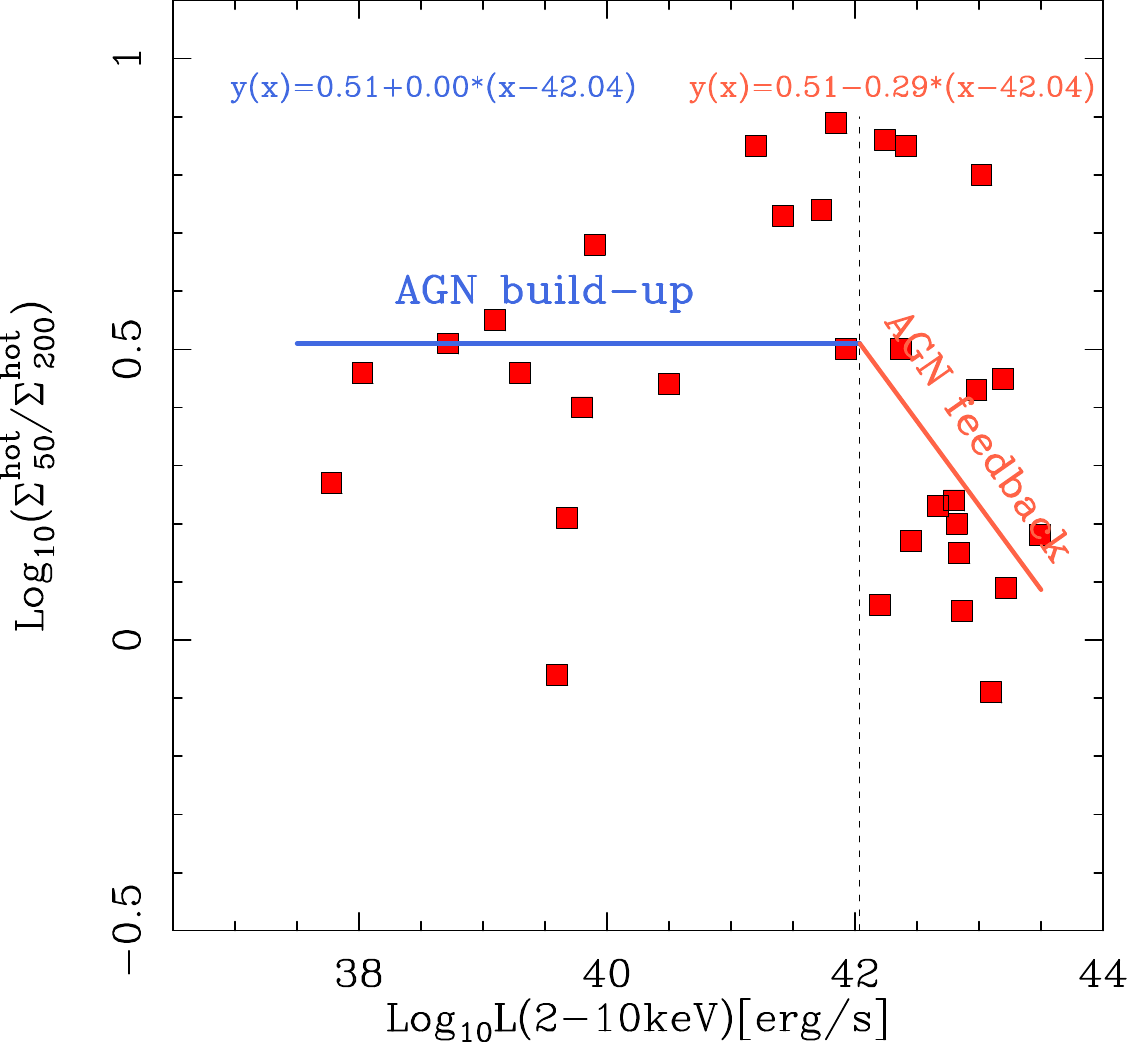}
        \caption{{\em Left panel}: Same as the {\em left panel} of Fig.~\ref{conc-CO} but showing the change with $L_{\rm X}$ of the concentration of `hot' molecular gas measured for the central regions of the galaxies in our sample derived  from the ratio (in log units) of the average H$_{\rm 2}$ surface densities estimated from the 2.1$\mu$m line emission at two spatial scales: $r\leq50$pc ($\Sigma^{\rm hot}_{\rm 50}$) and $r\leq200$pc ($\Sigma^{\rm hot}_{\rm 200}$).  The small (red) squares represent the individual galaxy measurements for AGN in the sample. Concentration indices for the 3 non-AGN targets are displayed as (gray) square symbols at log($L_{\rm X})$$=37$~erg~s$^{\rm -1}$. The big (green) squares stand for the median value of the concentration index estimated  for the 5 $L_{\rm X}$  bins that cover the range of AGN luminosities spanned by our sample, as  defined in  Sect.~\ref{NIR-ratios}. Symbols for errorbars are as in Fig.~\ref{conc-CO}. {\em Right panel}:~same as {\em left panel} but showing the two-branch linear solution found by the MARS algorithm to fit the distribution of concentration indices of `hot' molecular gas as a function of $L_{\rm X}$ for the AGN build-up phase (blue straight line) and the AGN feedback phase (red straight line).}  
   \label{conc-hot}
    \end{figure*}

  \begin{figure*}[tb!]
  \centering
    \includegraphics[width=0.77\textwidth]{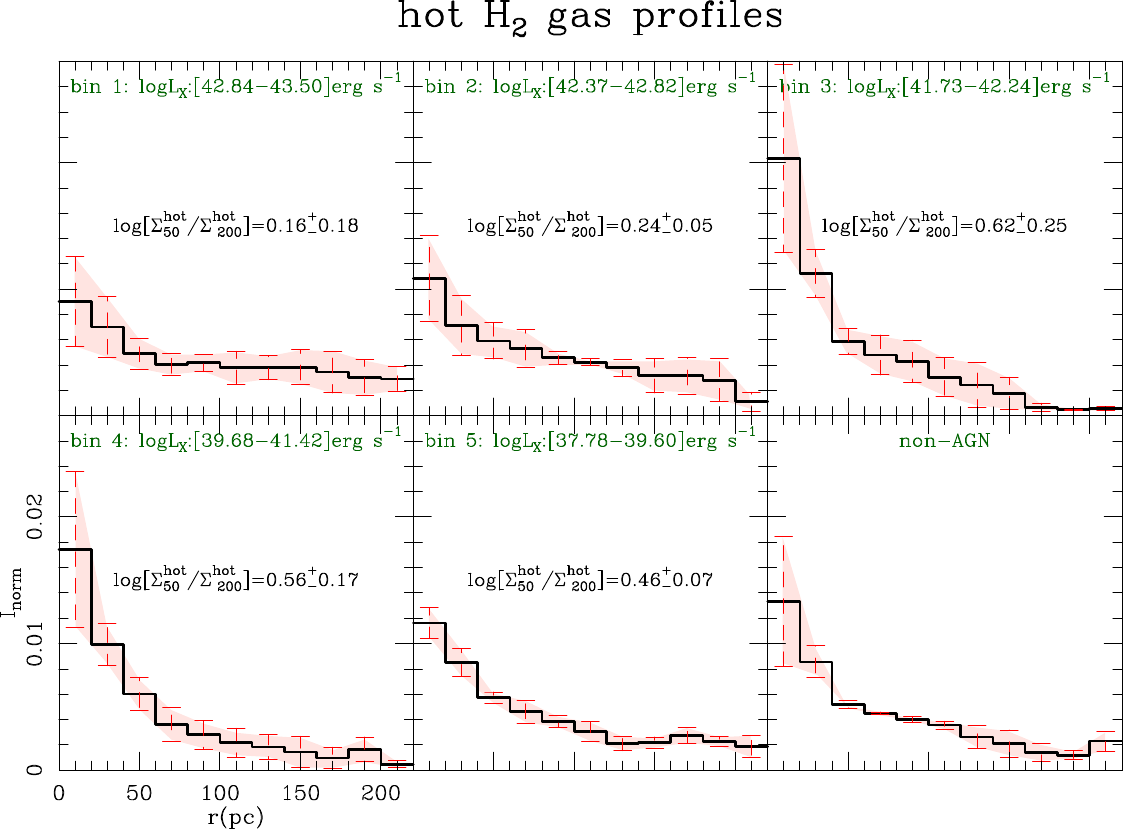}
       \caption{Same as Fig.~\ref{histo-median-all} but showing the average normalized radial distributions ($I_{\rm norm}$(r)) of `hot' molecular gas derived out to $r=210$~pc from the  2.1$\mu$m line emission for the 5 $L_{\rm X}$ bins  defined in  Sect.~\ref{NIR-ratios} and for the non-AGN galaxy subset. Symbols for histograms and errorbars  for the median and the median absolute deviation values for the `hot' molecular gas profiles are defined and shown as in Fig.~\ref{histo-median-all}.The median value of the concentration index of `hot' molecular gas and its associated uncertainty are shown in each panel.The common scales adopted for all bins in the x and y axes are shown in the lower left panel.}  
   \label{histo-NIR-median-all}
    \end{figure*}

  \begin{figure}[tb!]
  \centering
    \includegraphics[width=6cm]{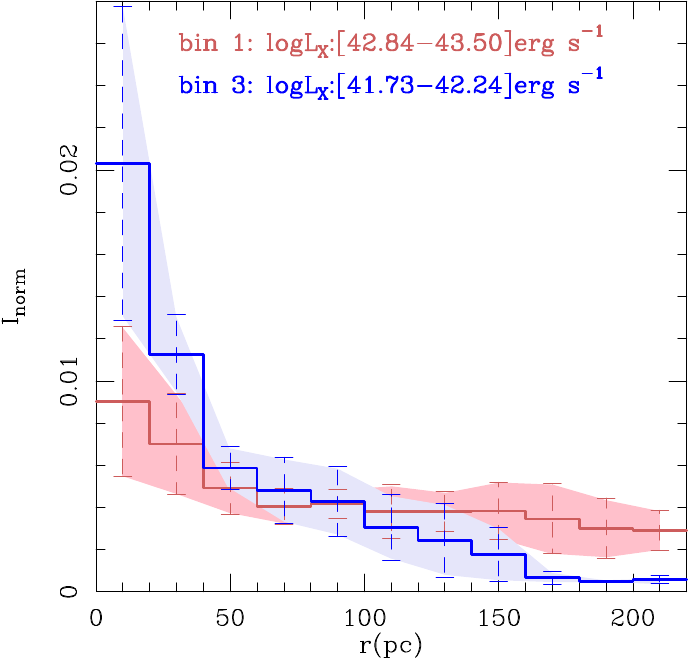}
    \includegraphics[width=6cm]{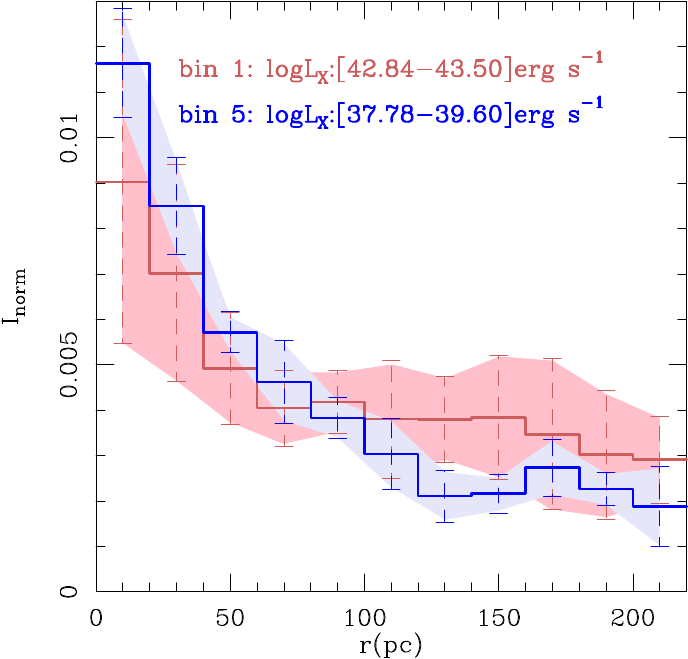}
       \caption{{\em Upper panel}:~Comparison between the average radial distributions of molecular gas for bins 1 and 3, which correspond to the subsets of galaxies showing the lowest and the highest concentration indices of `hot' molecular gas in the sample respectively, as defined in Sect.\ref{NIR-ratios}. {\em Lower panel}:~same as {\em upper panel} but comparing  the average radial distributions of molecular gas for bins 1 and 5 defined in Sect.\ref{NIR-ratios}, which represent  the highest and the lowest ends of the $L_{\rm X}$ distribution in our sample respectively.}    
    \label{histo-median-bin1-3-7}
    \end{figure}

\section{Distribution of hot molecular gas}\label{hot-H2}

More than half of the targets of the CO sample have available data from the H$_{\rm 2}$ 1–0 S(1) 2.12~$\mu$m  line emission, a tracer of `hot' molecular gas ($T_{\rm K}$$\simeq$a few 10$^3$~K). Using a similar methodology adopted  in Sects.~\ref{CO-ratios} and \ref{CO-profiles} to study the cold molecular gas, we use the NIR data to derive the concentration indices and the normalized radial distributions of hot molecular gas in the galaxies of our sample and investigate their potential trends with $L_{\rm X}$ in Sects.~\ref{NIR-ratios} and \ref{NIR-profiles}.

\subsection{Concentration indices}\label{NIR-ratios}

Figure~\ref{conc-hot} shows  the change with $L_{\rm X}$ of the concentration of hot molecular gas measured for the central regions of the galaxies of the sample as derived  from the ratio of the average H$_{\rm 2}$ surface densities estimated from the 2.1$\mu$m line emission at the two spatial scales used in this work:  nuclear ($r\leq50$pc; $\Sigma^{\rm hot}_{\rm 50}$) and CND scales 
($r\leq200$pc; $\Sigma^{\rm hot}_{\rm 200}$). The concentration index of hot molecular gas (hereafter referred to as $HCI$) is therefore defined as:

\begin{equation}
  HCI \equiv~{\rm log}_{\rm 10}(\Sigma^{\rm hot}_{\rm 50}/\Sigma^{\rm hot}_{\rm 200}). 
\end{equation}  
  
  By definition, the values of $HCI$ are insensitive to any potential galaxy-to-galaxy variation of the conversion factors implicitly assumed in Eq~\ref{hot-eq}. Table~\ref{Tab2} lists the values of $HCI$ for the sample.

 Figure~\ref{conc-hot} shows that the trend of $HCI$ with $L_{\rm X}$, while qualitatively similar, is significantly much less pronounced than the trend shown by the $CCI$ index in Fig.~\ref{conc-CO}. The distribution of median values of $HCI$ for the 5 luminosity bins listed in Table~\ref{Tab3} \footnote{The binning used along the $L_{\rm X}$ axis is chosen to have a similar number of 6-8 independent NIR data sets per luminosity bin.} shows only tentative evidence of a turnover at  $L_{\rm X}\simeq10^{42.0}$erg~s$^{-1}$. The position of this  break point is close to the turnover  identified in  the distribution of galaxies of the sample in the $CCI$-$L_{\rm X}$ parameter space of Fig.~\ref{conc-CO}.  AGN below $L_{\rm X}\leq10^{42.0}$erg~s$^{-1}$ in bins 3, 4, and 5 lie at a sequence characterized by median values of $HCI$ that monotonically increase with $L_{\rm X}$ from $\simeq+0.46$$\pm$0.07 to $\simeq+0.62$$\pm$0.25. However, we note that given the associated error bars, the reported increase in $HCI$ is not statistically significant. Beyond the turnover luminosity, AGN  in bins 1 and 2 lie at a descending  sequence qualitatively similar to the `AGN feedback branch' of Fig.~\ref{conc-CO}, which is characterized by median values of $HCI$ that decrease from  $\simeq+0.24$$\pm$0.05 down to $\simeq+0.16$$\pm$0.18.  These values are marginally different to the ones characterizing the peak phase of the AGN build-up branch. In particular, the 
concentration of hot molecular gas in the sample spans a 0.46~dex range between the galaxies lying in the peak phase of the AGN build-up branch and the most extreme  targets of the AGN feedback branch (see Fig.~\ref{conc-hot}), namely a factor of two smaller range than shown by the $CCI$ index.

Our working hypothesis in deriving $HCI$ for the galaxies in our sample is that the conversion factor used in Eq.~\ref{hot-eq} does not change between the nuclear and CND scales. Moreover, if anything, this conversion factor  would probably have to be reduced on the nuclear ($r<$50~pc) scales where gas is expected to be hotter in the highest luminosity AGN sources. In this scenario we would be underestimating the differences between the concentration of  hot molecular gas measured for AGN build-up and AGN feedback branch galaxies (see also a similar discussion regarding the trend of $CCI$ as a function of $L_{\rm X}$ in Sect.~\ref{CO-ratios}).

Following the same approach of Sect.~\ref{CO-ratios} we used the {\tt MARS} algorithm to find the position of the breaking point  and the slopes of the two AGN branches. The result of the fit is  displayed in the right panel of Fig.~\ref{conc-hot}. From this simulation we find a turnover for the distribution at log$_{\rm 10}L_{\rm X}=42.04$$\pm0.27$~erg~s$^{-1}$ with a fit value of $HCI=0.51$$\pm$0.04, 
which similarly splits the sample into the two branches identified above based on the trend shown by the median values of $HCI$ as a function of $L_{\rm X}$.  The power-law index found for the AGN build-up branch,  $N=+0.00$$\pm$0.05, is compatible with a flat trend for $L_{\rm X}\leq10^{42.04}$erg~s$^{-1}$. The power-law index found for the AGN feedback branch, $N=-0.29$$\pm$0.11, shows only a very marginal statistical evidence ($\simeq2.6\sigma$) of a decreasing trend of the concentration of hot molecular gas 
for galaxies lying beyond the turnover X-ray luminosity. This confirms the visual impression that the decrease in molecular gas concentration identified in galaxies belonging to the AGN feedback branch most severely affects the cold phase of the gas  probed by CO.

\subsection{Radial distributions}\label{NIR-profiles}

Figure~\ref{histo-NIR-median-all} shows the average normalized radial distributions ($I_{\rm norm}$(r)) of hot molecular gas derived out to $r=210$~pc from the  2.1$\mu$m line emission for the 5 $L_{\rm X}$ bins  defined in  Sect.~\ref{NIR-ratios}. The profiles show differences that reflect the trends displayed by the $HCI$ values as function of $L_{\rm X}$ in Fig.~\ref{conc-hot}.
Profiles appear to be nearly flat or with a moderate degree of central concentration for the higher X-ray luminosities characteristic of AGN feedback branch galaxies, grouped in $L_{\rm X}$ bins 1 and 2. Galaxies in the peak phase of the AGN build-up branch, grouped in  $L_{\rm X}$ bins 3, 4, and 5, tend to show higher concentrations of hot molecular gas.  

Figure~\ref{histo-median-bin1-3-7} shows that these differences are marginally statistically significant when we compare the profiles of the highest luminosity AGN feedback branch targets (bin 1) with those of the highest luminosity AGN build-up phase targets (bin 3). However, the differences in the profiles between galaxies belonging to the two extreme $L_{\rm X}$ classes are not statistically significant (bins 1 and 5).


  \begin{figure*}[tb!]
  \centering
    \includegraphics[width=12cm]{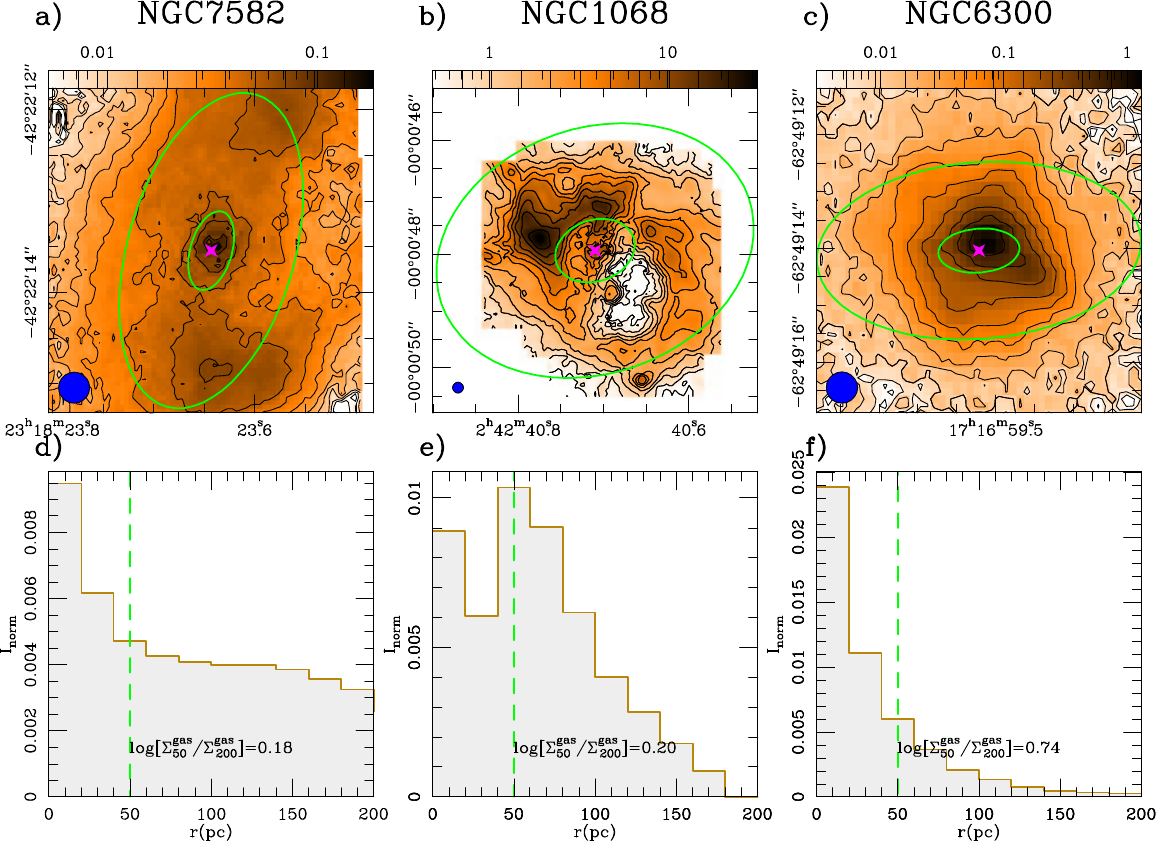}
       \caption{{\em Upper panels}:~velocity-integrated intensity maps derived from the 2.1~$\mu$m line for 3 representative targets selected from the galaxies of $L_{\rm X}$ bin~1 (NGC~7582; {\em a)}), bin 2 (NGC~1068; {\em b)}),  and bin~3 (NGC~6300; {\em c)}),  as defined in Sect.~\ref{NIR-ratios}. Contour levels have a logarithmic spacing from 2.5$\sigma$ to 90$\%$ of the peak 2.1~$\mu$m intensity inside the displayed field of view, which corresponds to 400~pc~$\times$~400~pc. {\em Lower panels}:~Normalized radial distributions of `hot' molecular gas derived out to $r=210$~pc for the galaxies shown in the {\rm upper panels}. The estimated `hot' molecular gas concentration indices are displayed in panels {\em d)}-to-{\em f)}. All symbols as in Fig.~\ref{maps-bins1-2-3}.}  
   \label{maps-bins1-2-3-NIR}
    \end{figure*}

  \begin{figure*}[tb!]
  \centering
    \includegraphics[width=12cm]{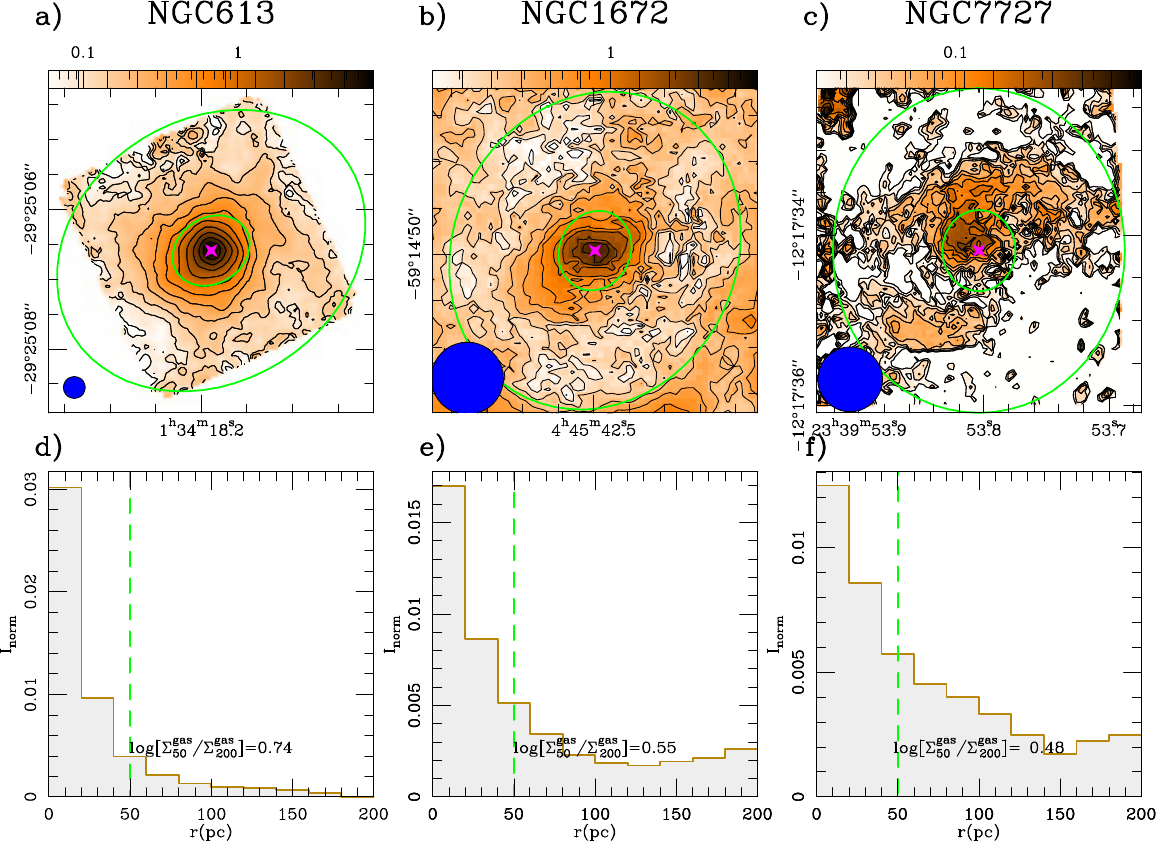}
       \caption{Same as the {\em upper} and  {\em lower} panels of Fig.~\ref{maps-bins1-2-3-NIR} but showing the maps and radial distributions of 3 representative targets selected from the galaxies of $L_{\rm X}$ bin~4 (NGC~613; {\em a) $\&$ d)}), bin 5 (NGC~1672; {\em b) $\&$ e)}),  and the non-AGN subset (NGC~7727; {\em c) $\&$ f)}), as defined in Sect.~\ref{NIR-ratios}. Symbols as in Fig.~\ref{maps-bins1-2-3-NIR}.}  
   \label{maps-bins4-5-6-NIR}
    \end{figure*}

Figures~\ref{maps-bins1-2-3-NIR} and ~\ref{maps-bins4-5-6-NIR} show the 2-dimensional and  normalized radial distributions of the hot molecular gas in the inner $r\leq200$~pc central regions of 6 galaxies representative of  the different AGN and non-AGN bin categories. These images help visualize the differences in the morphology of the distribution of the hot gas as a function of $L_{\rm X}$.
The latter is characterized by a low nuclear-scale ($r\leq50$~pc) concentration of hot molecular gas  in the AGN feedback branch targets, belonging to bins 1, 2 (NGC~7582 and NGC~1068 respectively). This is in contrast with the more concentrated nuclear-scale distribution of the gas in the targets  in the peak phase of the AGN build-up branch of bins 3, 4, and 5 (NGC6300, NGC~613, and NGC1672, respectively). The non-AGN target NGC7727 shows a concentration similar to that of NGC~1672.

  \begin{figure}[t!]
  \centering
    \includegraphics[width=0.47\textwidth]{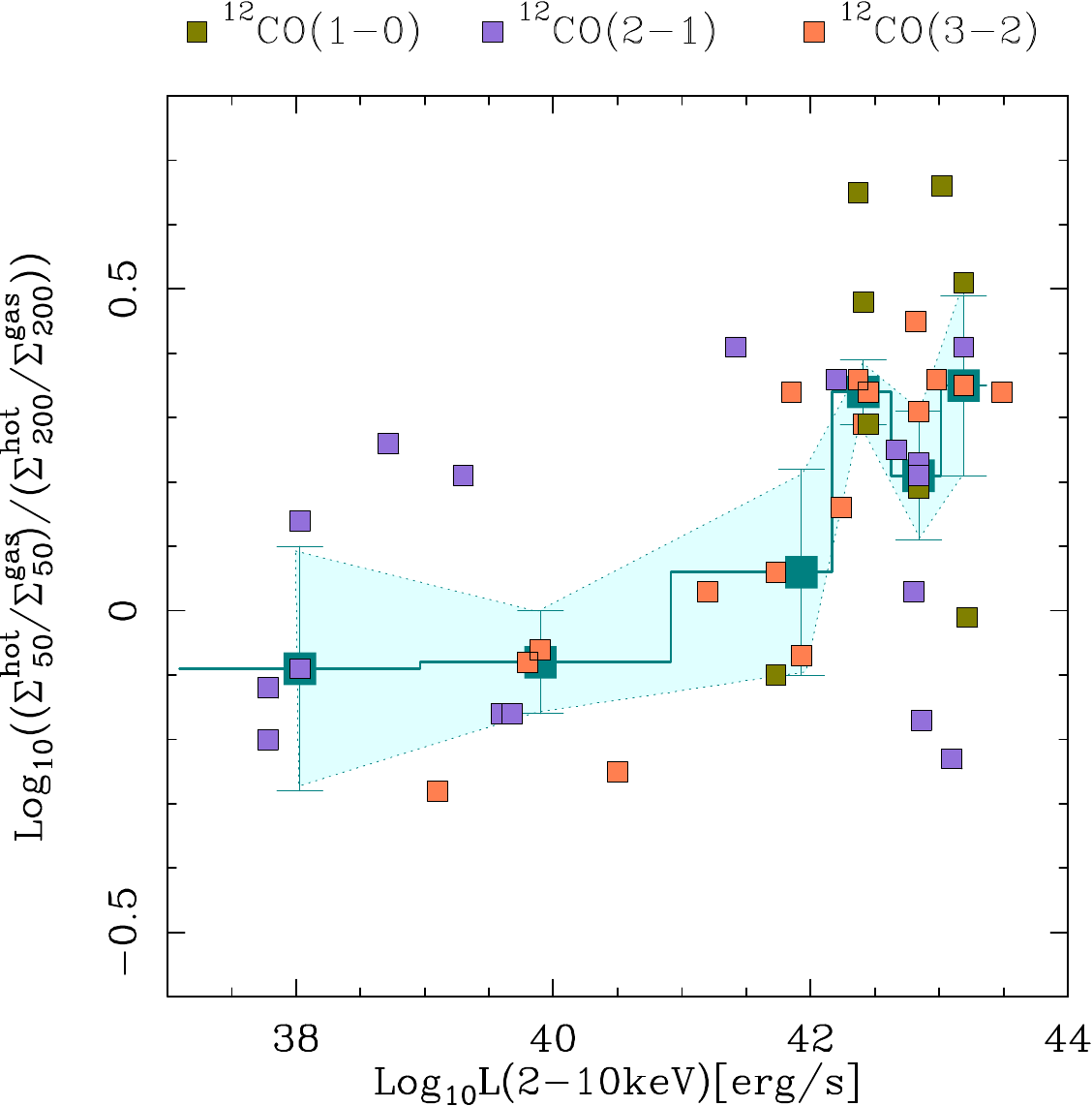}
     \includegraphics[width=0.47\textwidth]{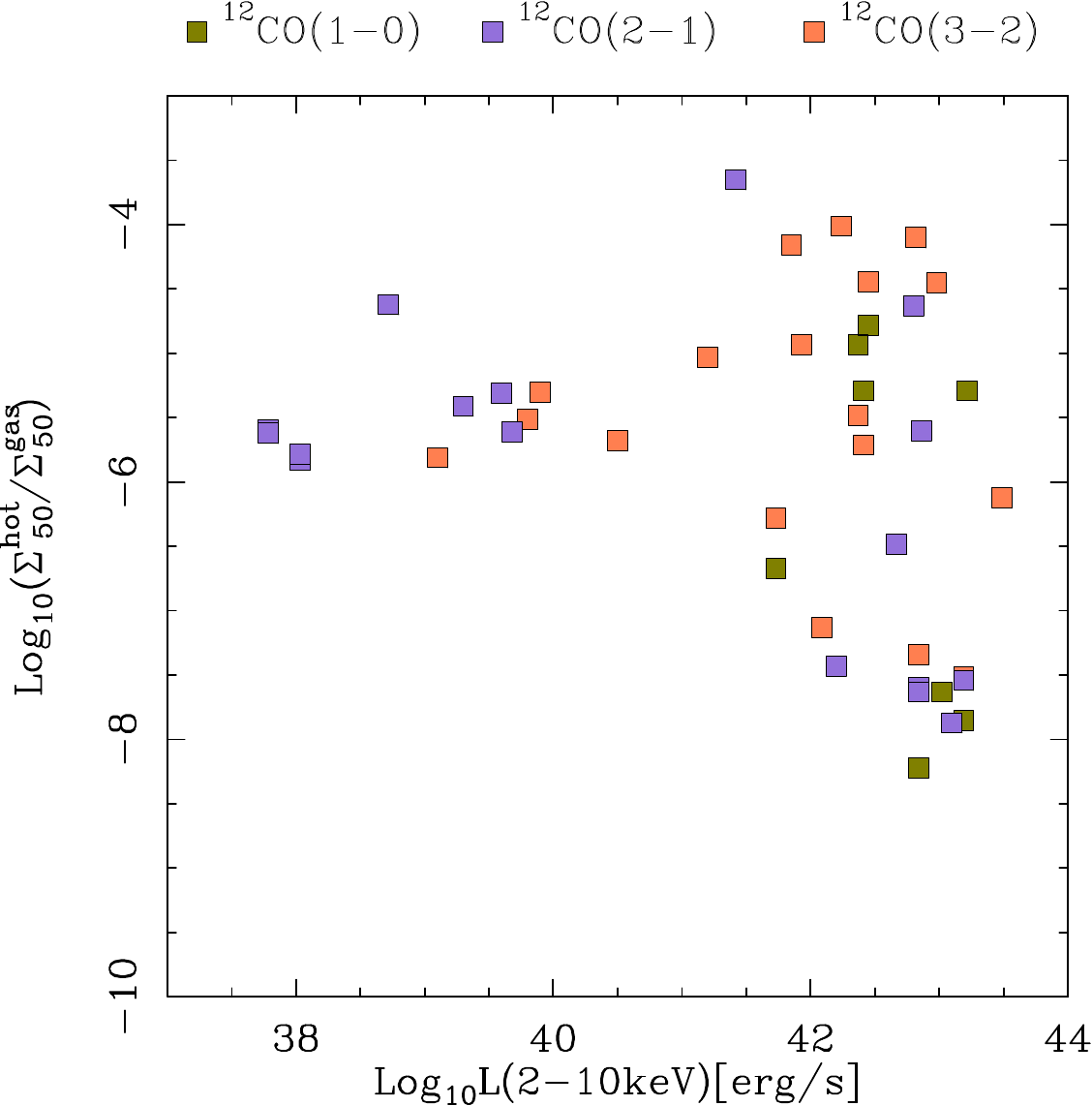}
       \caption{{\em upper panel}: Change of the ratio (in log units) of the hot-to-cold molecular gas mass  surface density ratios with $L_{\rm X}$ estimated from the CO and 2.1$\mu$m line emissions at the two spatial scales analyzed in this work: $r\leq50$pc ($\Sigma^{\rm hot}_{\rm 50}$/$\Sigma^{\rm gas}_{\rm 50}$) and 
$r\leq200$pc ($\Sigma^{\rm hot}_{\rm 200}$/$\Sigma^{\rm gas}_{\rm 200}$). {\em Lower panel}: hot-to-cold molecular gas mass  surface density ratio measured at $r\leq50$pc as a function of $L_{\rm X}$. All symbols as in Fig.~\ref{conc-CO}.}  
   \label{hot-to-cold}
    \end{figure}

  \begin{figure}[tb!]
 \centering
    \includegraphics[width=8cm]{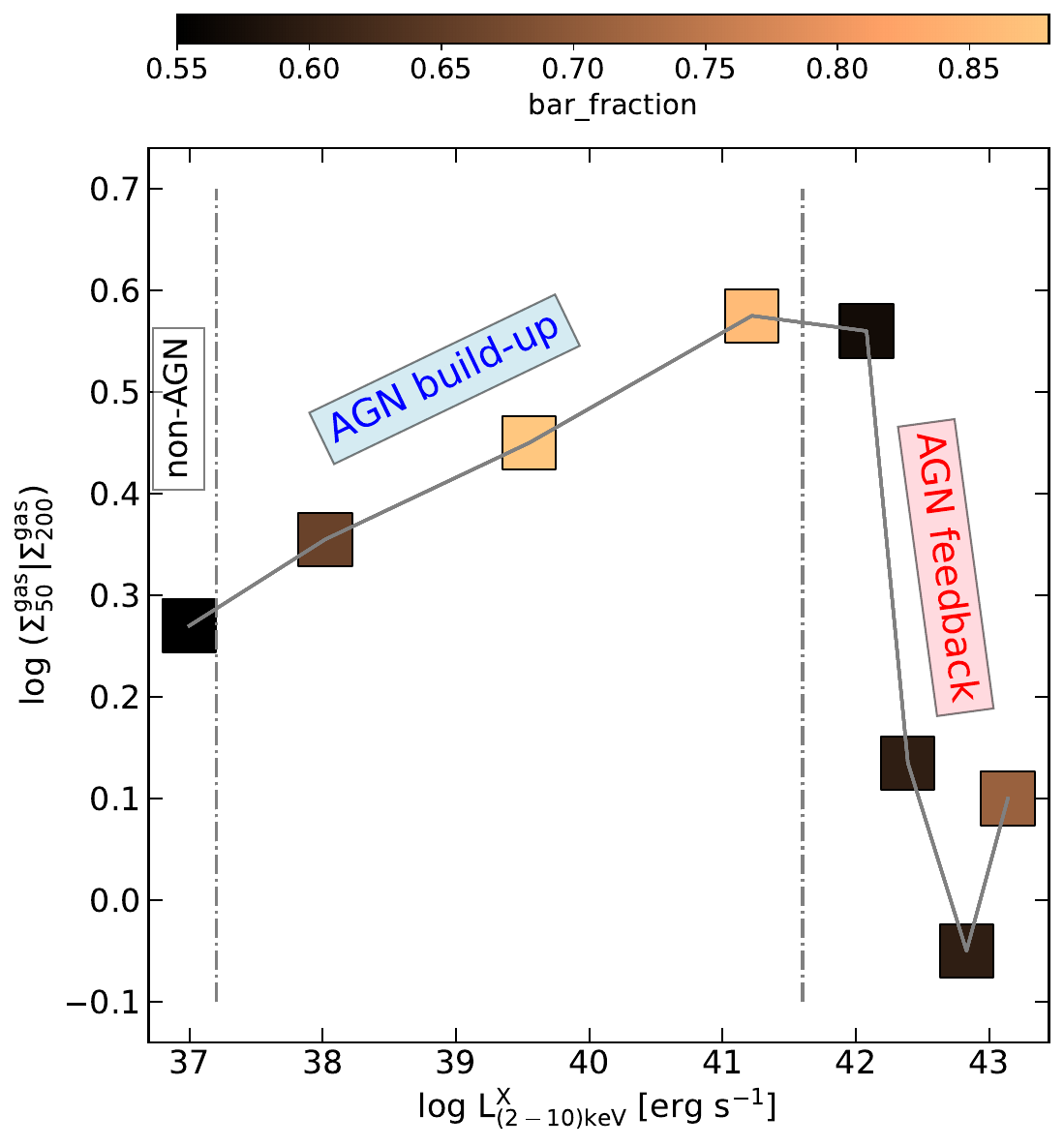}
       \caption{Median values of the concentration indices for the 7 $L_{\rm X}$  bins defined in  Sect.~\ref{CO-ratios} and for the non-AGN galaxies. Symbols are color-coded (in linear scale) to reflect the average fraction of large-scale bars present in the galaxies belonging to each luminosity bin. The ranges of $L_{\rm X}$ corresponding to the AGN build-up and AGN feedback branches are indicated.}  
   \label{bars}
    \end{figure}

  \begin{figure*}[tbh!]
 \centering
    \includegraphics[width=14cm]{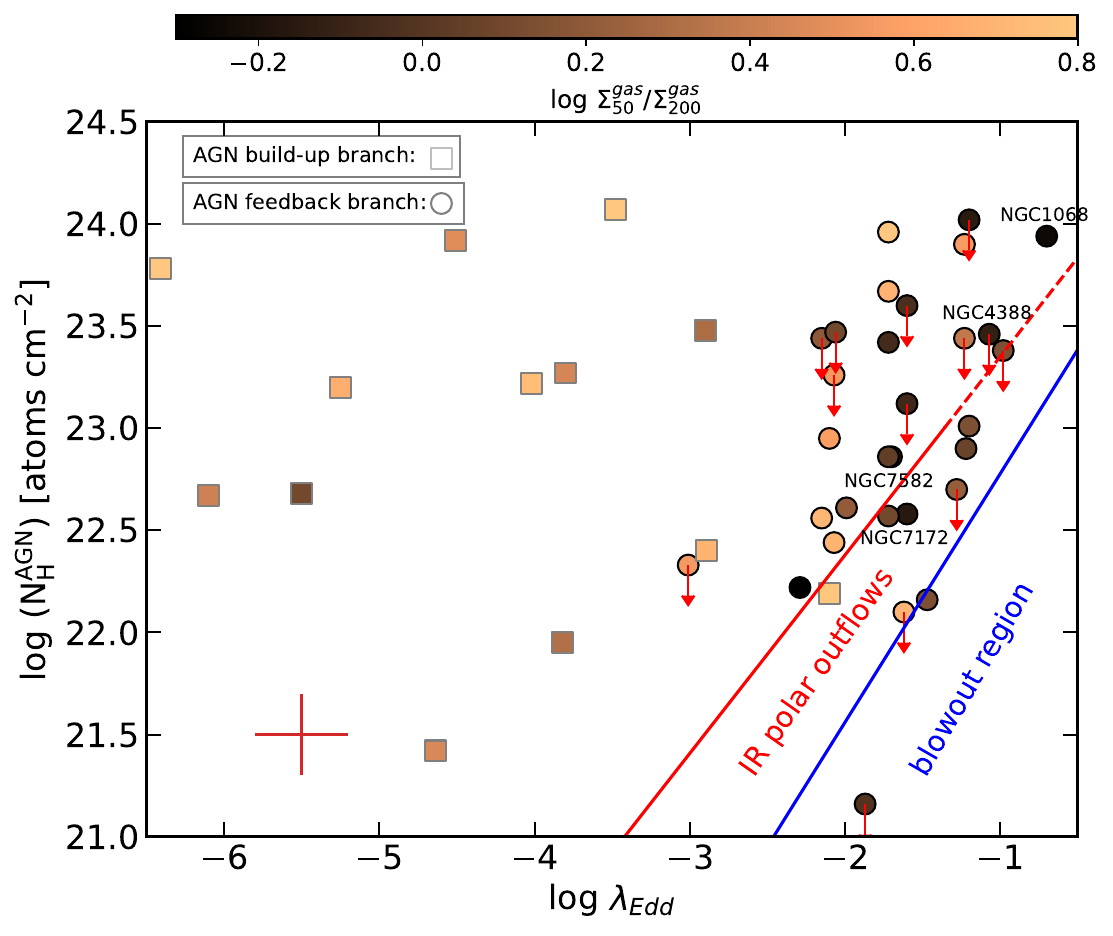}
       \caption{Diagram showing the hydrogen column densities measured towards the AGN ($N^{\rm AGN}_{\rm H}$ in log units) derived from CO as a function of the Eddington ratio ($\lambda_{\rm Edd}$ in log units) for the galaxies of our sample. Galaxies belonging to the AGN feedback and build-up branches as defined in Sect.~\ref{CO-ratios}  are shown by filled circles and squares respectively. In either case symbols are color-coded in terms of the cold molecular gas concentration index ($CCI \equiv$~log$_{\rm 10}(\Sigma^{\rm gas}_{\rm 50}/\Sigma^{\rm gas}_{\rm 200}$). The region below the blue solid curve defines the conditions conducive to the blowout of nuclear regions as defined by \citet{Fab08}. The region below the red line defines the conditions conducive to the launching of IR dusty outflows according to \citet{Ven20}. See also discussion in \citet{Alo21} and \citet{Gar22}. We identify in the diagram the location of the four AGN feedback branch sources NGC~1068, NGC~4388, NGC~7172, and NGC~7582 discussed in Sect.\ref{discussion}. Red arrows identify upper limits. }  
   \label{dusty}
    \end{figure*}

\section{Hot-to-cold molecular gas ratios}\label{hot-cold} 

Taking advantage of the significant overlap between the CO and NIR samples, which span 42 independent map pairs,  we also investigate  the changes in the hot-to-cold molecular gas ratio as a function of   $L_{\rm X}$.

The upper panel of Fig.~\ref{hot-to-cold} shows how the  hot-to-cold molecular gas mass  surface density ratios changes between the nuclear and CND scales as a function of $L_{\rm X}$ in the sample. We estimated the ratio from the CO and 2.1$\mu$m line emissions on the nuclear ($r\leq50$pc) ($\Sigma^{\rm gas}_{\rm hot}$/$\Sigma^{\rm gas}_{\rm cold}$[50~pc]) and CND scales $r\leq200$pc ($\Sigma^{\rm gas}_{\rm hot}$/$\Sigma^{\rm gas}_{\rm cold}$[200~pc]).   By definition, this ratio is insensitive to any galaxy-to-galaxy variation of conversion factors for the cold and hot molecular gas. The ratios,  listed in Table~\ref{Tab3} for 6 $L_{\rm X}$ luminosity bins \footnote{The binning used along the $L_{\rm X}$ axis is chosen to have a 7 independent data sets per luminosity bin.} indicate a trend suggestive of higher values of the hot-to-cold mass ratios on nuclear scales in the highest luminosity AGN sources. In particular,  AGN below the turnover luminosity identified in Sect.~\ref{NIR-ratios} ($L_{\rm X}\leq10^{42.0}$erg~s$^{-1}$) in bins 4, 5, and 6  show median values approximately  compatible with unity within the errors, indicating that for these range of luminosities there are no significant changes in the mass ratios between the two spatial scales.  In contrast, beyond the turnover luminosity, AGN  in bins 1, 2, and 3 show indications of a $\simeq1.5-2$ increase in the hot-to-cold molecular gas ratio on the nuclear scales compared to CND scales. The  increase in the hot-cold-ratio reflects the factor of $\simeq1.5-2$ corresponding to the ratio 10$^{(|\Delta[CCI]|-|\Delta[HCI]|)}$, where $\Delta[CCI]$ and $\Delta[CCI]$ represent the range of variation of $CCI$ and $HCI$ measured along the AGN feedback branch.

However, the suggested twofold increase of the nuclear-scale fraction of hot molecular gas mass  would not compensate for the nuclear-scale deficit of cold molecular gas mass identified in the highest-luminosity AGN sources of our sample. As illustrated in  the lower panel of Fig.~\ref{hot-to-cold}, the hot-to-cold molecular gas mass fraction measured on nuclear scales is $\leq10^{-4}$ in all galaxies of the sample.

\section{An evolutionary scenario  for the cycle of molecular gas at CND scales} \label{discussion}


The results obtained from the analysis of how the concentration and  normalized radial distribution of cold and hot molecular gas change as function of X-ray luminosity in our sample confirm on a more robust statistical basis the previous findings of \citet{GB21}. \citet{GB21} suggested the existence of two distinct categories  of Local Universe spiral galaxies classified as  AGN  grouping  them into two branches connected by an evolutionary sequence: the AGN build-up branch  and  the AGN feedback branch.  

\subsection{The AGN build-up phase}

 Local Universe disk galaxies are expected to spend most of their lives either in a non-active phase or transitioning from a quiescent phase to a nascent AGN phase, moving along the AGN build-up branch. A low-luminosity AGN (log$L_{\rm X}$~$<$~41.5-42~erg~s$^{-1}$) during its build-up phase  could start its activity cycle with a massive dusty molecular torus, which  could have assembled following a fueling episode driven  by the gas instabilities created by a stellar bar.  In particular,  the presence of a trailing nuclear spiral structure  has been identified  inside the ILR of some of the barred galaxies observed with ALMA  at  the highest spatial resolution ($\simeq4-9$~pc) in the sample:  NGC~613, NGC~1566, NGC~1672, and NGC~1808 \citep{Com14,Com19, Aud19, Aud21}. These galaxies lie at the peak phase of  the AGN build-up branch, which is characterized by a high nuclear-scale concentration of molecular gas ($CCI\geq$[+0.5,+0.8]).  The existence of a contrasting 2-arm CO trailing nuclear spiral structure in which the gas loses angular momentum has been interpreted as a smoking gun evidence of ongoing AGN feeding. Bars  that are able to pile up gas close to the sphere of influence of the supermassive black hole, where the precessing frequency $\Omega-\kappa/2$ decreases with radius, can develop trailing nuclear spirals and fuel the  AGN \citep {Wad94, But96, Pee06, Com14, Com21b, Com21a}. In this scenario, the gas can build up the gas reservoir of the molecular torus and its surroundings, which naturally explains the high nuclear-scale concentration index measured in these objects.

Figure~\ref{bars} shows how  the fraction of large-scale stellar  bars, identified in the Hubble SB or SAB  types, is related to the $CCI$ values measured in the galaxies of our sample. Within the limits of our  admittedly small statistics (eight galaxies for each of the eight $L_{\rm X}$ bins of AGN galaxies and 19 non-AGN galaxies; see Table~\ref{Tab3}), Fig.~\ref{bars} shows a nearly monotonic increase in both the fraction of large-scale bars and the  $CCI$ values along the AGN build-up branch. The fraction of large-scale bars  goes from $\simeq0.5-0.6$ for the non-AGN galaxies in our sample up to $\simeq0.9$ at luminosities just below the turnover. For galaxies belonging to the AGN feedback branch, the total fraction of large-scale bars decreases down to $\simeq$0.6-0.7, namely similar to the overall bar fraction derived from optical studies in spiral galaxies, regardless of their AGN or non-AGN classification \citep{deV91, Hun99}. Taken at face value, this result might suggest that while bar-driven secular evolutionary processes regulate the CND-scale distribution of molecular gas along the AGN buildup branch, the role of large-scale bars, while still present to a significant extent, does not appear to outweigh processes specifically related to AGN feedback in redistributing molecular gas beyond a certain value of $L_{\rm X}$.  However, the fact that large-scale bars are seen to be present in all galaxies in the sample at levels $\geq$2/3, suggests that the link between $CCI$ and the fraction of large-scale bars is weak.

Primary large-scale bars have timescales of $\simeq$Gyr and can eventually develop an ILR, and under certain circumstances a decoupled nuclear bar instability, a two-arm trailing spiral structure or a lopsided ($m=1$) mode can develop within this resonance. These smaller scale instabilities, with shorter associated timescales $\simeq10-100$~Myr, can help to drain the angular momentum of the gas and fuel the AGN.  Numerical simulations predict that large-scale bars can become rounder and eventually get destroyed after the cumulative effect of several massive gas inflow episodes \citep[e.g.,][]{Com00, Bou02, Hop10}. The timescales associated with the destruction of primary bars are $\simeq$10-100 longer than the corresponding timescales associated with the transit along the AGN feedback branch or with the black hole starvation phase, as estimated in Sect.~\ref{timescales}. Therefore,  given this significant timescale mismatch, $CCI$ and the fraction of large-scale bars are not expected to be strongly linked. However, the existence of a potentially tighter correlation between the $CCI$ trends and the presence of nuclear bars or nuclear spirals should be further investigated with a larger sample.

\subsection{The AGN feedback phase}

During the peak phase of the activity cycle the action of radiation pressure and/or jets could sweep away a significant fraction of the gas and dust `locally' in the torus.   These local (dusty) outflows could help regulate the feeding of the supermassive black hole and the growth of the  molecular torus.

Figure~\ref{dusty} shows the nuclear hydrogen column densities measured towards the AGN as a function of the Eddington ratio  for the galaxies in our sample. Using semi-analytical models \citet{Ven20} found that the launch of polar IR-driven dusty outflows would be more favorable under certain conditions, which can be ascribed to a well defined region in the  $N^{\rm AGN}_{\rm H}$-$\lambda_{\rm Edd}$ parameter space of Fig.~\ref{dusty}. We also plot in Fig.~\ref{dusty} the predicted location of the blowout region described by \citet{Fab08}, where the gas and dust are expected to be cleared by radiation-driven winds on nuclear scales. In particular, dusty winds can be launched from the torus  when the acceleration due to the AGN radiation pressure and the acceleration due to the gravitational force from the central BH are balanced \citep{Ven20}.  For a fixed value of $\lambda_{\rm Edd}$ column densities below the value associated with the balance between the expected pressures acting on dust are conducive to the launch of dusty outflows. This mechanism is expected to be effective above column densities $N^{\rm AGN}_{\rm H}\geq10^{22}$~cm$^{-2}$ which corresponds to an opacity equal to unity  in the near-IR. \citet{Alo21} analyzed the position of  a sample of 13 nearby AGN including the 10 galaxies of GATOS core sample \citep{GB21}, Circinus, NGC1365, and NGC~1068,  and found that six galaxies have Eddington ratios and nuclear column densities favorable to the launching of IR-driven polar dusty winds. The extended MIR polar components detected by \citet{Alo21} in these candidate galaxies seemed to confirm the first predictions of \citet{Ven20}.  Furthermore, the analysis of the nuclear IR SED of a sample of Seyfert galaxies published by \citet{Gar22}, found that galaxies close to the region favorable for the launch of dusty winds are best fit by torus models that include a wind component, while galaxies far from the IR-driven dusty wind region are best fit by classical torus models.

Figure.~\ref{dusty} shows a clear dichotomy in the location of  the galaxies belonging to the AGN build-up and feedback branches of our  enlarged sample in the  $N^{\rm AGN}_{\rm H}$-$\lambda_{\rm Edd}$ parameter space, where $N^{\rm AGN}_{\rm H}$ is the line-of-sight hydrogen column density derived from CO toward the AGN. In particular, galaxies in the AGN feedback branch  with the lowest nuclear-scale  molecular gas concentrations tend to be close to the region where IR-driven  polar dusty outflows would be more favorably launched. In contrast, the galaxies of the AGN build-up branch with the highest nuclear scale molecular gas concentrations are located far from the region conducive to the launch of dusty outflows.  
Although the spatial and temporal scales associated with the launch of local dusty outflows in the torus (by  the action of IR radiation pressure) and with the propagation of nuclear outflows (by the action AGN outflows and/or radio jets) are different, the distribution of the galaxies of our sample in Fig.~\ref{dusty} suggests that there is a causal connection between the two outflow phenomena.  

The vast majority of the sources belonging to the AGN feedback branch in our sample have ionized outflows and some have radio jets. At a later stage the 
propagation of AGN outflows and/or radio jets, which can be strongly coupled with the ISM of the host disk, can launch molecular outflows, entrain  and in some cases clear a large fraction of the  gas on larger spatial scales ($\sim$tens to hundreds  of pc) as the nuclear activity cycle reaches its maximum. However, other factors beside the AGN luminosity can play a critical role in shaping the distribution, kinematics and excitation of molecular gas: the presence of an AGN wind and/or radio jet does not necessarily lead to a nuclear-scale molecular gas deficit in the disk of the host galaxy, since the latter requires that the coupling between the wind and/or jet and the ISM in the disk is strong enough to affect the redistribution of molecular gas. The impact of geometry  on the coupling efficiency of AGN winds and the ISM has been shown by numerical simulations \citep{Gab14, Tor20, Mer23}. For a fixed AGN luminosity, we therefore expect the distribution of concentration indices to show a scatter reflecting the intrinsic variance of the geometrical factors we may encounter \citep[see also discussion in][]{Ram22}. 

 The cases of NGC\,613 and NGC\,1068 are a good illustration of the importance of geometry. \citet{Aud19} imaged a fast ($v_{\rm out}\geq300$~kms$^{-1}$) $\simeq50$~pc-sized molecular outflow in the central region of NGC~613, detected in several molecular species (CO, HCN, HCO$^+$, and CS).  \citet{Aud19} favored a scenario where the outflow is triggered by the 200~pc radio jet imaged by the VLA \citep{Hum92}.  Although the radio jet has triggered a small molecular outflow at its base, it has left the nuclear-scale in-plane distribution of the molecular gas virtually intact: NGC\,613 exhibits high values of the molecular gas concentration indices.  However, the low coupling efficiency of the jet with the ISM can be explained in a scenario where the jet is at a large angle relative to the disk of NGC\,613, in stark contrast to the strong coupling efficiency of the radio jet and AGN wind of NGC~1068, which are at a small angle relative to the galaxy disk.

\subsection{An estimate of timescales}\label{timescales}
 
The sharp decline of the cold molecular gas concentration indices beyond the turnover luminosity ($L_{\rm X}\simeq10^{41.5}$erg.~s$^{-1}$) shown 
by the distribution of $CCI$ values of Fig.\ref{conc-CO}, suggests that the transition from the initial stage to the  late-phase of the AGN feedback 
branch may occur on relatively short timescales. We can make a rough estimate  of these typical transition timescales for four of the galaxies belonging to 
the late-phase AGN feedback branch, which are characterized firstly, by showing very low (negative) $CCI$ values and, secondly, which  are known to 
harbor nuclear molecular outflows with $\dot{M}_{\rm out}$$\simeq$1 to a few tens of $M_{\sun}$~yr$^{-1}$ in the inner $r<50$~pc region \citep{GB14,GB19,GB21,Alo23}: NGC~1068 ($CCI=-0.25$), NGC~4388  ($CCI=-0.17$), NGC~7172 ($CCI=-0.16$) , and NGC~7582  ($CCI=-0.16$). Under the  hypothesis  that these Seyferts had initial $CCI$ values typical of that of the late-phase AGN build-up branch sources, such as NGC~613 ($CCI=+0.82$), the transit timescale ($t_{\rm transit}$) can be estimated as:

\begin{equation}
 t_{\rm transit}\simeq\pi\times r^{2}\times \Delta[{\rm concentration}] \times\Sigma^{\rm gas}_{\rm 200}\times \dot{M}_{\rm out}^{-1}.
 \label{transition}
 \end{equation}

 In Eq.~\ref{transition}, $\Delta[{\rm concentration}]=10^{CCI({\rm initial})}-10^{CCI({\rm final})}$ represents the concentration change for a particular galaxy, $r=50$~pc, and we assume that $\Sigma^{\rm gas}_{\rm 200}$ and $\dot{M}_{\rm out}^{-1}$ do not undergo a significant change during the transition. We derive for these sources: $t_{\rm transit}\simeq 1$-to a few 10 Myr, an estimate that confirms our expectation that the transition timescales are likely short.

 AGN fueling is expected to be hindered on intermediate spatial scales during the late phase of the activity cycle, represented by the sources of our sample that show the most extreme nuclear-scale deficits of the AGN feedback branch and also harbour molecular outflows in the inner $r<50$~pc region. The gas reservoir in the torus of these AGN would be therefore eventually exhausted. An exhaustion timescale ($t_{\rm exhaustion}$) can be approximately estimated for the molecular tori of  NGC~1068, NGC~4388, NGC~7172, and NGC~7582, using the masses of the tori  ($M_{\rm torus}$) derived by \citet{GB21} and \citet{Alo23} for these sources, assuming a canonical accretion efficiency $\epsilon\simeq0.1$, and a bolometric correction $k_{\rm bol}\simeq10$ in the expression:
 
 \begin{equation}
  t_{\rm exhaustion} \simeq M_{\rm torus} \times \epsilon \times c^2 \times (k_{\rm bol}\times L_{\rm X})^{-1}.
  \label{exhaustion}
 \end{equation} 
 
   We derive from Eq.~\ref{exhaustion} for these sources: $t_{\rm exhaustion}\simeq 1$-to a few 10 Myr. There is both observational and theoretical evidence that AGN can have much shorter {\it flickering} timescales ranging from a few years up to $\simeq$10$^5$ yr \citep{Sch15, Kin15}. With the values of $t_{\rm exhaustion}$ estimated above, AGN activity could still continue in these Seyfert galaxies for a  time span $\simeq10-100$ longer than the expected {\it flickering} timescale.

Once the gas reservoir in the torus is exhausted, a BH `starvation' or build-down phase should follow. During this phase we might expect, first, that the CND-scale  distribution of molecular gas does not change significantly (namely the $CCI$ index remains approximately constant) as AGN feedback is progressively turned off, and second, that the AGN luminosity decreases on short timescales. Thus, in the $L_{\rm X}-CCI$ parameter space, galaxies would move along quasi-horizontal trajectories during the BH starvation phase as they leave the feedback branch. The described evolutionary cycle could be restarted after the completion of another AGN fueling phase leading to a secular increase in the $CCI$ index and $L_{\rm X}$ along the AGN build-up phase. The fact that we see only one galaxy (ESO-093-G003, classified as non-AGN) in our sample with a low cold molecular gas concentration (negative $CCI$; $\sim$-0.02 in ESO-093-G003) and  a very low AGN luminosity ($L_{\rm X}<<$10$^{41.5}$~erg.~s$^{-1}$) suggests that the BH starvation phase should be very short.  Alternatively, the striking lack of targets in the lower left region of the $L_{\rm X}-CCI$ diagram could be explained if galaxies did not necessarily leave the AGN build-up branch during the build-down phase; in this scenario, half of the targets in the AGN build-up branch would actually be in the build-down evolutionary phase.

A proper dating of the transition timescales relevant to the parameter space of Fig~\ref{conc-CO} would require self-consistent numerical simulations of the gas cycle capable of following the evolution of the CND-scale distribution of cold molecular gas in realistic AGN environments similar to those characterizing the Seyfert galaxies in our sample. These simulations could be used to verify/falsify the validity of the different evolutionary scenarios.

\subsection{The hot molecular gas cycle}

That the hot molecular gas follows a qualitatively similar trend to that seen in the central concentration index of cold molecular gas can be explained in a scenario in which the hot molecular gas is entrained in the outflow as is suggested to be the case in NGC 5728 \citep{Shi19}.  In some galaxies, rather than AGN feedback removing all molecular gas within the central region, the cold molecular gas “hole” has been observed to be filled with hot molecular and/or highly ionized gas \citep[e.g.,][Davis et al. 2024, submitted]{Ros19,Fer20}.  However, as noted in Sect.~\ref{hot-cold}, the hot molecular gas within this CO “hole” does not account for the deficit of gas and thus to some extent the AGN feedback has successfully cleared a significant portion of the gas.   \citet{Shi19} found in NGC 5728 that within the cold molecular gas “holes” in the outflow region there is a highly stratified structure of highly ionized gas surrounded by an outer layer of hot molecular gas.  This stratified outflow is interpreted as an outflow which has successfully cleared the central region of molecular gas but is oriented such that this outflow disrupts the circumnuclear disk.  The hot molecular gas within the disk rotates into the outflow region and becomes entrained in it along the outer extent of the outflow.  This emphasizes the importance of the geometry of the disk and outflow orientations in that a coupling of the outflow with the ISM is necessary to effectively redistribute molecular gas.  In such a case both the cold and hot molecular gas will be cleared, although to varying degrees depending on the disk and outflow geometries. Whether the nuclear-scale deficit shared by the 'cold' and 'hot' phases of H$_{\rm 2}$ can be partly compensated by the 'warm' phase (T$_{k}$$\simeq$a few 100~K) of the molecular gas remains to be confirmed by future observations of the sample with facilities such as the James Webb Space Telescope. 
 
\section{Summary and conclusions}\label{summary}
 
We selected a sample of nearby ($D_{\rm L}=7-45$~Mpc) 45 AGN and 19 non-AGN galaxies that have high-resolution multiline CO observations obtained at millimeter wavelengths by the ALMA and/or PdBI arrays to study the distribution of cold molecular gas on CND scales ($r\leq200$~pc) in the galaxies of our sample. We analyzed whether the concentration and  normalized radial distribution of cold molecular gas  change as function of the X-ray luminosity to decipher the imprint potentially left by AGN feedback. We also incorporated NIR integral field spectroscopy data obtained for the H$_2$ 1-0 S(1) line available for a subset of 35 targets of the CO-based sample used in this work. These data allowed us to study  the concentration and normalized radial distribution of the hot molecular gas phase as well as the hot-to-cold-molecular gas mass ratio as a function of the X-ray luminosity in this subsample. The main results of this paper are summarized as follows:

\begin{itemize}

\item

 The cold molecular gas concentration, defined by the ratio  $CCI \equiv$~log$_{\rm 10}(\Sigma^{\rm gas}_{\rm 50}/\Sigma^{\rm gas}_{\rm 200}$),  shows  a statistically significant turnover at a luminosity breakpoint $L_{\rm X}\simeq10^{41.5}$erg~s$^{-1}$, which divides the sample into two branches in the $CCI$-$L_{\rm X}$ parameter space. AGN and non-AGN  below the  luminosity breakpoint, located along the `AGN build-up branch', have $CCI$ values that  increase with $L_{\rm X}$ from $\simeq+0.27$$\pm$0.13 to $\simeq+0.58$$\pm$0.23. Higher luminosity AGN beyond the luminosity breakpoint,  located along the  `AGN feedback branch', show a sharp decrease in the concentration of molecular gas from  $\simeq+0.56$$\pm$0.12 down to $\simeq-0.05$$\pm$0.07.  The cold molecular gas concentration spans a factor $\simeq$4.3 between the galaxies lying at the high end of the AGN build-up branch and the galaxies showing the most extreme nuclear-scale molecular gas deficits in the AGN feedback branch. These trends confirm, on a three times larger sample, previous evidence found in the context of the GATOS survey by \citet{GB21}.

\item

The normalized radial profiles ($I_{\rm norm}(r)$) of the cold molecular gas in the galaxies of our sample reflect the trends shown by the $CCI$ values as function of $L_{\rm X}$. The profiles appear as flat or inverted in the AGN feedback branch galaxies. In contrast, galaxies in the peak phase of the AGN build-up branch show highly concentrated molecular gas distributions. In addition,  lower  luminosity AGN galaxies and  non-AGN targets show  comparatively less concentrated molecular gas radial profiles.  

\item

We performed an independent  non-parametric analysis of the CND-scale distribution of the cold molecular gas in our sample by deriving the Gini coefficients associated with their radial profiles. The radial Gini coefficients are shown to be strongly correlated with the molecular gas concentration. On the other hand, the radial Gini coefficients are shown to be strongly anticorrelated with the flux-weighted mean radius of the distribution of the gas. The distribution of the radial Gini coefficients with $L_{\rm X}$  show a  similar turnover at a luminosity breakpoint $L_{\rm X}\simeq10^{41.5-42.0}$erg~s$^{-1}$.    

\item

 The hot molecular gas concentration, defined by the ratio  $HCI \equiv$~log$_{\rm 10}(\Sigma^{\rm hot}_{\rm 50}/\Sigma^{\rm hot}_{\rm 200}$) shows a  qualitatively similar but less pronounced turnover compared to the trend shown by the $CCI$ index. We identify a luminosity breakpoint $L_{\rm X}\simeq10^{42.0}$erg~s$^{-1}$ dividing the sample into the AGN build-up and feedback branches. The concentration of hot molecular gas in the sample spans a factor $\simeq2.9$ range between the galaxies lying in the peak phase of the AGN build-up branch and the most extreme AGN feedback branch targets. The normalized radial profiles ($I_{\rm norm}(r)$) of the hot molecular gas reflect the $HCI$-$L_{\rm X}$ trends.

\item

We find higher values of the hot-to-cold molecular gas mass ratios on nuclear scales in the highest luminosity AGN sources of the AGN feedback branch. In particular, beyond the turnover luminosity, AGN show indications of a twofold increase in the hot-to-cold molecular gas ratio on the nuclear scales compared to CND scales.  However, this twofold increase does not compensate for the nuclear-scale deficit of cold molecular gas mass identified in the highest luminosity AGN.

\item

 We present a tentative scenario describing an evolutionary link connecting the galaxies analyzed in this work. While most of galaxies are expected to spend most of their lives transitioning from a quiescent phase to a nascent AGN phase, moving along the AGN build-up branch, the sharp decline of the cold molecular gas concentration indices beyond the turnover luminosity ($L_{\rm X}\simeq10^{41.5}$erg.~s$^{-1}$) suggests that the transition from the initial stage to the  late-phase of the AGN feedback branch occurs on short timescales ($\simeq1$-to a few 10 Myr). At the end of the AGN feedback phase the gas reservoir in the torus  would be eventually exhausted.  Similarly, the timescales associated to the ensuing BH starvation phase are estimated to be very short.

\end{itemize}

 The evolutionary scenario described in this paper would outline the cycle of molecular gas at CND scales in spiral disk galaxies undergoing an AGN episode. However, the secular evolution processes in active galaxies classified as ETG (E and S0)  are expected to be very different from those governing the redistribution of molecular gas in later-type spiral galaxies, which are in all likelihood related to the action of stellar bars or spiral density waves absent  in E galaxies in particular. Furthermore, the CND-scale distribution of molecular gas in ETG may rather reflect the evolutionary stage of the gas settling process, which in E and S0 objects is often associated with an external accretion event during an interaction \citep[see also discussion in][]{Dav14, Dav17}. Therefore, we do not expect to find a direct evolutionary link that could connect bulge-dominated ETG to disk-dominated spiral galaxies in the scenario described in this paper. Taken together, this could explain that while the spiral  galaxies in \citet{Elf24} with X-ray luminosities higher than 10$^{42}$~erg~s$^{-1}$ closely follow the relation found by \citet{GB21} and this work, the fraction ($\simeq 60\%$) of the low-luminosity ETG analyzed by \citet{Elf24} would deviate from the general trend followed by spiral galaxies in the AGN build-up branch.

During the low-luminosity and low Eddington ratio phases, the kinetic feedback mode associated with radio jets can compete with, or even overtake, the radiative feedback mode that dominates at higher luminosities. Jets can sweep away interstellar gas provided that the coupling with the ISM in the disk of the host is sufficient. In some cases 
the jet ejected from the central engine can impact a significant fraction of the plane of the galaxy, due to the random orientation of the accretion disk and the tori, which are generally decoupled from the orientation of the large-scale disk \citep[e.g.,][]{Mor13, Mor15, GB14, Das15, Alo18, Ang21,Gar21, Ram22, Ruf22, Rao23, Aud23}. The potential role of jets in AGN feedback for different radio power regimes has been demonstrated by numerical simulations \citep[e.g.,][]{Wag11,Wag12, Muk18a, Muk18b, Mek22}. Unfortunately, the influence that radio jets can have on the distribution of molecular gas cannot be studied to the same extent as for $L_{\rm X}$ in most of the galaxies in our sample, because high spatial resolution and high sensitivity radio images are only available for a small fraction of our targets.

\begin{acknowledgements}
      We thank the referee for his thorough and constructive report. This paper
makes use of the following ALMA data, which have been processed as part of  1) the GATOS survey: ADS/JAO.ALMA
$\#$2016.1.00232.S, $\#$2016.1.00254.S., $\#$2017.1.00082.S, $\#$2018.1.00113.S, and $\#$ 2019.1.00618.S
2)the NUGA survey: $\#$2015.1.00404.S and $\#$2016.1.00296.S
3) the LLAMA survey: $\#$2019.1.01742.S 
4)the WISDOM survey: 
$\#$2017.1.00517.S, $\#$2018.1.00572.S, and $\#$2019.1.00363.S
4)the survey PI-ed by Claudio Ricci $\#$2019.1.01230.S and $\#$2021.1.00812.S
5) the PHANGS-ALMA CO(2-1)
survey:
ADS/JAO.ALMA$\#$2012.1.00650.S, ADS/JAO.ALMA$\#$2
013.1.00803.S, ADS/JAO.ALMA$\#$2013.1.01161.S, ADS/
JAO.ALMA$\#$2015.1.00121.S, ADS/JAO.ALMA$\#$2015.1.00
782.S, ADS/JAO.ALMA$\#$2015.1.00925.S, ADS/JAO.
ALMA$\#$2015.1.00956.S, ADS/JAO.ALMA$\#$2016.1.
00386.S, ADS/JAO.ALMA$\#$2017.1.00392.S, ADS/JAO.
ALMA$\#$2017.1.00766.S, ADS/JAO.ALMA$\#$2017.1.00886.
L, ADS/JAO.ALMA$\#$2018.1.00484.S, ADS/JAO.
ALMA$\#$2018.1.01321.S, ADS/JAO.ALMA$\#$2018.1.01
651.S, ADS/JAO.ALMA$\#$2018.A.00062.S, ADS/JAO.
ALMA$\#$2019.1.01235.S, ADS/JAO.ALMA$\#$2019.2.00129.S

ALMA is a partnership of ESO (representing its member states), NSF (USA), and NINS (Japan), together with NRC (Canada) and NSC 
and ASIAA (Taiwan), in cooperation with the Republic of Chile. The Joint ALMA Observatory is operated by ESO, AUI/NRAO, and NAOJ. 
The National Radio Astronomy Observatory is a facility of the National Science Foundation operated under cooperative
agreement by Associated Universities, Inc. 

This work is also based on observations carried out with the IRAM PdBI  Interferometer in the context of the NUGA (PIs: S. Garc\'{\i}a-Burillo \& and F. Combes) and  PAWS (PI: E. Schinnerer) projects. IRAM is supported by INSU/CNRS (France), MPG (Germany) and IGN (Spain). We acknowledge the IRAM staff from the Plateau de Bure and from Grenoble for carrying out the observations and help provided during the data reduction.
We acknowledge the use of the following data reduction and software packages: {\tt CASA} \citep{McM07}, {\tt astropy} \citep{Astropy2013, Astropy2018}, and  {\tt Rstudio package} \citep{Rcorteam}

SGB, AU, and MQ acknowledge support from the Spanish grant PID2022-138560NB-I00, funded by
MCIN/AEI/10.13039/501100011033/FEDER, EU. MPS acknowledges support from grant RYC2021-033094-I funded by
MICIU/AEI/10.13039/501100011033 and the European Union
NextGenerationEU/PRTR. IGB acknowledges support from STFC through grants ST/S000488/1 and ST/W000903/1. MTL acknowledges grant support from the Space Telescope Science Institute (ID: JWST-GO-01670.007-A). MS acknowledges support by the Ministry of Science, Technological Development and Innovation of the Republic of Serbia (MSTDIRS) through contract no. 451-03-66/2024-03/200002 with the Astronomical Observatory (Belgrade). AAH acknowledges support from grant PID2021-124665NB-I00 funded by  MCIN/AEI/10.13039/501100011033 and by ERDF A way of making Europe. EB acknowledges the María
Zambrano program of the Spanish Ministerio de Universidades
funded by the Next Generation European Union and is also partly
supported by grant RTI2018-096188-B-I00 funded by the
Spanish Ministry of Science and Innovation/State Agency of
Research MCIN/AEI/10.13039/501100011033. CRA acknowledges support from
project "Tracking active galactic nuclei feedback from parsec to kiloparsec
scales", with reference PID2022-141105NB-I00. DJR acknowledges support from the UK STFC through grant ST/X001105/1. CR acknowledges support from Fondecyt Regular grant 1230345 and ANID BASAL project FB210003. OGM acknowledges financial support from UNAM through the project PAPIIT IN109123 and CONAHCyT through the project "Ciencia de Frontera 2023" CF-2023-G-100.
\end{acknowledgements}

\bibliographystyle{aa}

\bibliography{aa-3}

\begin{appendix}

\section{Object tables}

For more details, readers can refer to Tables~\ref{Tab1} and \ref{Tab5}

\begin{table*}[h!]
\caption{Observational properties of the CO sample.}
\centering
\resizebox{.76\textwidth}{!}{ 
\begin{tabular}{lcccccccccc} 
\hline
\hline
\noalign{\smallskip} 
  Name        &  $\alpha_{\rm 2000}^{\rm (a)}$ &  $\delta_{\rm 2000}^{\rm (a)}$  & D$^{\rm (b)}$ &  log$_{\rm 10}$$L_{\rm X}^{\rm 2-10keV~(c)}$ & $\lambda_{\rm Edd}$$^{\rm (d)}$ & Hubble \& AGN type$^{\rm (e)}$& Hubble stage $T^{\rm (e)}$ & $PA^{\rm (f)}$ & $i^{\rm (f)}$ \\
\noalign{\smallskip}  
\hline
\noalign{\smallskip}
   ----        &  $^{\rm h}$  $^{\rm m}$ $^{\rm s}$  & $^{\circ}$ $\arcmin$ $\arcsec$ & Mpc   & erg~s$^{-1}$ &  ---- &  ---- &  ---- &  $^{\rm o}$  &  $^{\rm o}$ \\
\noalign{\smallskip}  
	   \hline
	   \hline
 \noalign{\smallskip} 
 \multicolumn{8}{c}{AGN sample: CO(3-2)} \\
 	   \hline
	   \hline
 \noalign{\smallskip} 
      NGC1566    &  04:20:00.395    &   -54:56:16.60    &   7.2    &   40.50   &  -2.89   &  (R'$_{1}$)SAB(rs)bc; Sy1.5 &  4.0 &  44	   & 49 \\
      NGC1808    &  05:07:42.329    &   -37:30:45.85    &   9.3    &   39.80   &  -4.51   &  (R'$_{1}$)SAB(s:)b; Sy2 &  1.2 &  146	   & 83 \\ 
      NGC1672     &  04:45:42.496   &   -59:14:49.92  &  11.4   &   39.10   &  -6.41   &   (R'$_{1}$)SB(r)bc; Sy2 &  3.3 &  155	   & 29 \\
      NGC1068    &   02:42:40.709   &   -00:00:47.94   &  14.0    &   42.82  &  -0.70  & (R)SA(rs)b; Sy2 &  3.0 &   289  & 41 \\
      NGC6300    &  17:16:59.543.   &   -62:49:14.04  &  14.0    &  41.73   &  -1.72   & SB(rs)b Sy2  &  3.1 & 95   &  57 \\
      NGC1326    &  03:23:56.416    &   -36:27:52.68  &  14.9    &   39.90   & -4.02    &   (R$_1$)SB(rl)0/a;  LINER   &  -0.7 &  71 &  53	\\
      NGC5643    &  14:32:40.699   &  -44:10:27.93  &  16.9    &   42.41   & -1.23    &  SAB(rs)c; Sy2 &  5.0 &  301 &  30	\\
        NGC613      &  01:34:18.189         &  -29:25:06.59&   17.2    &  41.20   &  -3.48  & SB(rs)bc; HII Sy  &  4.0 &  122 &	36 \\
      NGC7314    &  22:35:46.201  &  -26:03:01.58  &   17.4    &  42.18   &  -2.07   &  SAB(rs)bc; Sy1.9   & 4.0  & 191 & 55 	\\
      NGC4388    &  12:25:46.781&  +12:39:43.75   &   18.1   &  42.45  &  -1.07 &  SA(s)b: sp; Sy2 Sy1.9   & 2.8   & 82  & 79 	\\
      NGC1365    &  03:33:36.369  & -36:08:25.50   &   18.3  &  42.09 &  -2.15 & (R')SBb(s)b; HII Sy1.8  & 3.2  & 40  & 41 \\
      NGC4941    &   13:04:13.103 &  -05:33:05.73   &   20.5    &  41.40  & -2.10   &  (R)SAB(r)ab:; Sy2  & 2.1 & 212 &  41 \\
      NGC7213    &  22:09:16.209  &   -47:10:00.12  &   22.0   &   41.85  &  -3.01   &   SA(s)a:;  LINER Sy1.5  &  0.9 & 133  & 35	\\
      NGC7582    &  23:18:23.643  &  -42:22:13.54  &   22.5    &  43.49   &  -1.70   &   (R'$_1$)SB(s)ab;  Sy2  & 2.1   &  344 & 59	\\
      NGC6814    &  19:42:40.587  &  -10:19:25.10   &   22.8    &   42.24   & -1.62     &   SAB(rs)bc; HII Sy1.5  &  4.0	 &  84   & 57  \\
      NGC3227    &   10:23:30.577 & +19:51:54.28    &   23.0    &  42.37    &  -1.20    &   SAB(s) pec;  Sy1.5  &  1.5	& 152 & 52  \\
      NGC5506    &   14:13:14.878  &  -03:12:27.66   &   26.4    &  42.98  &  -1.22   &  Sa pec sp ; Sy1.9  & 1.2  & 275 & 80	\\
      NGC7465    &  23:02:00.961 & +15:57:53.21     &   27.2    &  41.93   & -2.10  &    (R')SB(s)0;  Sy2  & -1.8  & 66 & 54.5 	\\
      NGC7172    &  22:02:01.891   &  -31:52:10.48 &   37.0    &  42.84  &  -1.60    & Sa pec sp; Sy2 HII &  0.6 & 92 & 85 	\\
      NGC5728    &    14:42:23.872   & -17:15:11.01     &   44.5    & 43.19    &  -1.72   &  (R$_1$)SAB(r)a; HII Sy2  &  1.2  & 15	& 59  \\ 
   \noalign{\smallskip}
  \hline
  \hline
 \noalign{\smallskip}  
   \multicolumn{8}{c}{AGN sample: CO(2-1)} \\
 	   \hline
	   \hline
	 NGC4826    &  12:56:43.643  &   +21:40:59.30   &   4.4   & 37.78   &  -6.66   &  (R)SA(rs)ab; HII Sy2    &  2.2	& 112	& 60  \\
         NGC5236    &   13:37:00.94    &  -29:51:56.16    &   4.9    & 38.70    & ----    &   SAB(s)c; HII Sbrst  &  5.0 & 225  &  24  \\
         NGC2903    &    09:32:10.10 &   +21:30:02.88   &   10.0    &   38.00   &  -6.62   & SAB(rs)bc;  HII & 4.0 & 204 & 67 \\
         NGC3351    &    10:43:57.731 &  +11:42:13.35     &   10.0    &   38.03    &   -6.75   &   SB(r)b; HII Sbrst   &  3.1 & 193 & 45  \\
         NGC1637    &    04:41:28.10 &  -02:51:28.80 & 11.7    &  38.04   &  ---- &  SAB(rs)c; AGN   & 5.0  &	21 & 31 \\
         NGC4501    &  12:31:59.220    &  14:25:12.69    &   14.0   &    39.68  & -5.78   &  SAb -SA(rs)b; HII Sy2    &  3.3 & 135 & 59 \\
         NGC4438    &  12:27:45.675  &   13:00:31.18  &   16.5   &   38.72   &   -6.35  &  SA(s)0/a pec:; LINER   & 2.8    &  30    &  60 \\
         NGC4569    &    12:36:49.80  &   +13:09:46.30   &   16.8    &  39.60    &  -5.50    & SAB(rs)ab; LINER Sy & 2.4  &  23   &  70 \\
         NGC4579    &     12:37:43.58   &  +11:49:02.49     &   16.8    &  41.42     &  -3.82    & SAB(rs)b; LINER Sy1.9 & 2.8  & 95  & 36  \\
         NGC3718    &  11:32:34.880  &    +53:04:04.32    &   17.0    &   40.64   &  -4.64   &   SB(s)a pec; Sy1 LINER   &  1.1 &.  120 & 60  \\
         NGC3368    &    10: 46:45.50 &    11:49:12.00  &   18.0  &  39.30   &  -5.53  & SABab SAB(rs)ab;Sy LINER & 2.1  & 165  & 60  \\ 
         NGC2110    &   05:52:11.377  &   -07:27:22.48    &   34.8   &    42.67  &  -1.87   &   SAB0-; Sy2    & -3.0  & 175  & 46 \\
         NGC2782    &   09:14:05.111  &   40:06:49.24.  &   35.0    &  39.50   &  -6.10  &  SAB(rs)a; Sy1 Sbrst  & 1.1  & 75	& 20	 \\
         NGC7172    &  22:02:01.891   &  -31:52:10.48 &   37.0    &  42.84  &  -1.60    & Sa pec sp; Sy2 HII &  0.6 & 92 & 85 	\\
         MCG-06-30-15    &  13:35:53.770 &   -34:17:44.16  &   38.3   &  42.86  &  -1.28   &   S?;  Sy1.2  & 2.0  & 116	& 59	 \\
         NGC2992    & 09:45:41.943  &  -14:19:34.57  &   39.2   &  42.20 & -2.29    &  Sa pec; Sy1.9   &   0.9 & 29  & 80  \\
         NGC3081    & 09:59:29.546    &  -22:49:34.78   &   40.3    & 43.10     & -1.47    & ( R$_1$)SAB(r)0/a; Sy2 & 0.0 & 71   & 60  \\
         ESO137-34  &  16:35:13.996   &  -58:04:47.77  &  41.5    &  42.80 & -1.99   &  SAB(s)0/a? Sy2  & 0.6 & 18 & 41  \\
         NGC5728    &    14:42:23.872   & -17:15:11.01     &   44.5    & 43.19    &  -1.72   &  (R$_1$)SAB(r)a; HII Sy2  &  1.2  & 15	& 59  \\ 
         ESO21-g004    &  13:32:40.621   & -77:50:40.40   &   45.1   &  42.32  & ----  &   SA(s)0/a:  & 0.20 &  100  & 65 \\
 \noalign{\smallskip}
  \hline
  \hline
 \noalign{\smallskip}  
   \multicolumn{8}{c}{AGN sample: CO(1-0)} \\
 	   \hline
	   \hline
                   M51   &   13:29:52.68    &   +47:11:42.72   &   8.6   &   39.00   & -5.25   &  SA(s)bc pec; HII Sy2.5   &  4.0 & 173 & 21 \\
     	 NGC6300    &  17:16:59.543.   &   -62:49:14.04  &  14.0    &  41.73   &  -1.72   & SB(rs)b Sy2  &  3.1 & 95   &  57 \\
         NGC4321    &12:22:54.954   &  +15:49:20.49    &   16.8    &   40.40     & -3.80    &  SAB(s)bc; LINER HII  & 4.0  &  153  & 32 \\ 
         NGC5643    &  14:32:40.699   &  -44:10:27.93  &  16.9    &   42.41   & -1.23    &  SAB(rs)c; Sy2 &  5.0 &  301 &  30	\\
         NGC7314    &  22:35:46.201  &  -26:03:01.58  &   17.4    &  42.18   &  -2.07   &  SAB(rs)bc; Sy1.9   & 4.0  & 191 & 55 	\\
	 NGC4388    &  12:25:46.781&  +12:39:43.75   &   18.1   &  42.45  &  -1.07 &  SA(s)b: sp; Sy2 Sy1.9   & 2.8   & 82  & 79 	\\
         NGC6221    &   16:52:46.346     &   -59:13:01.08   &   22.9    &  41.26    &  -2.90  &   Sb: pec ; HII LIRG  &  3.1 &  1   & 51  \\
         NGC3227    &   10:23:30.577 & +19:51:54.28    &   23.0    &  42.37    &  -1.20    &   SAB(s) pec;  Sy1.5  &  1.5	& 152 & 52  \\
         NGC4180    &   12:13:03.072   & +07:02:19.95   &   36.0    &   42.08   &  ----   &  Sab:; Sy LINER   & 2.0  &  21   & 80 \\
         NGC7172    &  22:02:01.891   &  -31:52:10.48 &   37.0    &  42.84  &  -1.60    & Sa pec sp; Sy2 HII &  0.6 & 92 & 85 	\\
         NGC4593    &  12:39:39.444    &   -05:20:39.03   &   41.8    &   43.02   & -0.98    & (R)SB(rs)b Sy1  &  3.0  & 38 &  33 \\
         NGC1125    &   02:51:40.459   &   -16:39:02.34   &   42.6   &  42.65     &  -2.15   &  (R')SAB(rl:)0$^+$; Sy2    & 0.0   & 54    & 75  \\
         NGC5728    &    14:42:23.872   & -17:15:11.01     &   44.5    & 43.19    &  -1.72   &  (R$_1$)SAB(r)a; HII Sy2  &  1.2  & 15	& 59  \\ 
         NGC3281    &    10:31:52.082      &   -34:51:13.38    &   52.0    & 43.22     &  -2.06   &    SAB(rs+)a;  Sy2   &  2.4 & 140 & 60  \\       
  \hline
  \hline
  \noalign{\smallskip}  
   \multicolumn{8}{c}{non-AGN sample: CO(2-1)} \\
  \hline
\hline
     NGC5068    &  13:18:54.700 &   -21:02:19.32   &   5.2   &  ----    & ----  &  SB(s)d    & 6.0  &  342   &  36 \\
     NGC3621    &   11:18:16.296  &   - 32:48:45.36   &   7.1   & ----   &  ----  &  SA(s)d; HII  & 6.9  & 344    & 66 \\ 
     IC5332    &    23:34:27.480	 &   -36:06:03.9   &  9.0    &  ----  &  ----   &   SA(s)d &  6.8  & 74 & 27 \\
     NGC3596    &  11:15:06.100   &  +14:47:13.56  &   11.3   &  ----    & ----    &  SAB(rs)c; HII   & 5.2  & 169   & 15 \\
     NGC4781    &  12:54:23.800  &   -10:32:13.56 &   11.3    &   ----   & ----    &  SB(rs)d; HII  & 7.0  & 290  & 59 \\
     NGC2835    &  09:17:52.800 &   -22:21:16.92  &   12.2  &  ----  & ----  &  SAB(rs)c; HII  & 5.0  & 1  & 41 \\
     NGC5530    & 14:18:27.300   &  -43:23:17.88   &   12.3    &  ----   &  ---- & SA(rs)bc   &  4.2 &  305  & 62 \\
     NGC1947    &  05:26:47.545 &  -63:45:35.82   &   17.2  &   ----   &  ----  & S0- pec; LINER& -2.8 & 119  & 29  \\
     NGC1079    &   02:43:44.337  & -29:00:11.75 &   18.9    & ----    & ----   & (R)SAB(rs)0/a pec   &  0.6  & 90  & 56 \\
     NGC3717    &  11:31:31.878  &   -30:18:27.78  &   19.1   &  ----     &  ----   &  SAb: sp; HII  & 3.1  &  32   & 79  \\
     NGC7727    &   23:39:53.804  &    -12:17:34.18  &   21.0   & ----   & ----  &  SAB(s)a pec   &   1.1 &  0   & 26 \\
     NGC5921    &   15:21:56.487    &   +05:04:14.27  &  21   &  ----    &  ----  &   SB(r)bc; LINER & 4.0  & 140 & 50 \\
     NGC3175    &  10:14:42.111    &  -28:52:19.42  &   21.0   &  ----   &  ----  &   SAB(s)b; HII & 2.0   &  55  & 76  \\
     NGC718    &   01:53:13.282   &  +04:11:44.93  &   21.4    &   ----   & ----     & SAB(s)a    & 1.0   &  150   & 50 \\
     NGC5845    & 15:06:00.787    & +01:38:01.77     &   27.0  &   ----   &  ----   &   E   &  -4.9 & 153     & 50  \\
     ESO-093-g003    &  10:59:26.062   &   -66:19:58.31    &   29.8    &  ----   & ---- & SAB(r)0/a?   & 0.3  &  145  & 66 \\
     NGC5037    &   13:14:59.357  &  -16:35:24.99  &   35.0  &  ----  &  ---- & SA(s)a  & 1.2  & 44 & 74 \\
     NGC3749    &   11:35:53.192  &    -37:59:50.90     &   41.0  &  ----   & ----   & SA(s)a pec sp; HII  & 1.1 & 109 & 65 \\
     NGC4224    &    12:16:33.782  &  +07:27:43.54   &   41.0   &  ----   & ----   &  SA(s)cd: sp; HII  &  6.0 &  56  & 59 \\
\noalign{\smallskip} 
\hline 
\hline
\end{tabular}}
\tablefoot{$^{\rm (a)}$ Positions of the AGN continuum sources derived from the ALMA and PdBI CO mages taken from the references listed in Sect.~\ref{sample}.  $^{\rm (b)}$ Median value of redshift-independent distances taken from the Nasa Extragalactic Database (NED). $^{\rm (c)}$ Luminosities of hard X-rays in the 2-10~keV range are taken from \citet{Ric17a, Com19, GB21} or redetermined in this work (see Appendix~\ref{Lx});  luminosities are intrinsic (corrected for absorption) and re-scaled to the adopted distances. $^{\rm (d)}$ Eddington ratios adopted from \citet{GB21} and references therein, except for NGC~1068 where we used the most recent determination of the BH mass published by \citet{Gal23} and the bolometric AGN luminosity derived by the torus model of \citet{GB14}. $^{\rm (e)}$ Hubble classification, stage type $T$-parameter, and AGN classification (when available) are taken from NED. $^{\rm (f)}$ Position and inclination angles used in the deprojection of the images are taken from HyperLEDA or redetermined in this work using the software package {\tt kinemetry} \citep{Kra06} for the galaxies published by \citet{GB21}.} \label{Tab1} 
\end{table*}

\begin{table*}[h!]
\caption{NIR observations.}
\centering
\resizebox{8.8cm}{!}{ 
\begin{tabular}{llll} 
\hline
\hline
\noalign{\smallskip} 
  Name        &  Filter &  Plate Scale  & Reference or ESO Prog. ID \\
\noalign{\smallskip}  
\hline
\noalign{\smallskip}
   ----        &   ----        & arcsec	& 	   ----           \\
\noalign{\smallskip}  
	   \hline
	   \hline
 \noalign{\smallskip} 
 \multicolumn{4}{c}{AGN sample} \\
 	   \hline
	   \hline
 \noalign{\smallskip} 
ESO137-34	&	H+K	&	0.050	&	\citet{Caglar2020}	\\
MCG-05-23-016	&	H+K	&	0.050	&	\citet{Caglar2020}	\\
MCG-06-30-015	&	H+K	&	0.125	&	\citet{Caglar2020}	\\
NGC613	&	K	&	0.050	&	0102.B-0365 (PI: Erwin)	\\
NGC628	&	K	&	0.125	&	\citet{Hic13}	\\
NGC1068	&	H+K	&	0.050	&	\citet{FMS2009}	\\
NGC1326	&	K	&	0.125	&	\citet{Fazeli2020a}	\\
NGC1365	&	H+K	&	0.050	&	\citet{Bur15}	\\
NGC1433	&	K	&	0.125	&	\citet{Smajic2015}	\\
NGC1566	&	K	&	0.125	&	\citet{Bur15}	\\
NGC1672	&	K	&	0.125	&	\citet{Fazeli2020b}	\\
NGC1808	&	K	&	0.125	&	\cite{Busch2017}	\\
NGC2110	&	H+K	&	0.050	&	\citet{Caglar2020}	\\
NGC2992	&	H+K	&	0.050	&	\citet{Caglar2020}	\\
NGC3081	&	H+K	&	0.050	&	\citet{Caglar2020}	\\
NGC3227	&	K	&	0.125	&	\citet{Davies2007}	\\
NGC3281	&	H+K	&	0.050	&	\citet{Bur15}	\\
NGC3351	&	K	&	0.125	&	\citet{Mazzalay2013}	\\
NGC3368	&	K	&	0.050	&	\citet{Nowak2010}	\\
NGC3783	&	H+K	&	0.050	&	\citet{Caglar2020}	\\
NGC4261	&	K	&	0.125	&	\citet{Bur15}	\\
NGC4388	&	K	&	0.050	&	\citet{Bur15}	\\
NGC4435	&	K	&	0.050	&	\citet{Bur15}	\\
NGC4438	&	K	&	0.050	&	\citet{Bur15}	\\
NGC4501	&	K	&	0.050	&	\citet{Bur15}	\\
NGC4569	&	K	&	0.050	&	\citet{Bur15}	\\
NGC4579	&	K	&	0.050	&	\citet{Bur15}	\\
NGC4593	&	H+K	&	0.050	&	\citet{Caglar2020}	\\
NGC4826	&	K	&	0.125	&	078.B-0289 (PI: Israel)	\\
NGC5236	&	K	&	0.050	&	078.B-0505 (PI: Davies)	\\
NGC5506	&	H+K	&	0.050	&	\citet{Caglar2020}	\\
NGC5643	&	K	&	0.125	&	\citet{Hic13}	\\
NGC5728	&	H+K	&	0.050	&	\citet{Caglar2020}	\\
NGC6300	&	K	&	0.125	&	\citet{Hic13}	\\
NGC6814	&	H+K	&	0.050	&	\citet{Caglar2020}	\\
NGC7172	&	H+K	&	0.050	&	\citet{Caglar2020}	\\
NGC7213	&	H+K	&	0.050	&	\citet{Caglar2020}	\\
NGC7465	&	Kbb	&	0.035	&	\citet{FMS2018}	\\
NGC7582	&	H+K	&	0.050	&	\citet{Caglar2020}	\\
  \hline
  \hline
  \noalign{\smallskip}  
   \multicolumn{4}{c}{non-AGN sample} \\
  \hline
\hline						
IC5332	&	K	&	0.125	&	095.B-0660 (PI: Georgiev)	\\
NGC718	&	H+K	&	0.050	&	\citet{Caglar2020}	\\
NGC1079	&	H+K	&	0.125	&	0102.B-0365 (PI: Erwin)	\\
NGC3175	&	H+K	&	0.050	&	\citet{Caglar2020}	\\
NGC3621	&	K	&	0.050	&	60.A-9800	\\
NGC5845	&	H+K	&	0.050	&	\citet{Caglar2020}	\\
NGC7727	&	H+K	&	0.050	&	\citet{Caglar2020}	\\
\hline 
\hline
\end{tabular}} \label{Tab5}
\end{table*}

\section{Estimate of $L_{\rm X}$ for a fraction of the lower luminosity sample galaxies}\label{Lx}

In order to complete the X-ray luminosities for the sources in our sample, we checked in the \emph{XMM}-Newton and \emph{Chandra} archives for observations when the objects were not included in the BASS sample. We found \emph{XMM}-Newton observations for five objects (namely NGC\,3718, NGC\,4321, NGC\,4569, NGC\,4579 and NGC\,6951) and \emph{Chandra} for six sources (namely NGC\,2782, NGC\,3718, NGC\,4321, NGC\,4569, NGC\,4579, and NGC\,5953). Four targets have been observed with both satellites. Unfortunately, after processing the data, the observations of NGC\,5953 and NGC\,6951 resulted in less than 5 ksec that were not enough to obtain a spectrum for these two sources. Only one \emph{XMM}-Newton observations was available for each of the targets. Among \emph{Chandra} observations, we chose the one with the longest exposure time when more than one observation is available. Table~\ref{Tab4} contains the information of the five sources with available spectra together with the spectral fitting results.

\emph{Chandra} and \emph{XMM}-Newton observations were processed using the standard pipelines CIAO and SAS, respectively. We extracted the nuclear spectrum, coincident with the coordinates of the targets using a 0.5 and 20 arcsec radial aperture, respectively. A background region was manually selected in a source-free region close to the target extraction. We grouped the spectra with a minimum of 20 counts per bin using the command line {\sc grppha} within {\tt Heasoft} software. We then used the latest version of the  {\tt XSPEC} software to perform spectral fitting.

Our simplest model accounts for partial observed intrinsic emission. This in the nomenclature of the software is written as \emph{pcfabs $\times$ powerlaw} (M1 in Table~\ref{Tab4}). Clear contribution from extended host diffuse emission is 
found for most of the targets when using \emph{XMM}-Newton observations due to the poorer spatial resolution compared to \emph{Chandra}. To account for this emission, we added a thermal component to the model when needed. This 
new baseline model is written in the nomenclature of the software as \emph{apec +   pcfabs $\times$ powerlaw} (M2 in Table~\ref{Tab4}). In two sources (NGC\,2782 and NGC\,4579), we found clear signatures of the $\rm{FeK\alpha}$ emission 
line. In these two case we added to M2, a reflection component to account for this line. This final baseline model is written in the software nomenclature of the software as \emph{pexmon + apec +  pcfabs $\times$ powerlaw} (M3 in 
Table~\ref{Tab4}). When both satellites are available, we link all the parameters between observations except for the normalizations of the intrinsic continuum (to allow plausible changes in flux). No significative variations are found in our analysis. 
The photon index was poorly constrained in three sources. In these cases, we fixed it to the standard value of $\rm{\Gamma = 1.8}$. Although some objects do show a complex behavior that might require additional components, this 
analysis is good enough to reproduce the data in order to obtain the X-ray observation corrected luminosity used in this work.

The existence of the $\rm{FeK\alpha}$ line and/or the reflection component confirm the AGN nature except for NGC~4321 and NGC~4569, where the luminosity and spectral shape could be consistent with a stellar nature. Indeed, NGC~4321 was classified as non-AGN by \citet{Gon09}. Therefore, the AGN nature of these two sources is still uncertain.

\begin{table*}[tbh!]
\caption{$L_{\rm X}$ values obtained from X-ray spectral fitting.}
\centering
\resizebox{1\textwidth}{!}{ 
\begin{tabular}{lccccccccc} 
\hline
\hline
\noalign{\smallskip} 
       Source & OBSID             &  Exptime   & Model &  log$_{\rm 10}(L_{\rm X})$           &  kT(keV)                       & $\rm{\Gamma}$ 		& $N_{\rm H}$            	 &	 $f_{\rm cov}$     &  $\rm{\chi / dof}$ \\ 
NGC2782  &  3014               &  29.58       & M3      & ${39.5_{0.5}^{0.3}}$        &  ${1.02\rm0.05}$          &   1.8$^{*}$        		& $\rm{3.6_{-3.5}^{+6.6}}$ & 	$\rm{0.8\pm0.2}$  & 13.6/19    \\
NGC3718  &  0795730101/3993  & 16.14/0.5 & M1    & ${40.64\pm 0.01}$/${40.84\pm 0.01}$      &       &            $\rm{1.7\pm0.1}$   &$\rm{0.90\pm0.05}$       & $\rm{0.98\pm0.01}$ & 182.0/200  \\ 
NGC4321  &  0106860201/14230 & 1.0/78.8  & M2    & ${40.4\pm 0.2}$/${39.2_{-0.2}^{+0.7}}$   &  ${0.65_{-0.03}^{+0.09}}$ &   1.8$^{*}$        &$\rm{76_{-38}^{+113}}$   & $\rm{0.6\pm0.2}$   & 107.1/86   \\ 
NGC4569  &  0200650101/5911  & 40.0/29.9 & M2    & ${39.64\pm 0.06}$/${39.22\pm 0.07}$      &  ${0.76\rm0.01}$          &   1.8$^{*}$        &$\rm{16.8\pm0.5}$        & $\rm{0.4\pm0.1}$   & 221.3/152 \\ 
NGC4579  &  0790840201/807   & 17.1/31.4 & M3    & ${41.42\pm 0.01}$/${41.29\pm 0.01}$      &  ${0.84\rm0.06}$          & $\rm{1.76\pm0.01}$ &$\rm{1.20\pm0.05}$       & $\rm{0.68\pm0.01}$ & 1081.0/994 \\ 
\noalign{\smallskip} 
\hline 
\hline
\end{tabular}}
\tablefoot{Table of the results for the X-ray spectral fitting. $N_{\rm H}$ are in units of $\rm{10^{22}cm^{-2}}$. M1: \emph{pcfabs~$\times$~powerlaw}, M2: \emph{apec +   pcfabs~$\times$~powerlaw}, M3: \emph{apec + pexmon +   pcfabs$\times$~powerlaw}.} \label{Tab4} 
\end{table*}

\section{Molecular gas surface densities and concentration indices}

We list in Table~\ref{Tab2} the cold and hot molecular gas surface densities and concentration indices ($CCI$ and $HCI$) obtained for the galaxies of the sample used in this paper.

\begin{table*}[h!]
\caption{Gas surface densities and concentration indices in the galaxies of our combined sample.}
\centering
\resizebox{8.6cm}{!}{ 
\begin{tabular}{lccccccc} 
\hline
\hline
\noalign{\smallskip} 
  Name        &  D  & log$_{10}$$\Sigma^{\rm gas}_{50}$ &  log$_{10}$$\Sigma^{\rm gas}_{200}$  & $CCI$ & log$_{10}$$\Sigma^{\rm hot}_{50}$   & log$_{10}$$\Sigma^{\rm hot}_{200}$ & $HCI$ \\
\noalign{\smallskip}  
\hline
\noalign{\smallskip}
   ----        &  Mpc  & M$_{\sun}$pc$^{-2}$  &  M$_{\sun}$pc$^{-2}$  &  ---- &   M$_{\sun}$pc$^{-2}$  &  M$_{\sun}$pc$^{-2}$  & ---- \\
\noalign{\smallskip}  
	   \hline
	   \hline
 \noalign{\smallskip} 
 \multicolumn{8}{c}{AGN sample: CO(3-2)} \\
 	   \hline
	   \hline
 \noalign{\smallskip} 
      NGC1566    &   7.2    &   2.58    &   1.89    &   0.69    &  -3.10    &  -3.54    &   0.44 	\\
      NGC1808    &   9.3    &   2.68    &   2.20    &   0.48    &  -2.83    &  -3.23    &   0.40 	\\
      NGC1672    &  11.4    &   3.16    &   2.22    &   0.83    &  -2.65    &  -3.20    &   0.55 	\\
      NGC1068    &  14.0    &   2.69    &   2.94    &  -0.25    &  -1.41    &  -1.61    &   0.20 	\\
      NGC6300    &  14.0    &   3.09    &   2.42    &   0.68    &  -3.19    &  -3.93    &   0.74 	\\
      NGC1326    &  14.9    &   2.51    &   1.77    &   0.74    &  -2.79    &  -3.47    &   0.68 	\\
      NGC5643    &  16.9    &   3.17    &   2.61    &   0.56    &  -2.54    &  -3.38    &   0.85 	\\
      NGC613      &  17.2    &   3.22    &   2.51    &   0.82    &  -1.81    &  -2.65    &   0.85 	\\
      NGC7314    &  17.4    &   2.22    &   1.53    &   0.69    & ---    & ---    & --- 	\\
      NGC4388    &  18.1    &   1.97    &   2.13    &  -0.17    &  -2.47    &  -2.64    &   0.17 	\\
      NGC1365    &  18.3    &   2.22    &   1.52    &   0.70    &  -4.91    &  -4.18    &  -0.73 	\\
      NGC4941    &  20.5    &   2.21    &   1.42    &   0.79    & ---    & ---    & --- 	\\
      NGC7213    &  22.0    & <1.18    &   0.63    &   <0.55    &  -2.98    &  -3.87    &   0.89 	\\
      NGC7582    &  22.5    &   2.77    &   2.93    &  -0.16    &  -3.35    &  -3.53    &   0.18 	\\
      NGC6814    &  22.8    &   1.66    &   0.96    &   0.70    &  -2.35    &  -3.21    &   0.86 	\\
      NGC3227    &  23.0    &   2.88    &   2.74    &   0.14    &  -2.60    &  -3.11    &   0.50 	\\
      NGC5506    &  26.4    &   2.42    &   2.34    &   0.07    &  -2.03    &  -2.47    &   0.43 	\\
      NGC7465    &  27.2    &   2.65    &   2.07    &   0.57    &  -2.28    &  -2.78    &   0.50 	\\
      NGC7172    &  37.0    &   1.81    &   1.97    &  -0.16    &  -5.53    &  -5.68    &   0.15 	\\
      NGC5728    &  44.5    &   2.65    &   2.55     &   0.10    &  -4.86    &   -5.31   &   0.45      \\ 
   \noalign{\smallskip}
  \hline
  \hline
 \noalign{\smallskip}  
   \multicolumn{8}{c}{AGN sample: CO(2-1)} \\
 	   \hline
	   \hline
      NGC4826    &   4.4   &   2.67    &   2.28    &   0.39    &  -2.92    &  -3.19    &   0.27 \\
      NGC5236    &   4.9    &   2.87    &   2.62    &   0.25    & ---    & ---    & --- \\
      NGC2903    &  10.0    &   2.40    &   2.31    &   0.09    & ---    & ---    & --- \\
      NGC3351    &  10.0    &   2.52    &   2.21    &   0.32    &  -3.31    &  -3.77    &   0.46 \\
      NGC1637    &  11.7    &   2.68    &   2.20    &   0.48    & ---    & ---    & --- \\
      NGC4569    &  16.8    &   2.41    &   2.31    &   0.10    &  -2.90    &  -2.84    &  -0.06 \\
      NGC4579    &  16.8    &   1.76    &   1.45    &   0.32    &  -1.89    &  -2.62    &   0.73 \\
      NGC3718    &  17.0    &   1.26    &   0.80    &   0.46    & ---    & ---    & --- \\
      NGC2110    &  34.8    &   <1.39    &   1.41    &  <-0.02    &  -5.09    &  -5.32    &   0.23 \\
      NGC2782    &  35.0    &   2.77    &   2.34    &   0.42    & ---    & ---    & --- \\
      NGC7172    &  37.0    &   2.07    &   2.16    &  -0.08    &  -5.53    &  -5.68    &   0.15 \\
      MCG-06-30-15    &  38.3    &   2.27    &   2.05    &   0.22    &  -3.33    &  -3.38    &   0.05 \\
      NGC2992    &  39.2   &   1.70    &   1.99    &  -0.30    &  -5.73    &  -5.79    &   0.06 \\
      NGC3081    &  40.3    &   2.03    &   1.88    &   0.14    &  -5.84    &  -5.75    &  -0.09 \\
      ESO137-34    &  41.5    &   2.42    &   2.21    &   0.21    &  -2.21    &  -2.45    &   0.24 \\
      NGC5728    &  44.5    &   2.68    &   2.64    &   0.04    &  -4.86    &  -5.31    &   0.45 \\ 
      ESO21-g004    &  45.1    &   1.70    &   1.57    &   0.13    & ---    & ---    & --- \\
      NGC4826    &   4.4    &   2.71    &   2.23    &   0.47    &  -2.92    &  -3.19    &   0.27 \\
      NGC3351    &  10.0    &   2.50    &   1.91    &   0.55    &  -3.31   &  -3.57    &   0.46 \\
      NGC4501    &  14.0    &   2.73    &   2.36    &   0.37    &  -2.88    &  -3.09    &   0.21 \\
      NGC4438    &  16.5    &   2.75    &   2.49    &   0.25    &  -1.87    &  -2.38    &   0.51 \\
      NGC3368    &  18.0    &   2.93    &   2.67    &   0.25    &  -2.48    &  -2.94    &   0.46 \\ 
      NGC7172    &  37.0    &   2.10    &   2.16    &  -0.06    &  -5.53    &  -5.68    &   0.15 \\
 \noalign{\smallskip}
  \hline
  \hline
 \noalign{\smallskip}  
   \multicolumn{8}{c}{AGN sample: CO(1-0)} \\
 	   \hline
	   \hline
                 M51   &   8.6    &   2.84    &   2.18    &   0.66    & ---    & ---    & --- \\
      NGC6300    &  14.0    &   3.48    &   2.64    &   0.84    &  -3.19    &  -3.93    &   0.74 \\
      NGC4321    &  16.8    &   3.35    &   2.91    &   0.44    & ---    & ---    & --- \\ 
      NGC5643    &  16.9    &   2.75    &   2.38    &   0.37    &  -2.54    &  -3.38    &   0.85 \\
      NGC7314    &  17.4    &   2.86    &   2.31    &   0.55    & ---    & ---    & --- \\
      NGC4388    &  18.1    &   2.31    &   2.43    &  -0.12    &  -2.47    &  -2.64    &   0.17 \\
      NGC6221    &  22.9    &   3.33    &   3.03    &   0.30    & ---    & ---    & --- \\
      NGC3227    &  23.0    &  < 2.33    &   2.49    &  <-0.15    &  -2.60    &  -3.11    &   0.50 \\
      NGC4180    &  36.0    &   3.22    &   2.79    &   0.43    & ---    & ---    & --- \\
      NGC7172    &  37.0    &   2.69    &   2.72    &  -0.04    &  -5.53    &  -5.68    &   0.15 \\
      NGC4593    &  41.8    &   2.78    &   2.65    &   0.14    &  -4.85    &  -5.65    &   0.80 \\
      NGC1125    &  42.6    &  < 2.79    &   2.60    &   <0.19    & ---    & ---    & --- \\
      NGC5728    &  44.5    &   2.99    &   3.05    &  -0.06    &  -4.86    &  -5.31    &   0.45 \\ 
      NGC3281    &  52.0    &   3.00    &   2.89    &   0.10    &  -2.29    &  -2.38    &   0.09 \\       
  \hline
  \hline
  \noalign{\smallskip}  
   \multicolumn{8}{c}{non-AGN sample: CO(2-1)} \\
  \hline
\hline
      NGC5068    &   5.2    &   0.75    &   0.29    &   0.46    & ---    & ---    & --- \\
      NGC3621    &   7.1    &   0.68    &   0.54    &   0.14    &  ---    &  ---    & --- \\
       IC5332    &   9.0    &   0.25    &  -0.01    &   0.25    &  ---    &  ---    & --- \\
      NGC3596    &  11.3    &   1.67    &   1.58    &   0.09    & ---    & ---    & --- \\
      NGC4781    &  11.3    &   1.38    &   1.11    &   0.27    & ---    & ---    & --- \\
      NGC2835    &  12.2    &   1.24    &   0.99    &   0.25    & ---    & ---    & --- \\
      NGC5530    &  12.3    &   1.54    &   1.13    &   0.41    & ---    & ---    & --- \\
      NGC1947    &  17.2    &   3.46    &   2.70    &   0.76    & ---    & ---    & --- \\
      NGC1079    &  18.9    &   2.12    &   1.63    &   0.48    &  ---   &  ---    & --- \\
      NGC3717    &  19.1    &   2.88    &   2.49    &   0.39    & ---    & ---    & --- \\
      NGC7727    &  21.0    &   2.36    &   2.02    &   0.34    &  -2.97    &  -3.45    &   0.48 \\
      NGC5921    &  21.0    &   2.73    &   2.26    &   0.47    & ---    & ---    & --- \\
      NGC3175    &  21.0    &   2.67    &   2.50    &   0.17    &  -3.30    &  -3.55    &   0.25 \\
       NGC718    &  21.4    &   2.18    &   1.78    &   0.41    &  -2.98    &  -3.77    &   0.79 \\
      NGC5845    &  27.0    &   1.73    &   1.01    &   0.62    &  ---    &  ---    & --- \\
      ESO-093-g003    &  29.8    &   1.97    &   1.99    &  -0.02    & ---    & ---    & --- \\
      NGC5037    &  35.0    &   2.59    &   2.34    &   0.25    & ---    & ---    & --- \\
      NGC3749    &  41.0    &   2.84    &   2.63    &   0.20    & ---    & ---    & --- \\
      NGC4224    &  41.0    &   1.86    &   1.70    &   0.16    & ---    & ---    & --- \\
\noalign{\smallskip} 
\hline 
\hline
\end{tabular}}
\tablefoot{Columns (3) and (4) list the surface densities of the cold molecular derived from different CO lines on the nuclear ($r\leq50$~pc; $\Sigma^{\rm gas}_{50}$) and circumnuclear  ($r\leq200$~pc; $\Sigma^{\rm gas}_{200}$) disk scales of the AGN and non-AGN galaxies of the sample. The target names and their distances are listed in columns (1) and (2).  The concentration indices of cold molecular gas ($CCI$) are listed in column (5). Columns (6), (7) and (8) list the corresponding surface densities and concentration indices ($\Sigma^{\rm hot}_{50}$, $\Sigma^{\rm hot}_{200}$, and $HCI$) derived for the hot molecular gas component.} \label{Tab2} 
\end{table*}

\section{Other trends of the cold molecular gas concentration index}

\begin{figure}[ht!]
  \centering
   \includegraphics[width=8.5cm]{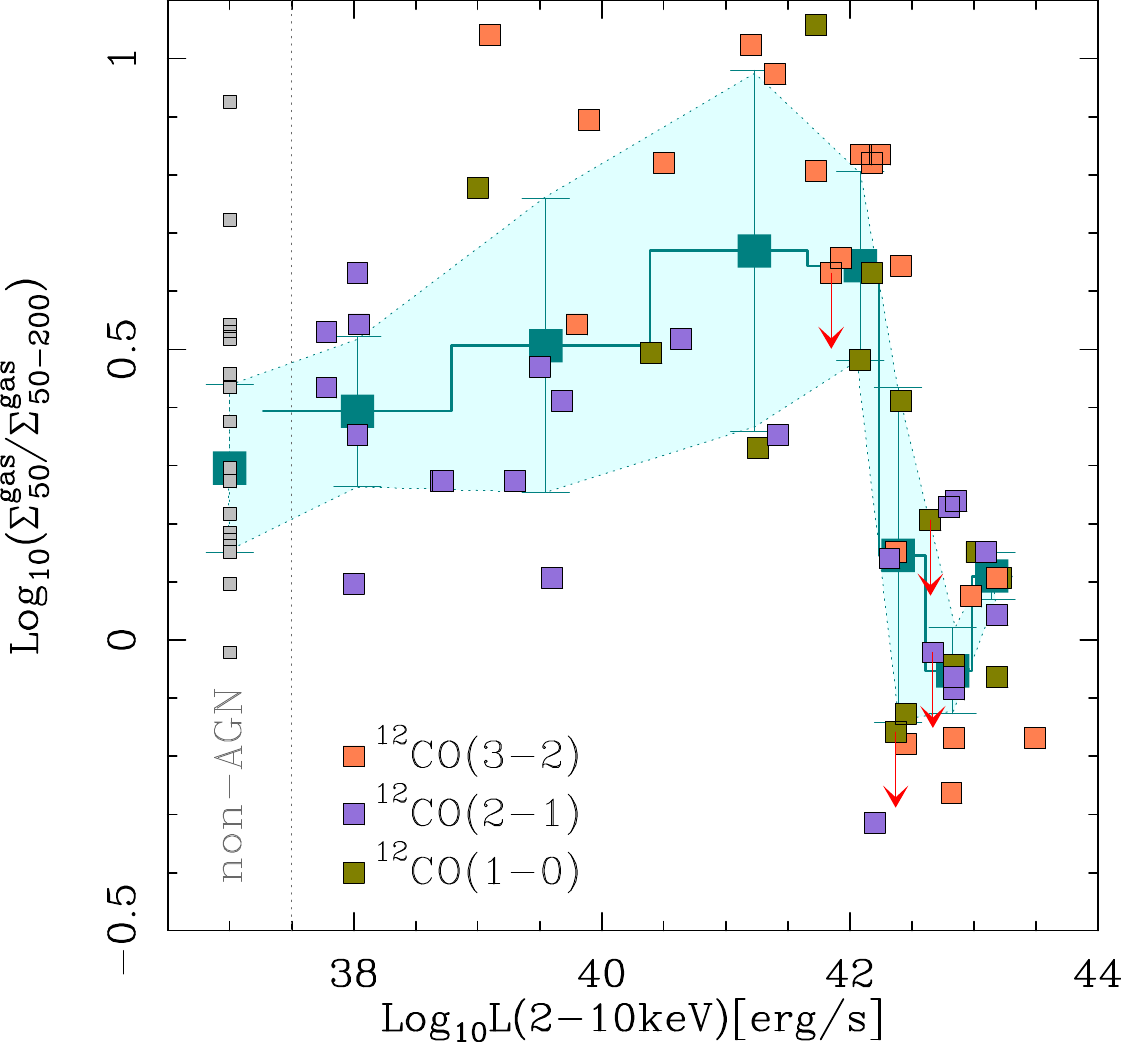}
    \includegraphics[width=8.5cm]{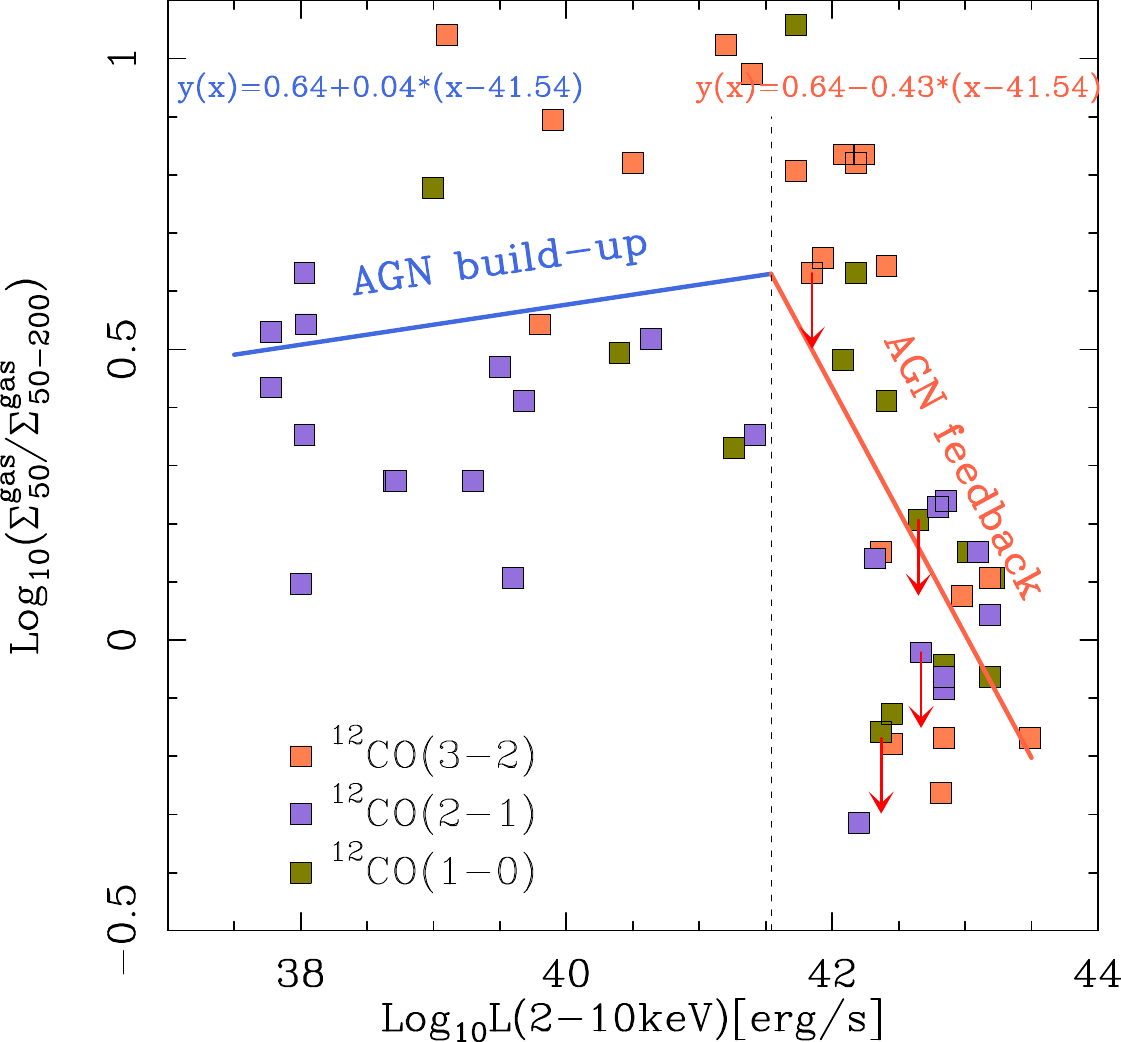}
      
       \caption{Same as Fig.\ref{conc-CO} but defining the  circumnuclear  scales as the outer corona 
$r\leq50-200$pc ($\Sigma^{\rm gas}_{\rm 50-200}$).}  
   \label{conc-CO-bis}
    \end{figure}

Fig.~\ref{conc-CO-bis} shows the trend with $L_{\rm X}$ of an alternative definition of the concentration of cold molecular gas, which adopts the region defined by the outer corona of the CND ($50$~pc$\leq r \leq 200$~pc) to  normalize  the molecular surface density of the nuclear region (r$\leq50$~pc).

\begin{figure}[tb!]
  \centering
    \includegraphics[width=8.5cm]{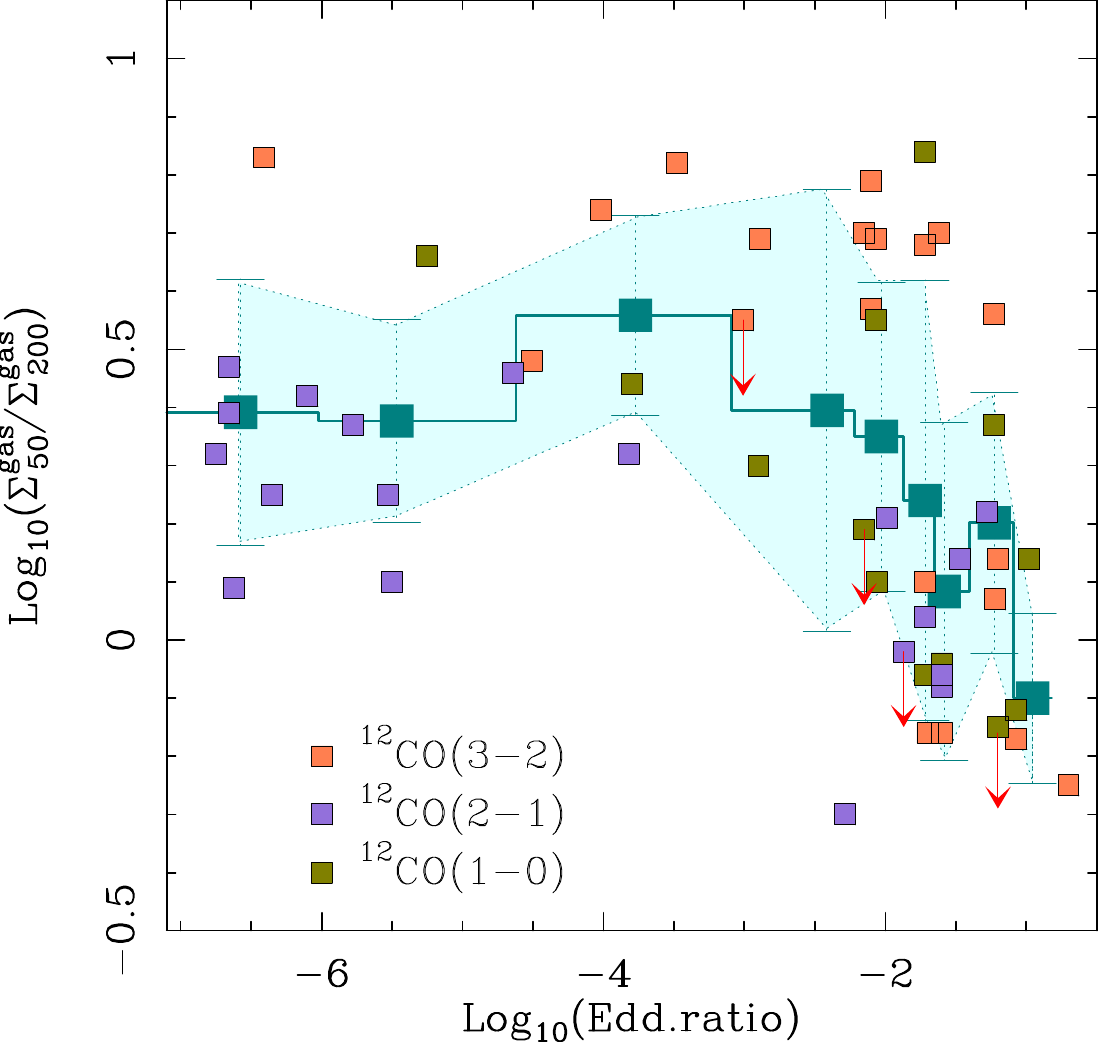}
      
       \caption{Same as Fig.\ref{conc-CO} but as a function of the Eddington ratio.}
   \label{Edd_ratio}
    \end{figure}

Fig.~\ref{Edd_ratio} shows the $CCI$ values as a function of the Eddington ratio for the 27 galaxies in our sample  for which an estimate of $\lambda_{\rm Edd}$ is available. 

\section{CND-scale distribution of molecular gas from different CO line transitions}\label{CO-lines}

  \begin{figure}[tb!]
  \centering
    \includegraphics[width=9cm]{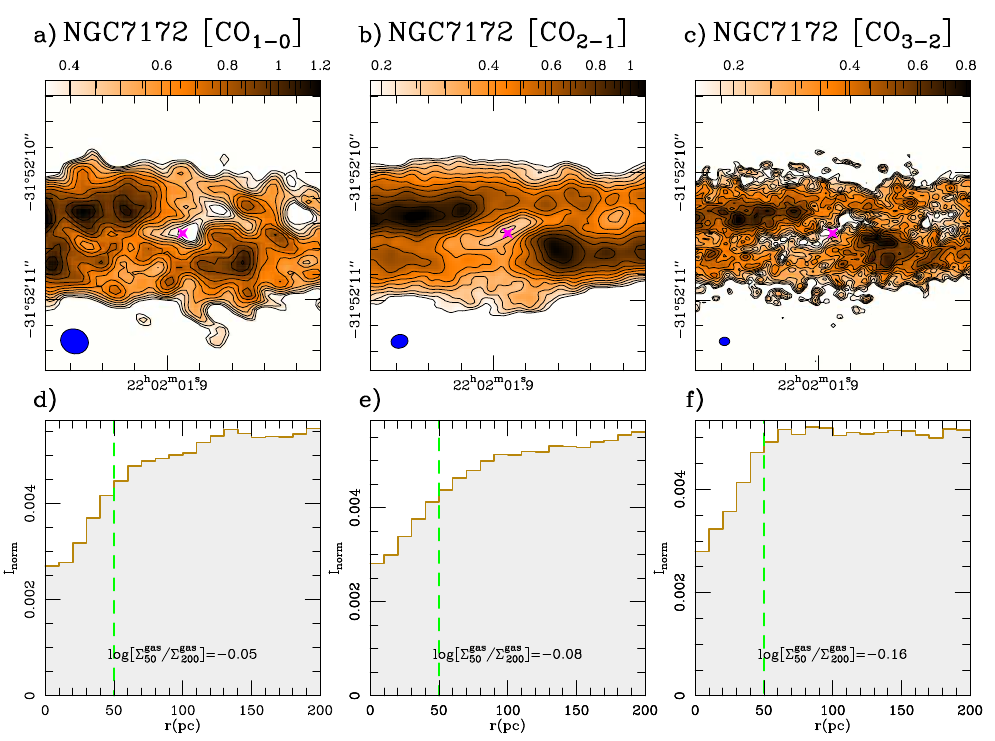}
       \caption{{\em Upper panels}:~velocity-integrated intensity maps derived from the  CO(1--0), CO(2--1), and CO(3--2) lines in the central 400~pc$\times$400~pc region of NGC~7172.   Contour levels have a logarithmic spacing from 2.5$\sigma$ to 90$\%$ of the peak CO intensity inside the displayed field of view.   The (blue) filled ellipses  at the bottom left corners represent the beam sizes of the observations.   {\em Lower panels}:~Normalized radial distributions of molecular gas derived out to $r=200$~pc from the 3 CO maps of NGC~7172 shown in the {\rm upper panels}. The estimated molecular gas concentration indices are displayed in panels {\em d)}-to-{\em f)}.}  
   \label{maps-ngc7172}
    \end{figure}
  \begin{figure}[tb!]
  \centering
    \includegraphics[width=9cm]{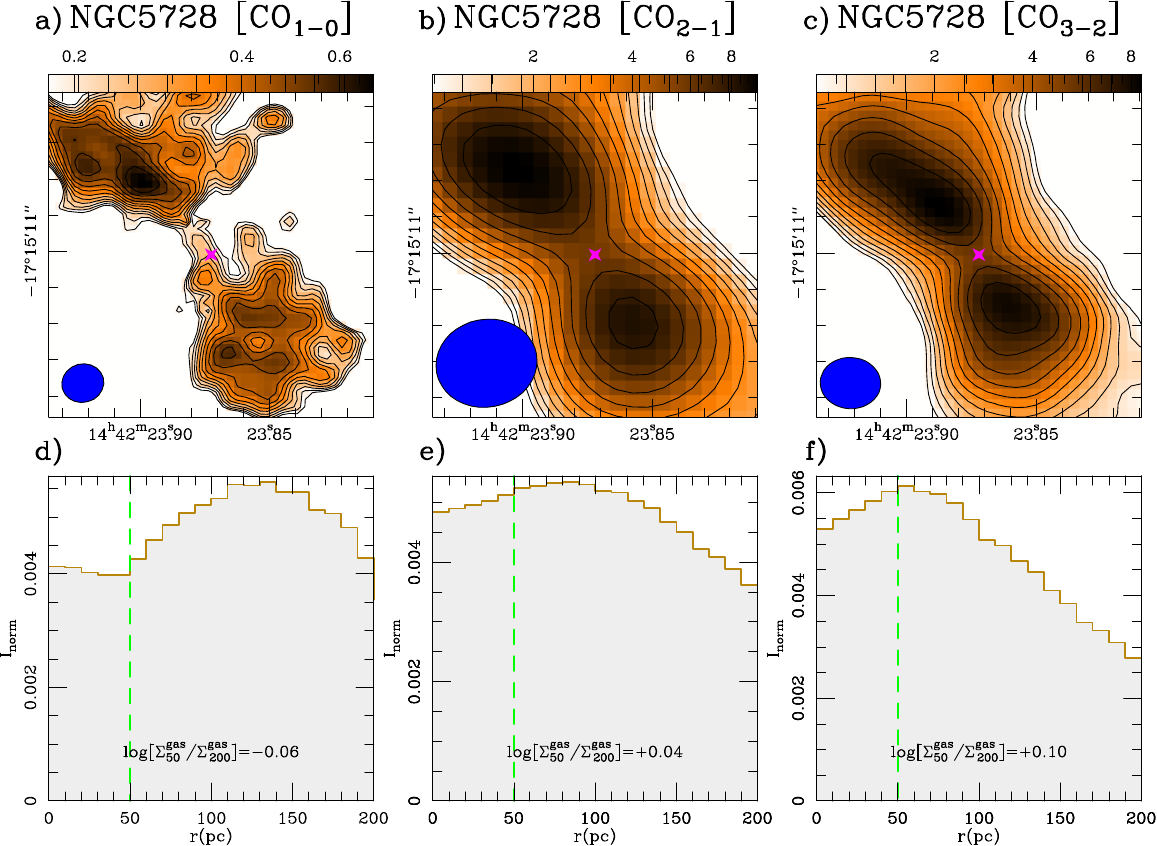}
       \caption{Same as Fig.~\ref{maps-ngc7172} but showing the maps and normalized radial distributions for NGC~5728.}  
   \label{maps-ngc5728}
    \end{figure}

Figures~\ref{maps-ngc7172} and \ref{maps-ngc5728} show the images of the CND of the AGN feedback branch targets NGC~7172 and NGC~5728 obtained in the 3--2, 2--1, and 1--0 transitions of CO.  In spite of the different spatial resolutions used in these observations, the derived 2D distributions, normalized radial profiles, and $CCI$ values show a weak dependence on the CO transition used to image the distribution of molecular gas  in these sources.


\end{appendix}

\end{document}